\documentclass[twocolumn]{svjour3} 
\usepackage{url}
\usepackage{graphicx}
\usepackage{balance}
\usepackage{subfigure}
\usepackage{color}
\usepackage{times,amsmath,epsfig}
\usepackage{amssymb}
\usepackage{textcomp}
\usepackage[linesnumbered,algoruled,boxed,lined]{algorithm2e}
\usepackage{cite}
\usepackage{todonotes} 

\newcommand{\y}[1]{{\color{blue} #1}\normalcolor}
\newcommand{\yy}[1]{{\color{magenta} #1}\normalcolor}

\newcommand{\remove}[1]{}

\newcommand{\ignore}[1]{}

\usepackage{booktabs} % For formal tables

\begin{document}
	
\title {ParIS+: Data Series Indexing on Multi-Core Architectures}
%
%\author{Botao PENG}
%\affiliation{%
%  \institution{LIPADE, Paris Descartes University}
% \streetaddress{45 rue des saints peres}
%  \city{Paris}
%  \state{France}
%  \postcode{75006}
%}
%\email{botao.peng@parisdescartes.fr}

%\author{Panagiota Fatourou}
%\affiliation{%
%  \institution{Department of Computer Science, University of Crete \& FORTH ICS}
 % \streetaddress{45 rue des saints peres}
  %\city{Paris}
  %\state{France}
  %\postcode{75006}
%}
%\email{faturu@csd.uoc.gr}
%\author{Themis Palpanas}
%\affiliation{%
% \institution{LIPADE, Paris Descartes University}
% \streetaddress{45 rue des saints peres}
 %\city{Paris}  
 %\state{France}
 % \postcode{75006}
%}
%\email{themis@mi.parisdescartes.fr}

%\numberofauthors {3}

%author{
%	\alignauthor 
%	Botao PENG\\
%	\affaddr{LIPADE, Universit{\'e} de Paris}\\
	%\affaddr{45 rue des saints peres}\\
	%\affaddr{Paris, France}\\
%	\email {\large botao.peng@hotmail.com}
%	\alignauthor 
%	Panagiota Fatourou\\
%	\affaddr{Department of Computer Science, University of Crete \& FORTH ICS}\\
	%\affaddr{45 rue des saints peres}\\
	%\affaddr{Paris, France}\\
%	\email {\large faturu@csd.uoc.gr}
%	\alignauthor
%	Themis Palpanas\\
%	\affaddr{LIPADE, Universit{\'e} de Paris}\\
	%\affaddr{45 rue des saints peres}\\
	%\affaddr{Paris, France}\\
%	\email{\large themis@mi.parisdescartes.fr}
%}

\author{Botao Peng         \and
	Panagiota Fatourou\and
	Themis Palpanas
}


\institute{F. Author \at
	LIPADE, Universit{\'e} de Paris \\
	\email{botao.peng@u-paris.fr}           %  \\
	\and
	P. Fatourou \at
	FORTH ICS
	\email { faturu@csd.uoc.gr}
	\and
	T. Palpanas \at
		LIPADE, Universit{\'e} de Paris \\
	\email{ themis@mi.parisdescartes.fr} 
}

\date{Received: date / Accepted: date}

\maketitle

\begin{abstract}
Data series similarity search is a core operation for several data series analysis applications across many different domains. 
Nevertheless, even state-of-the-art techniques cannot provide the time performance required for large data series collections.
%In this work, 
We propose ParIS and ParIS+, the {\em first} disk-based data series indices carefully designed to inherently take advantage of 
%both adaptive indexing techniques and 
%modern hardware parallelization, 
multi-core architectures, in order to accelerate similarity search processing times. 
Our experiments demonstrate that ParIS+
completely removes the CPU latency during index construction 
for disk-resident data, %. In terms of exact query answering, ParIS+ 
and for exact query answering is up to $1$ order of magnitude faster 
than the current state of the art index scan method, 
and up to $3$ orders of magnitude faster
than the optimized serial scan method. 
ParIS+ (which is an evolution of the ADS+ index) owes its efficiency to the effective use of multi-core 
and multi-socket architectures, 
in order to distribute and execute in parallel both index construction and query answering, and to the exploitation of the Single Instruction Multiple Data (SIMD) capabilities 
of modern CPUs, in order to further parallelize the execution of %individual 
instructions inside each core. 
\end{abstract}


%\begin{CCSXML}
%<ccs2012>
%%  <concept_id>10010520.10010553.10010562</concept_id>
 % <concept_desc>Computer systems organization~Embedded systems</concept_desc>
 %% <concept_significance>500</concept_significance>
 %</concept>
%  <concept_id>10010520.10010575.10010755</concept_id>
%  <concept_desc>Computer systems organization~Redundancy</concept_desc>
%  <concept_significance>300</concept_significance>
%  <concept_id>10010520.10010553.10010554</concept_id>
%  <concept_desc>Computer systems organization~Robotics</concept_desc>
%  <concept_significance>100</concept_significance>
%  <concept_id>10003033.10003083.10003095</concept_id>
%  <concept_desc>Networks~Network reliability</concept_desc>
%  <concept_significance>100</concept_significance>
%</ccs2012>
%\end{CCSXML}

%%\ccsdesc[500]{Computer systems organization~Embedded systems}

%\keywords{Time Series}

\section{introduction}
\label{sec:intro}
\noindent{\bf{[Motivation]}}
An increasing number of applications across many diverse domains continuously 
produce very large amounts of data series\footnote{A data series, or data sequence, 
is an ordered sequence of data points. If the ordering dimension is time then we talk about time series, 
though, series can be ordered over other measures
(e.g., angle in astronomical radial profiles, mass in mass spectroscopy, 
position in genome sequences, etc.).} (such as in finance, environmental sciences, 
astrophysics, neuroscience, engineering, multimedia, 
etc.~\cite{DBLP:journals/sigmod/Palpanas15,fulfillingtheneed,Palpanas2019,DBLP:journals/dagstuhl-reports/BagnallCPZ19}),
%~\cite{kashino1999time,ye2009time,shasha1999tuning,huijse2014computational,raza2015practical}), 
which makes them one of the most common types of data. 
When these sequence collections are generated (often times composed of a large number of short series~\cite{Palpanas2019}), users may need to query 
and analyze them as soon as they become available.
%Often times, this is part of an exploratory process, where users ask a query, review the results, and then decide what their subsequent queries, or analysis tasks should be.
%The above process is
This process is heavily dependent on data series similarity search 
(which apart from being a useful query in itself, also lies at the core 
of several machine learning methods, such as, clustering, classification, 
motif and outlier detection, etc.)~\cite{tan2017indexing,lernaeanhydra,norma,series2graph}.
The brute-force approach for evaluating similarity search queries 
is by performing a sequential pass over the complete dataset.
However, as data series collections grow larger, scanning the complete dataset 
becomes a performance bottleneck, taking hours or more to complete~\cite{zoumpatianos2016ads}. 
This is especially problematic in exploratory search scenarios, where 
%batch execution of queries is impossible because the 
every next query 
depends on the results of previous queries.
Consequently, we have witnessed an increased interest in developing indexing techniques 
and algorithms 
%that can efficiently support 
for similarity %search~\cite{ding2008querying,rakthanmanon2012searching,wang2013data,dallachiesa2014top,isax2plus}. 
search~\cite{shieh2008sax,rakthanmanon2012searching,wang2013data,isax2plus,zoumpatianos2016ads,DBLP:conf/icdm/YagoubiAMP17,DBLP:journals/pvldb/KondylakisDZP18,ulisseicde,ulissevldb,coconutpalm,dpisaxjournal,lernaeanhydra,lernaeanhydra2,evolutionofanindex}.

\noindent{\bf [Scalability problem]}
Nevertheless, the continued increase in the rate and volume 
of data series production
%~\cite{DBLP:journals/sigmod/Palpanas15} 
with collections that grow to several terabytes~\cite{DBLP:journals/sigmod/Palpanas15}
%~\cite{url:adhd,url:sds,DBLP:journals/pvldb/PelkonenFCHMTV15}, 
renders existing data series indexing technologies inadequate. 
For example, the current state-of-the-art index, ADS+~\cite{zoumpatianos2016ads,DBLP:conf/edbt/GogolouTPB19}, 
requires more than 4min to answer any single exact query on a moderately sized 250GB sequence collection. 
Moreover, index construction time also becomes a significant bottleneck in the analysis process~\cite{zoumpatianos2016ads}, especially in cases where new data arrive frequently and need to be indexed~\cite{Palpanas2019}.
Thus, traditional solutions and systems are inefficient at, 
or incapable of managing and processing the voluminous sequence collections 
that already exist in several domains. %and applications.
Finally, we note that, given the evolution of CPU performance, 
where the processor clock speed is not increasing due to the power wall constraint, 
efforts for algorithmic speedups now exploit the parallelism opportunities 
offered by modern 
%hardware~\cite{akhter2006multi,herlihy2011art,blanas2011design,xiao2013parallelizing,ailamaki2015databases}. 
hardware~\cite{xiao2013parallelizing,ailamaki2015databases,DBLP:conf/ieeehpcs/Palpanas17,messi,seriesgoneparallel}. 

\noindent{\bf [Parallel Indexing]}
In this work, we propose the Parallel Index for Sequences (ParIS), 
the first data series index that inherently takes advantage of modern hardware parallelization, 
and incorporate the state-of-the-art techniques 
in sequence indexing, in order to accelerate processing times. 
ParIS, which is a disk-based index based on the principles of ADS+, takes advantage of multi-core 
and multi-socket architectures, in order to distribute and execute in parallel the computations needed for both index construction and query answering. 
Moreover, ParIS uses the Single Instruction Multiple Data (SIMD) CPU instructions, 
in order to further parallelize the execution of individual instructions inside each core. 
Overall, ParIS achieves very good overlap of the CPU computation with the required disk I/O. 
To completely remove the CPU cost during index creation,  
we present ParIS+, an alternative of ParIS that results in index creation that is purely I/O bounded.
ParIS+ is 2.6x faster than the current state-of-the-art approach~\cite{zoumpatianos2016ads}.
ParIS and ParIS+ employ the same algorithmic techniques for query answering.  
The experiments also demonstrate their effectiveness in exact query answering:
they are up to 1 order of magnitude faster than the state-of-the-art index scan method~\cite{zoumpatianos2016ads}, and up to 3 orders of magnitude faster 
than the state-of-the-art optimized serial scan~\cite{rakthanmanon2012searching}.
We also note that ParIS and ParIS+ have the potential to deliver more benefit as we move to faster storage media.
%and that it has the potential to deliver even more benefit when we move 
%to faster storage mediums for dataset storage.

%in order to achieve these goals, we had to make careful design choices in the coordination 
In developing ParIS+ (and ParIS), we made careful design choices in the coordination 
of the compute and I/O tasks, and consequently, developed new algorithms for the construction 
of the index and for answering similarity search queries on this index.
%For query answering in particular, we studied alternative solutions, and evaluated the tradeoff between execution speed and the amount of communication among the parallel worker threads, which affects the effectiveness of each individual worker.
%Our study shows that each one of the alternative algorithms is suitable for different hardware platforms. 
%\here{Y: Do we provide justification below? ?If yes, where?}
%to deduce our experimental results we performed a comprehensive experimental analysis, based on a variety of synthetic benchmarks and real datasets.

We note that even though scaling out to multiple machines is also a valid research direction~\cite{DBLP:conf/ieeehpcs/Palpanas17,DBLP:conf/icdm/YagoubiAMP17,dpisaxjournal}, 
in this work, we focus on addressing the problem in the context of a single machine, so as to maximize the benefit we can get out of the hardware.
Our results can be combined with a scale-out solution.
Examining other hardware solutions, like GPUs and FPGAs are also very promising directions, but ouf of the scope of this work.
	
\noindent{\bf [Contributions]}
Our contributions\footnote{A preliminary version of this work appeared in~\cite{DBLP:conf/bigdataconf/PengFP18}.} are summarized as follows:\\
%
%\begin{itemize}
%\item
%we propose paris, a parallel adaptive data series index, designed for multi-core systems which can almost entirely remove CPU time cost, i.e., mask it behind the disk i/o cost. 
%\item
%\begin{itemize}
%\vspace*{-.2cm}
%\item
$\bullet$
We propose ParIS, the first data series index designed for multi-core architectures. 
We describe parallel algorithms for index creation and \emph{exact} query answering, which employ parallelism in reading the data from disk and processing them in the CPU. 
Moreover, we propose ParIS+, a ParIS alternative that completely masks out the CPU cost when creating the index. 
ParIS+ results in improved performance during index creation 
%in comparison to that of ParIS 
in systems that support a reasonable level of parallelism (more than four cores).\\
%at the same time, they achieve a balanced workload distribution among the worker threads, and enable these threads to exchange information (while executing) in order to reduce the amount of work to be done. 
%\item
%\vspace*{-.2cm}
%\item 
$\bullet$
In order to further speedup query answering, we exploit SIMD for complex vectorial 
computations: we develop  novel vectorized implementations for computing lower bounding distances between the Piecewise Aggregate Approximation (PAA)~\cite{keogh2001dimensionality} 
and indexable Symbolic Aggregate Approximation (iSAX)~\cite{shieh2008sax} representations.\\
%\item
%\vspace*{-.2cm}
%\item 
$\bullet$
Finally, we experimentally evaluate ParIS and ParIS+ using a variety of synthetic and real datasets. 
The results demonstrate the efficiency of the proposed approach, which is orders 
of magnitude faster for exact query answering than the state-of-the-art methods.
Moroever, the results show that, in settings of more than $4$ cores, ParIS+ completely hides the CPU time during index creation. 

%paris and paris+ have the potential to deliver more benefit as we move to faster storage mediums. % for dataset storage.
%\here{yf: i prefer if the last sentence is removed. it is mentioned earlier, so i do not see very serious reasons
%to repeat it here.}
%\end{itemize}
%\end{itemize}

%\noindent{\bf Paper Organization.}
%The rest of this paper is organized as follows.
%In Section~\ref{sec:prelim}, we provide the necessary background for the rest of this paper.
%Section~\ref{sec:indexcreation} presents our index creation algorithm 
%%for ParIS, and Section~\ref{sec:query} describes 
%and the corresponding query answering techniques. 
%In Section~\ref{sec:experiments}, we present the experimental evaluation.
%We discuss related work in Section~\ref{sec:related}, and conclude in Section~\ref{sec:conclusions} .

\section{Preliminaries}
\label{sec:prelim}

We now provide some necessary definitions, and introduce the related work on state-of-the-art data series indexing.

\subsection{Data Series and Similarity Search}

\noindent{\bf [Data Series]}
A data series, $S=\{p_1, ..., p_n\}$, is %defined as 
a sequence of points, 
where each point $p_i=(v_i,t_i)$, $1 \le i \le n$,
is associated to a real value $v_i$ and a position $t_i$.
The position corresponds to the order of this value in the sequence. 
%(in the case of time series, positions are expressed in terms of time, i.e., timestamps).
We call $n$ the \emph{size}, or \emph{length} of the data series.
We note that all the discussions in this paper are applicable 
to high-dimensional vectors, in general.
(In the case of streaming series, we first create subsequences of length $n$ using a sliding window, and then index those.)

\begin{figure}[tb]
	\centering
	\includegraphics[page=1,width=0.9\columnwidth]{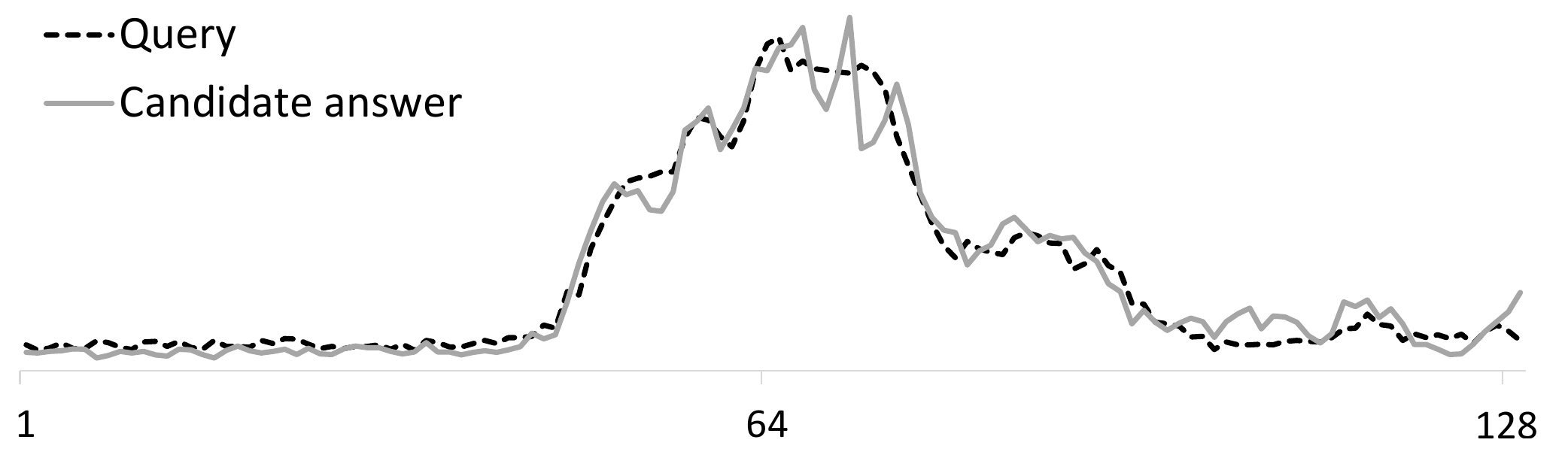}
	\caption{Query series and candidate answer (length 128; SALD dataset)}
	\label{fig:23nn}
\end{figure}

\noindent{\bf [Similarity Search]}
Analysts perform a wide range of data mining tasks on data series 
including clustering~\cite{rakthanmanon2011},
%~\cite{keogh1998,liao2005,rodrigues2008,rakthanmanon2011}, 
classification and deviation detection~\cite{Shieh2009,Shandola2009}, 
and frequent pattern mining~\cite{DBLP:journals/datamine/MueenKZCWS11}.
%~\cite{DBLP:journals/datamine/MueenKZCWS11,DBLP:journals/tkdd/GrabockaSS16}.
Existing algorithms for executing these tasks rely on performing fast similarity search 
across the different series.
Thus, efficiently processing nearest neighbor (NN) queries is crucial 
for speeding up the above tasks.
NN queries are formally defined as follows: given a query series $S_q$ of length $n$,  
and a data series collection $\mathcal{S}$ of sequences of the same length, $n$, 
we want to identify the series $S_c \in \mathcal{S}$ 
that has the smallest distance to $S_q$ among all the series in the collection $\mathcal{S}$.
Figure~\ref{fig:23nn} depicts an example of a query series and a candidate answer (the 1-NN, in this case).

Common distance measures for comparing data series are Euclidean Distance (ED)~\cite{Agrawal1993} 
and dynamic time warping (DTW)~\cite{rakthanmanon2012searching}.
While DTW is better for most data mining tasks, 
the error rate using ED converges to that of DTW 
as the dataset size grows~\cite{shieh2008sax}.
%~\cite{Ratanamahatana2005, Xiaopeng2006, shieh2008sax}.
Therefore, data series indices for massive datasets use ED 
as a distance metric~\cite{shieh2008sax,rakthanmanon2012searching,wang2013data,isax2plus,zoumpatianos2016ads}, 
though simple modifications can be applied to make them compatible with DTW~\cite{shieh2008sax}.
%~\cite{shieh2008sax,Kate2016}. 
Euclidean distance is computed as the sum of distances between the pairs of corresponding points in the two sequences.
%, where normalizing the sequences for alignment and length is a pre-processing step~\cite{shieh2008sax,Shieh2009,Zoumpatianos2014,Zoumpatianos2015rinse,ZoumpatianosIP16}.
%In all cases, data are z-normalized by subtracting the mean and dividing by the standard deviation (
Note that minimizing ED on z-normalized data (i.e., a series whose values have mean 0 and standard deviation 1) is equivalent to maximizing their Pearson's correlation coefficient~\cite{MueenNL10}. 

%\subsection{Distance calculation in SIMD}
%\textbf{Distance calculation in SIMD: }
\noindent{\bf [Distance calculation in SIMD]}
Single-Instruction-Multiple-Data (SIMD) refers to a parallel architecture that allows the execution of the same operation on multiple data simultaneously~\cite{lomont2011introduction}. 
Using SIMD, we can reduce the latency of an operation, because the corresponding instructions are fetched once, and then applied in parallel to multiple data.
%All 
%Intel
Modern CPUs support 256-bit wide SIMD vectors, 
%(AVX and AVX2) and 
%while some of them (e.g., Intel Xeon Phi) support 512-bit wide SIMD~\cite{intel2011advanced,firasta2008intel}. 
which means that some floating point (or other 32-bit data) computations can be up to 8 times faster when executed using SIMD~\cite{lomont2011introduction}.
%Figure~\ref{fig:SIMDexplain} is an example of how SIMD works for computing the Euclidean distance between two vectors (e.g., data series).
%
%\textcolor{red}{Obviously, we have a strong motivation to accelerate the typical functions by SIMD instruction. }
Even though no SIMD solutions have been proposed so far for data series indices, this idea has been exploited for the computation of distance functions~\cite{tang2016exploit}.
%% (namely, Euclidean Distance, and Dynamic Time Warping)~\cite{tang2016exploit}, which are at the core of data series similarity search.
%%\textcolor{red}{Last year, Bo Tang et al. used SIMD instruction optimize some typical function in time series (like lower-bound and distance functions)~\cite{tang2016exploit}. }
%%\textcolor{red}{By the way, their implementation also support to the early abandon. 
%%They calculate the accumulated distance sum every bloc and compare it with BSF. 
%%It means that we do the early abandoning every 8 cycles. }
%This is a natural use of SIMD, since these distance functions are in their essence vectorial computations between two subsequences.
%%Figure~\ref{fig:SIMDexplain} illustrates how SIMD works for computing the Euclidean distance between two vectors (i.e., data series). 
In our study, we take an extra step, 
%in addition to using SIMD for this purpose, 
and we also use SIMD for operations related to the proposed data series index structure (i.e., for conditional branch calculations during the computation of the lower bound distances; see Section~\ref{sec:query}).

%\begin{figure}[tb]
%\centering
%\hspace*{-0.5cm}
%\includegraphics[width=1.07\columnwidth]{SIMDexplain}
%\caption{ Using SIMD for distance computation}
%\label{fig:SIMDexplain}
%\end{figure}

\begin{figure}[tb]
\centering
%\subfigure[index tree of iSAX.\label{fig:saxtree}] {
%    \hspace{-1em}
%    \includegraphics[width=0.7\columnwidth]{saxtree}
%}

	\subfigure[\y{a} raw data series\label{fig:saxa}] {
	\hspace{-1em}
	\includegraphics[page=1,width=0.42\columnwidth]{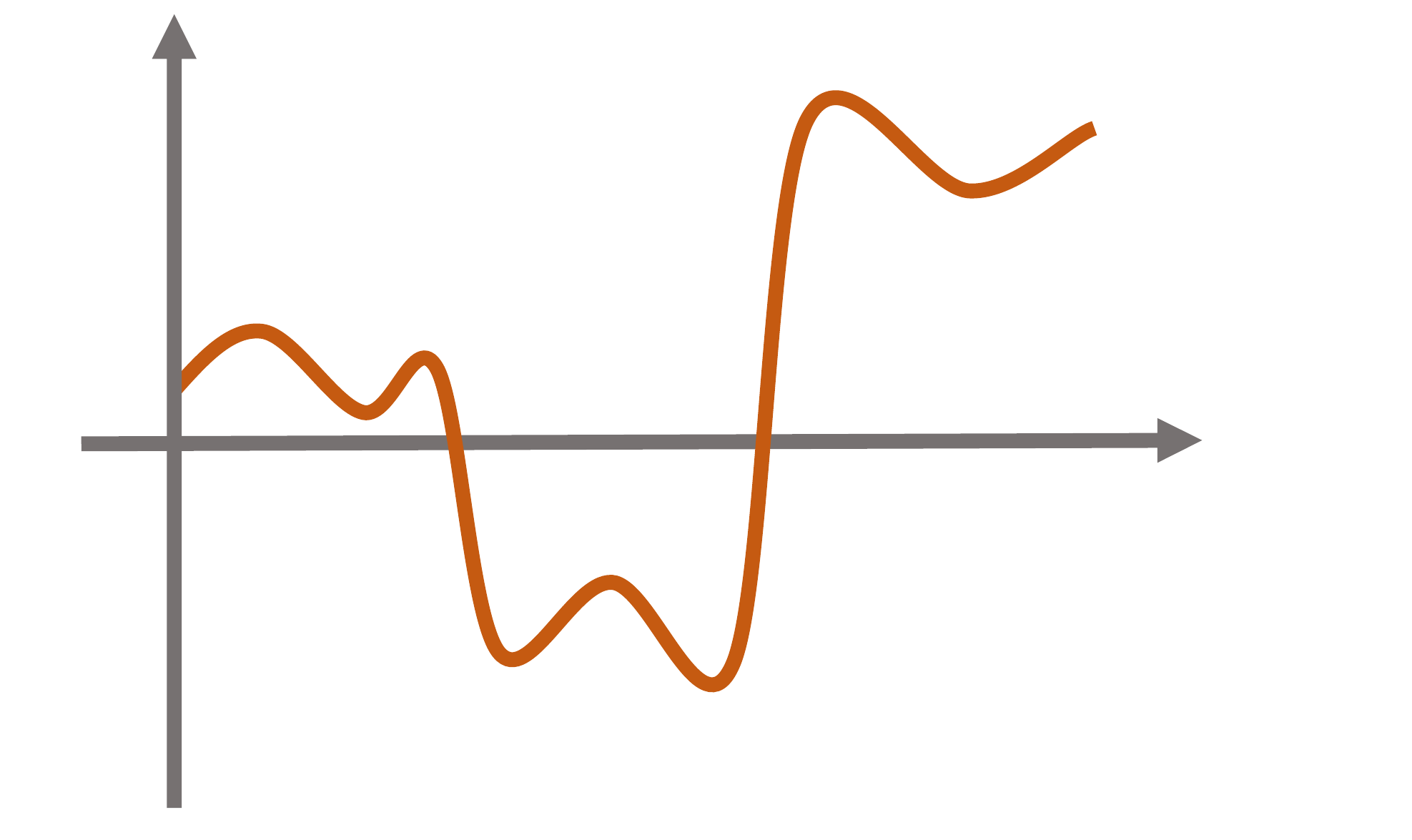}
}
\hspace*{0.5cm}
	\subfigure[\y{its} PAA representation\label{fig:saxb}] {
	\hspace{-1em}
	\includegraphics[page=1,width=0.42\columnwidth]{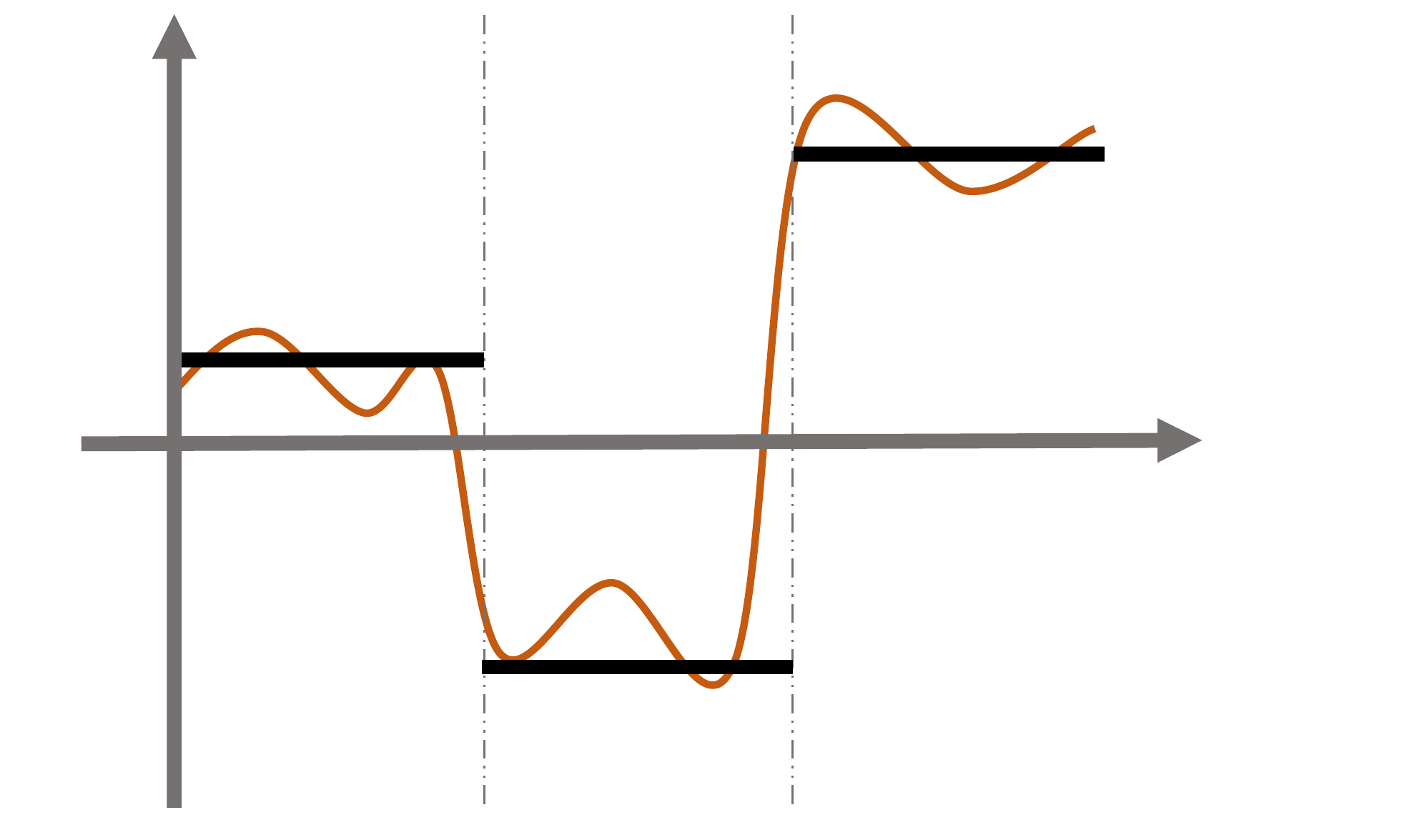}
}\\
	\subfigure[\y{its} iSAX representation\label{fig:saxc}] {
	\hspace{-1em}
	\includegraphics[page=1,width=0.42\columnwidth]{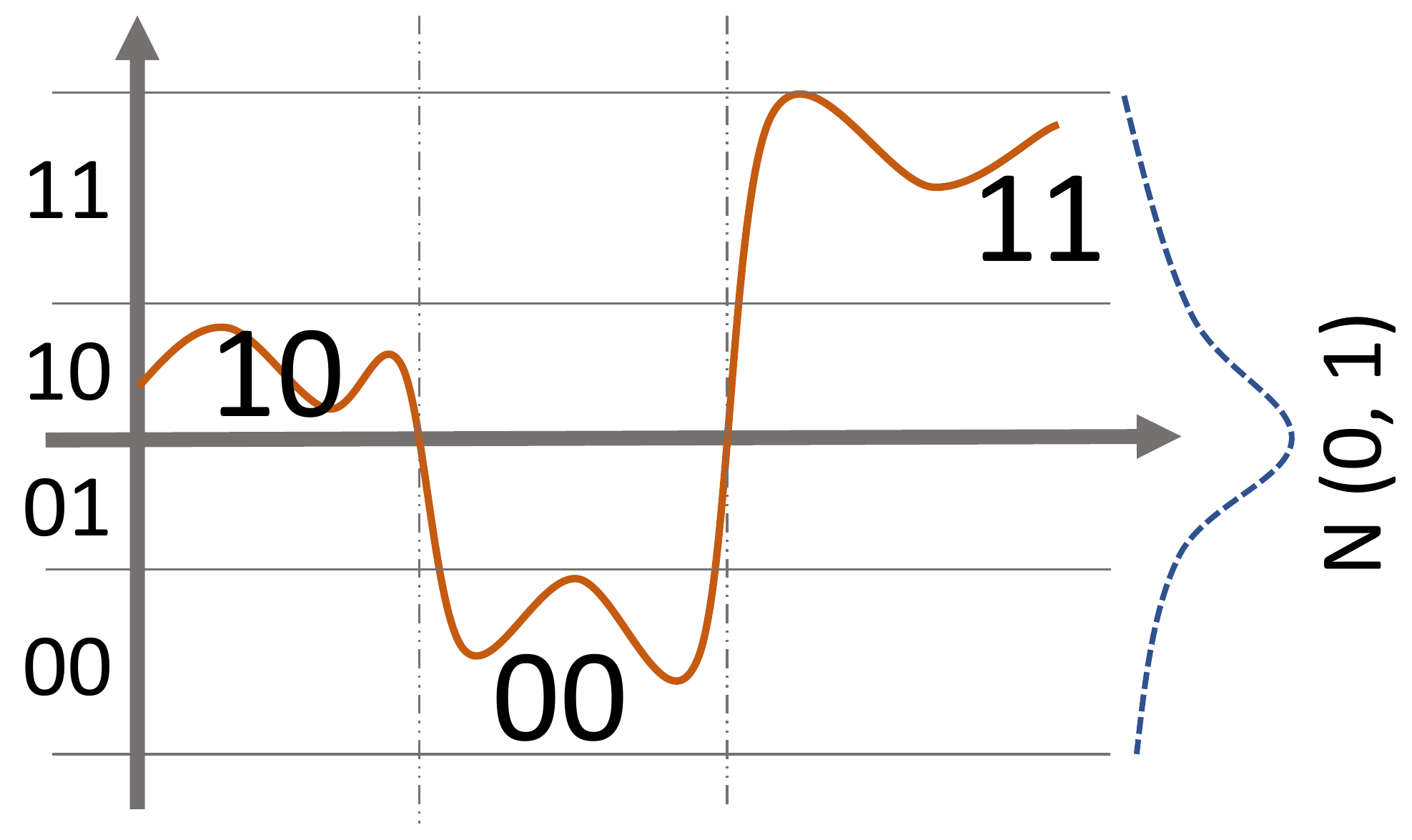}
}
\hspace*{0.5cm}
\subfigure[ADS+ index\label{fig:ads}] {
	\hspace{-1em}
	\includegraphics[width=0.5\columnwidth]{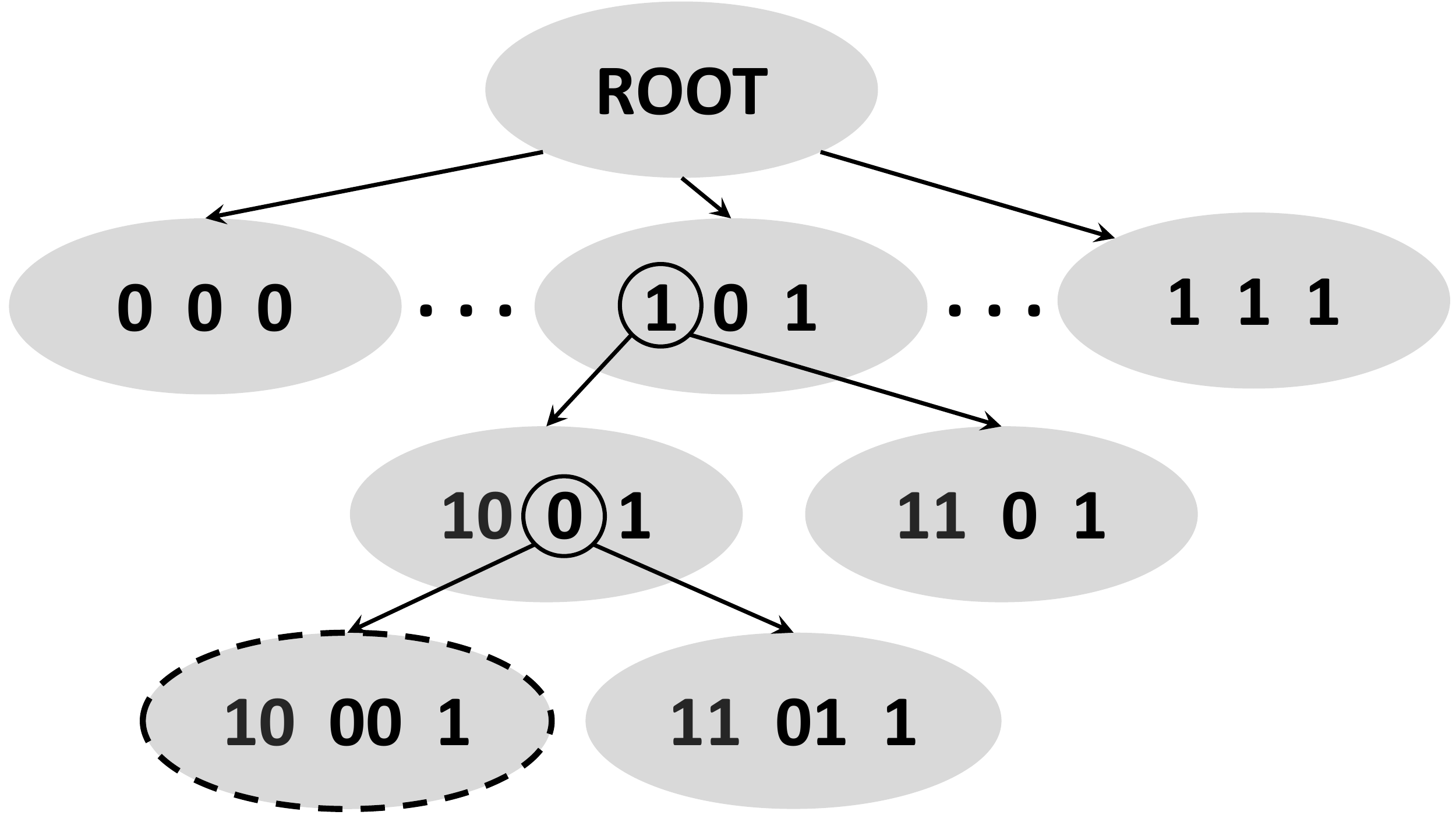}
}
\caption{The iSAX representation, and the ADS+ index structure}
\end{figure}

\subsection{iSAX Representation and ADS+ Index}

\noindent{\bf [iSAX Representation]}
The iSAX representation is based on the 
%idea of segmented means, or Piecewise Aggregate Approximation (PAA)~\cite{yi2000fast,keogh2001dimensionality}. The 
Piecewise Aggregate Approximation (PAA) representation~\cite{keogh2001dimensionality}, which divides the data series in {\em segments} of equal length, and uses the mean value of the points in each segment in order to summarize a data series. 
Figure~\ref{fig:saxb} depicts an example of PAA representation with three segments (depicted with the black horizontal lines), for the data series depicted in Figure~\ref{fig:saxa}. 
Based on PAA, the indexable Symbolic Aggregate approXimation (iSAX) representation was proposed in~\cite{shieh2008sax}.

%~\cite{lin2003symbolic,shieh2008sax}. 
This method first divides the (y-axis) space in different regions, and assigns a bit-wise symbol to each region.
In practice, the number of symbols is small: 
%previous work has shown that 
iSAX achieves very good approximations with as few as 256 symbols, the maximum alphabet cardinality, 
%$|alphabet|$, 
which can be represented by $8$ bits~\cite{isax2plus}.
It then represents each segment of the series
%not by the real value of the PAA, but 
with the symbol of the region the PAA falls into, forming the word $10_2 00_2 11_2$ shown in Figure~\ref{fig:saxc} (subscripts denote the number of bits used to represent the symbol of each segment). 
%Therefore, iSAX further reduces the size of the data series summarization, and more importantly it leads to a bit-wise representation. 

For an overview of iSAX-based indices, see~\cite{evolutionofanindex}.

%\begin{figure}[tb]
%\centering
%\hspace*{-0.5cm}
%\includegraphics[width=1.1\columnwidth]{sax}
%\caption{Example of PAA and iSAX representations}
%\label{fig:sax}
%\end{figure}

%\subsection{The ADS+ Index}

\noindent{\bf [ADS+ Index]}
Based on this representation, the state-of-the-art ADS+ index was developed~\cite{zoumpatianos2016ads}. 
It makes use of variable cardinalities (i.e., variable degrees of precision for the symbol of each segment; see Figure~\ref{fig:saxc}) in order to build a hierarchical tree index (see Figure~\ref{fig:ads}), consisting of three types of nodes:
(i) the root node points to several children nodes, $2^w$ in the worst case (when the series in the collection cover all possible iSAX representations), where $w$ is the number of segments;
(ii) each inner node contains the iSAX representation of all the series below it, and has two children; and 
(iii) each leaf node contains both the iSAX representation \emph{and} the raw data of all the series inside it (in order to be able to prune false positives and produce exact, correct answers). 
When the number of series in a leaf node becomes greater than the maximum leaf capacity, the leaf splits: it becomes an inner node and creates two new leaves, by increasing the cardinality of the iSAX representation of one of the segments (the one that will result in the most balanced split of the contents of the node to its two new children~\cite{isax2plus,zoumpatianos2016ads}). 
The two refined iSAX representations (new bit set to \textit{0} and \textit{1}) are assigned to the two new leaves. 
In our example, the series of Figure~\ref{fig:saxc} will be placed in the outlined node of the index (Figure~\ref{fig:ads}).

The ParIS and ParIS+ indices use the iSAX representation and basic ADS+ index structure, but implement techniques and algorithms specifically designed for multi-core architectures.

\section{Proposed Solution: ParIS and ParIS+}
\label{sec:indexcreation}

In this section, we describe our approach, called Parallel Indexing of Sequences (ParIS), for parallel index construction and query answering. We then present, ParIS+, an improved version of ParIS 
(in settings with reasonable levels of parallelism). 

\begin{figure*}[tb]
\centering
	\includegraphics[page=1,width=\textwidth]{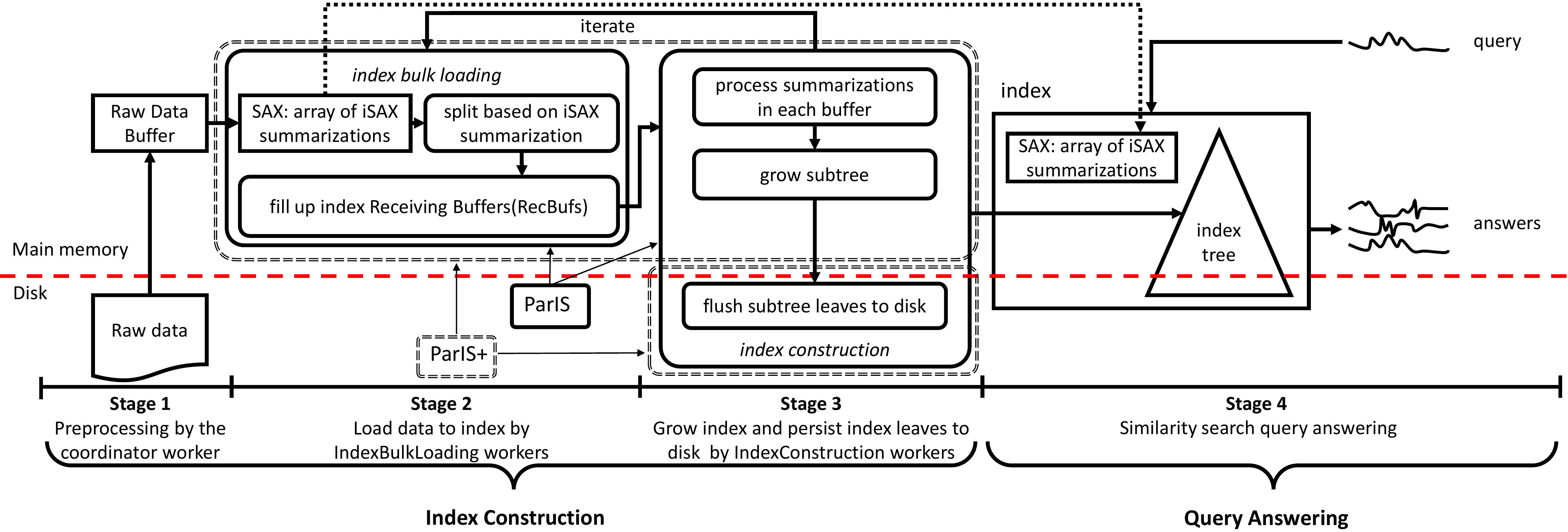}
\caption{Overview of the pipeline for creating the ParIS/ParIS+ index, and using the index for query answering.}
\label{fig:workflow}
\end{figure*}

Figure~\ref{fig:workflow} provides a high level overview
of the entire pipeline of how the ParIS index is created and then used for query answering. 
This pipeline is comprised of four stages. In Stage~$1$,
a thread, called the {\em Coordinator} worker, reads raw data series from the disk 
and transfers them into the {\em raw data buffer} in main memory.
In Stage~$2$,
a number of {\em IndexBulkLoading} workers, process the data series in the raw data buffer
to create their iSAX summarizations.  
Each iSAX summarization determines to which root subtree of the tree index the series belongs.
Specifically, this is determined by the first bit of each of the $w$ segments of the iSAX summarization. 
The summarizations are then stored in one of the index Receiving Buffers (RecBufs) in main memory. 
There are as many RecBufs as the root subtrees of the index tree, each one storing
the iSAX summarizations that belong to a single subtree. This number is usually a few tens of thousands and at most $2^w$, where $w$ is the number of segments in the iSAX representation of each time series ($w$ is fixed to $16$ in this paper, as in previous studies~\cite{zoumpatianos2016ads}).
The iSAX summarizations are also stored 
in the array SAX (used during query answering).

When all available main memory is full, Stage~$3$ starts. In this stage, 
%a dedicated {\em IndexConstruction worker} is assigned to each one of the \textcolor{black}{RecBufs}: 
a pool of {\em IndexConstruction} workers processes the contents of RecBufs;
every such worker is assigned a distinct RecBuf at each time: it reads the data stored in it
and builds the corresponding index subtree. 
%In the (usual) case, where we have a smaller number of workers than RecBufs, 
%the same worker is reused until all RecBufs are processed. 
So, root subtrees are built in parallel. The leaves of each subtree is flushed to the disk at the end of the %index 
tree construction process. 
This results in free space in main memory. 
These 3 stages are repeated until all raw data series are read from the disk,
the entire index tree is constructed, and the SAX array is completed. 
The index tree together with SAX form the ParIS index,
which is then used in Stage~$4$ for answering similarity search queries.

In the following, we elaborate on the stages of this pipeline.

\subsection{Index Construction: ParIS}

The main challenge in devising an algorithm 
for the creation of our index in parallel 
is that a significant part of time is required for disk I/O (i.e., \y{for} reading the raw data and writing the index leaves).
In order to address this challenge, we concentrate our efforts in two directions: 
execute the CPU computations so as to achieve the largest possible
overlap with the required disk I/O, and reduce the number of random accesses to disk as much as possible.
We achieve these by maintaining the synchronization cost among different threads as low as possible.

\subsubsection{Index Initialization}

\begin{figure*}[tb]
\centering
\subfigure[Create index {\em Coordinator} \& {\em IndexBulkLoading} workers\label{fig:inca}] {
    \includegraphics[page=1,width=0.95\columnwidth]{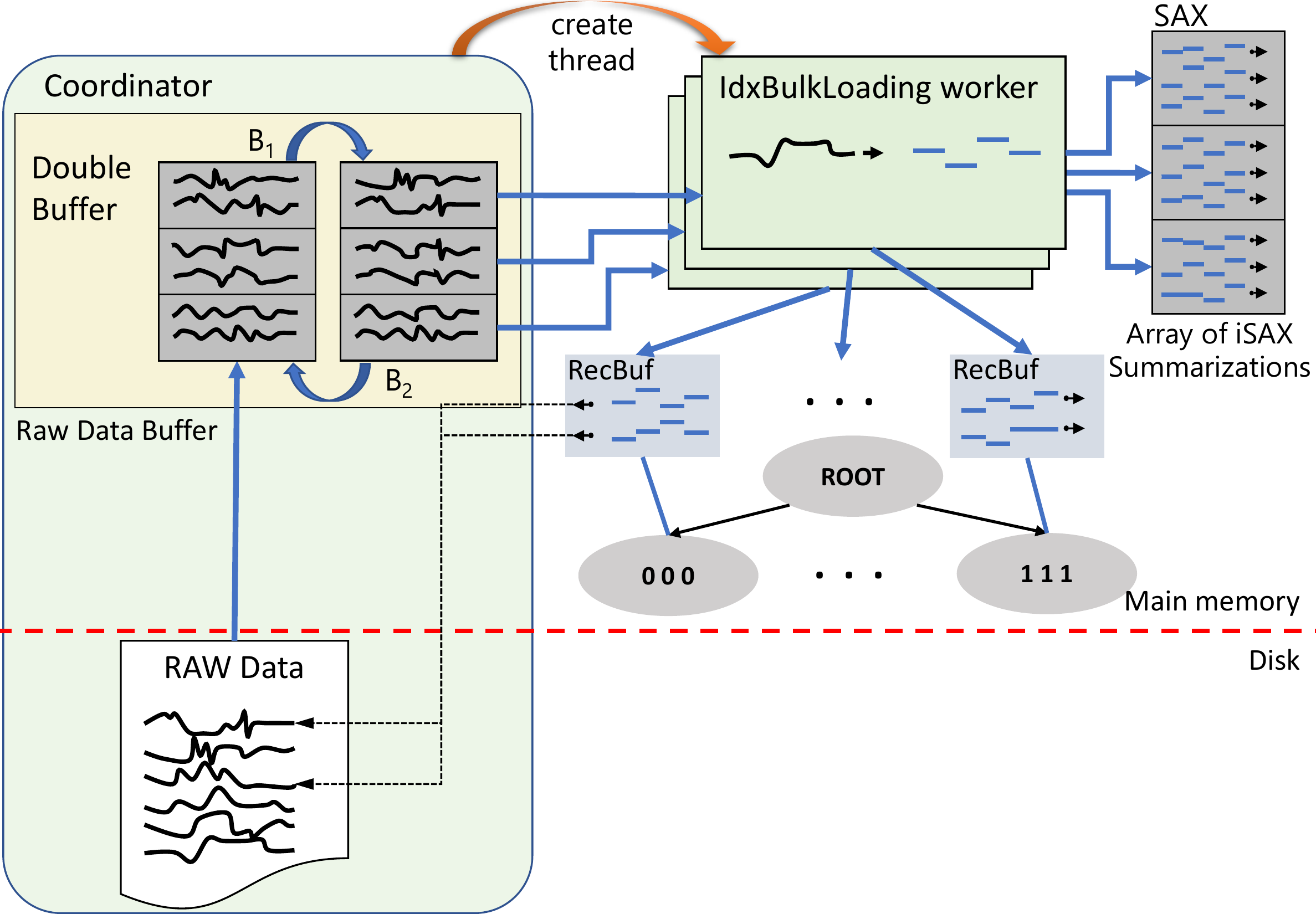}
    }
\hspace*{0.2cm}
    \subfigure[{\em IndexConstruction} workers\label{fig:incb}] {
    \includegraphics[page=1,width=0.95\columnwidth]{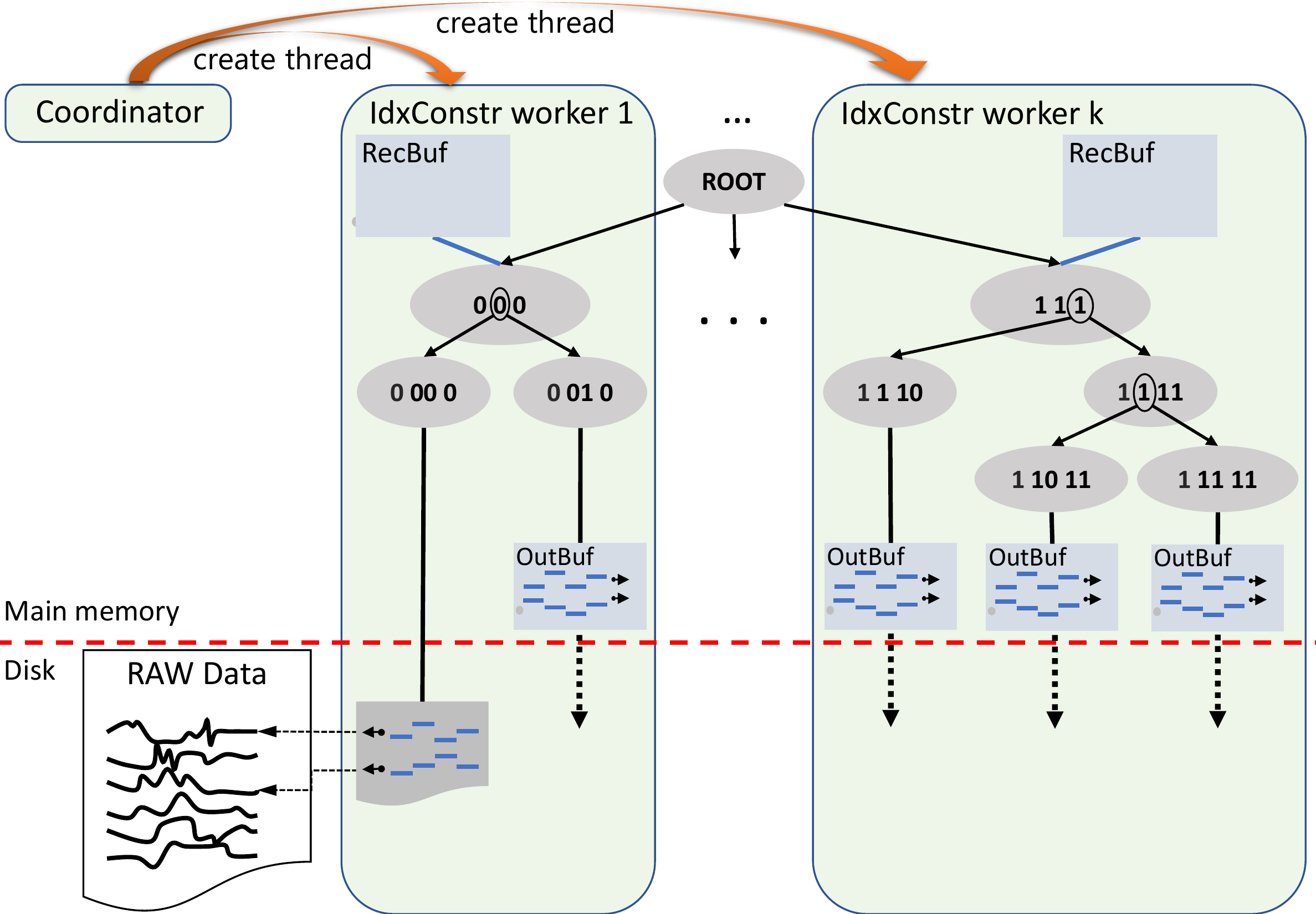}
    }
\caption{Workflow and algorithms relevant to index creation.}
\label{fig:inc}
\end{figure*}

In this section, we describe Stages~1 and~2.
Figure~\ref{fig:inca} summarizes how the coordinator and IndexBulkLoading workers work.

The raw data buffer is implemented using double buffering. 
So, it is comprised of two parts, one on which the {\em Coordinator} works, and another on which the {\em IndexBulkLoading} workers work.
In this way, the data the {\em Coordinator} is accessing
and the data the {\em IndexBulkLoading} workers are handling form two independent sets.
So, all these threads work in parallel (as much as possible). 
Our tuning experiments (refer to Figure~\ref{fig:newin3}) showed that setting the size of the double buffer to 2MB
results in the best performance (the time cost reduces as the buffer size increases until we reach 2MB and then it stabilizes).

The pseudocode for the {\em Coordinator} worker is shown in Algorithm~\ref{algo1}.
We assume that the $index$ variable is a data structure containing all buffers,
a pointer to the root of the tree index, some arrays of locks
that are needed for synchronizing access to RecBufs, and SAX.
In this algorithm, $B_{1}$ and $B_{2}$ are pointers to the two parts, $TS[0]$ and $TS[1]$, 
of the raw data buffer.
Moreover, we denote by $n_t$ the number of {\em IndexBulkLoading} workers 
%and the number of IndexConstruction workers 
that are created by the coordinator (see discussion below
about the value of $n_t$).
The algorithm works as follows. 
The {\em Coordinator} worker first fills in the part of the raw data buffer pointed to by $B_{1}$ (line~\ref{al1:readd}).
%\here{Y: Botao, check the line number here.}
Then, the {\em Coordinator} worker creates the $n_t$ {\em IndexBulkLoading} 
worker threads (lines~\ref{al1:callworkers}). These threads create the iSAX
summarizations of the data in the raw data buffer part pointed to by $B_{1}$ 
and place them in the appropriate RecBufs and in SAX (see Figure~\ref{fig:inca}); for each data series, we also store in RecBuf its offset in the raw data file. 
While the {\em IndexBulkLoading} workers are performing this task, 
the {\em Coordinator} concurrently fills in the other part of the raw data buffer (line~\ref{al1:readd2}).
This process is repeated until the main memory is exhausted. 

The {\em Coordinator} worker is aware of the current memory usage by monitoring the number of data series that it has processed.  
When the available memory is (nearly) exhausted\footnote{Note that we only need a small amount of additional memory for creating new index nodes in the subtree of the root currently being processed, which can have a maximum depth of $w(|alphabet|-1)$~\cite{zoumpatianos2016ads}, where $|alphabet|$ is the cardinality of the alphabet. 
Moving data inside the index (e.g., from RecBuf to OutBuf, as we will discuss later) does not require extra memory: we reallocate the same memory addresses between the buffers.}
(line~\ref{al1:fullmemory}), then the {\em Coordinator} creates the {\em IndexConstruction} worker threads (lines~\ref{al1:flushfbl}),
%where $n$ is the number of cores in the system.
which build the part of the index that corresponds to the iSAX summarizations
stored in the RecBuf, and flush the leaf nodes of the tree to disk. 

The pool of {\em IndexBulkLoading} workers could be as big as the number
of cores in our machine (minus one which is reserved for the {\em Coordinator}).
%(i.e., $2^w$ = $2^{16}$ in our case).
The {\em IndexBulkLoading} workers are assigned each RecBuf one-at-a-time in round-robin fashion,
by using either an atomic fetch and increment primitive, or a lock. 
As we will discuss later (in Section~\ref{sec:experiments}), 
%Figure~\ref{fig:res1} demonstrates that
for ParIS we see the best performance when we use five {\em IndexBulkLoading} workers 
and six {\em IndexConstruction} workers; 
%
%are enough to completely mask out the CPU latency 
%at this stage;
note that these numbers are orders of magnitude less than the number of the index root subtrees
(usually tens of thousands).
Note that because of the small number of {\em BulkIndexLoading} 
(and {\em IndexConstruction} workers),
the use of locks for synchronizing access to RecBufs 
(or the assignment of subtrees) does not result in any synchronization bottlenecks.
Moreover, because the computation is heavily I/O bounded at this stage,
the performance does not degrade even if the {\em Coordinator} creates 
the {\em IndexBulkLoading} workers from scratch each time it fills up a part of the raw data buffer.
%(doing so simplifies the implementation).
%The reason is that(and therefore the cost
%of periodically creating/destroying threads is negligible).
For the same reason, techniques like thread pinning does not improve performance.

\begin{algorithm}[tb]
	{	\footnotesize
		\SetAlgoLined
		\KwIn{\textbf{File*} $file$, \textbf{Index} $index$, \textbf{Integer} $n_t$}
		% 	\KwOut{$index$}
		% 	\tcp*[f]{create data/iSAX buffers for worker threads}\\
		\vspace*{.1cm}
		\textbf{Pointer} $B_{1} \leftarrow index.TS[0], B_{2} \leftarrow index.TS[1]$\;
		\textbf{Integer} p  = 0\;
		\vspace*{.1cm}
		% 	\textbf{shared iSAX summarizations} $SAX$\;
		%\textbf{sax\_type} $B_{s1s}$[],$B_{s2}$[]\;
		%\tcp*[f]{prepare the data of first round}\\
		%Move data from $file$ to part of raw data buffer pointed to by $B_{1}$\;\label{al1:doublebuffer1}
		
		$B_{1}\leftarrow$ read data from $file$\; \label{al1:readd}
		%{\footnotesize store in $B_1$, pairs of the form $\langle ts, p \rangle$, where $ts$ is a time series from $file$ and $p$ is its position in it} \\  
		\While{not reached end of $file$}
		{
			\For{$i$ $\leftarrow$ $0$ \emph{\KwTo} $n_t-1$} 
			{
				%\tcp*[f]{call workers to generate index}\\
				create a thread to execute an instance of $IndexBulkLoading$($index$,$B_{1}+i*chunksize$, $p+i*chunksize$); \label{al1:callworkers}
			}
			$B_{2}$ $\leftrightarrow$ $B_{1}$\;\label{al1:exd}
			$B_{1}\leftarrow$ read data from $file$ \; \label{al1:readd2}
			Wait for IndexBulkLoading workers to finish\;
			\If{ main memory is full}
			{
				\label{al1:fullmemory}
				\For{$i$ $\leftarrow$ $1$ \emph{\KwTo} $n_t+1$} 
				{
					create a thread to execute an instance of $IndexConstruction$($index$)\; \label{al1:flushfbl}
				}
				Wait for IndexConstruction workers to finish\;
			}
			$p \leftarrow p + n_t * chunksize$\;
		}
		%\tcp*[f]{store iSAX summaries of last round}\\
		%disk$\leftarrow$ $B_{s1}$\;\label{al1:writes2}
	}
	\caption{$Coordinator$}
	\label{algo1}
\end{algorithm}

The pseudocode that the IndexBulkLoading workers execute is shown in Algorithm~\ref{algo2}. 
%They simply convert the raw values of the data series to the iSAX summarizations (line~\ref{al2:cover}),
%and write these summarizations in SAX. They also store them in the corresponding FBL buffer 
%by inserting the (iSAX, position)-pair into it (line~\ref{al2:insert}).
Each such worker has been assigned a chunk, of size $chunksize$, in each part of the raw data buffer
(therefore, the size of the raw data buffer is $2*chunksize*n_{t}$). 
Each worker operates only on its chunk.
In this way, no synchronization is needed between the {\em IndexBulkLoading} workers for accessing the
raw data buffer. 
Each {\em IndexBulkLoading} worker reads the data series in its chunk one after the other (line~\ref{al2:cover}), and calculates the iSAX summarization for each of them 
by calling the  ConvertToSAX() function (line~\ref{al2:cover}). 
These summaries are stored in SAX, the {\em Array of iSAX Summarizations} (line~\ref{al2:cover}), and in the appropriate RecBuf (line~\ref{al2:insert}; refer also to Figure~\ref{fig:inca}).
Recall that each RecBuf gathers together all data that must be stored into the same root subtree.
These data may exist in chunks of the raw data buffer that are associated to different {\em IndexBulkLoading} workers.
So, more than one such workers may require to concurrently access the same RecBuf.
Therefore, synchronization is needed.
This synchronization is achieved by using a lock for each such buffer,
stored in array $RecBufLock[]$ of $index$.
%(we assume that there are $r$ RecBufs, i.e., the number of root subtrees in the index tree is $r$). 

To eliminate the need for synchronization between the {\em IndexBulkLoading} workers in accessing SAX,
the iSAX summarization of the data series stored in the $p$ position of the raw data file,
is  stored in the $p$ position of SAX. 
%\here{Youla: I am not sure that I fully understand this. I made a few changes to the code of Alg 2. Please check that they are correct. }

\begin{algorithm}[tb]
	%\DontPrintSemicolon
	{	\footnotesize
		\SetAlgoLined
		\KwIn{\textbf{Index} $index$, \textbf{Raw data buffer} $TS$[], \textbf{Integer} $p$}
		%, \textbf{iSAX summarizations} $SAX$[]}
		%, \textbf{Initial file position} $p$}
		\vspace*{.1cm}
		% 	\textbf{thread\_lock} $RecBufLock[]$; {\footnotesize // a lock for each RecBuf} \\
		\vspace*{.1cm}
		\For{$i$ $\leftarrow$  $0$ \emph{\KwTo} $chunksize-1$}
		{
			%\y{		\tcp*[f]{convert the data series to its iSAX summarization and store  this summarization in RecBuf and SAX}}\\
			$index.SAX[p+i]$ = $ConvertToSAX$ ($TS[i]$)\;\label{al2:cover}
			%\tcp*[f]{insert the (isax,position) pair into RecBuf}\\
			acquire appropriate  lock from $index.RecBufLock[]$\;
			$InsertIntoRecBuf$ ($\langle index.SAX[p + i], p+i \rangle$)\;\label{al2:insert}
			release the acquired lock; \\
		}
	}
	\caption{$IndexBulkLoading$}
	\label{algo2}
\end{algorithm}

\subsubsection{Subtree Construction and Leaf Materialization}

We now describe Stage~3, where the index is gradually constructed and 
its leaves materialized. 
On top of the raw data buffer and the RecBufs,
ParIS makes use of an additional set of main memory buffers, the
{\em Output Buffers} (OutBufs). 
Each OutBuf is associated to one leaf of the index tree and stores the iSAX representations of the data series and pointers to them in the raw data file. 
%that 
% \textcolor{red}{to arrange the data need to store on disk}.

The {\em Coordinator} worker 
creates a number of {\em IndexConstruction} workers when
it discovers that the main memory is exhausted.
(Based on our experiments, the best number of {\em IndexConstruction} workers is 6.)
%In our experiments, the number of workers is set to the number of cores in the system. 
%Note that since the need for synchronization among workers is minimal, the corresponding communication overhead is insignificant.
These workers process the data in the RecBufs in order to grow the corresponding subtree, 
until the data end up in the OutBufs of that subtree. 
Finally, the OutBufs are flushed to disk.
This process is illustrated in Figure~\ref{fig:incb},
(where we have assumed that the contents of the OutBuf for the leftmost leaf have been flushed to disk, whereas the rest OutBufs have not). 

All {\em IndexConstruction} workers process different root subtrees, so  they work independently and no synchronization is needed. 
A worker that finishes its work on one subtree gets assigned to a new RecBuf, until all RecBufs are processed. 
%Parallelism in this case is achieved at the top level of the tree.
In order to maintain the scheme simple and efficient, 
we have chosen not to parallelize processing 
inside each one of the index root subtrees since that would require 
a lot of synchronization (due to node splitting).
Our experiments have shown that this decision does not have any negative impact in the performance of our scheme.

The pseudocode that the {\em IndexConstruction} workers execute is shown in Algorithm~\ref{algo3}. 
An {\em IndexConstruction} worker first selects one of the RecBufs to process in an atomic way (line~\ref{al3:gotnode}). 
This can be done by using either an atomic \emph{fetch and increment} primitive 
($n_b$ in Algorithm~\ref{algo3}), or a lock.
%To do so in an atomic way, it acquires $glock\_RecBuf$, a lock that all IndexConstruction workers share.  
Then, it moves the data to the appropriate OutBuf in the index (line~\ref{al3:putdata}), and if necessary (i.e., if the leaf node is full), it (repeatedly) performs node splitting (line~\ref{al3:split}).
When node splitting is performed, the iSAX summarizations 
(i.e., the contents of the leaf node to be split) are read
from disk and they are placed in the appropriate OutBuf 
(if they have already been flushed). 
Then, the leaf node is split to  
two new leaf nodes,
%at the best point where the original node can be divided into 2 most balanced parts.
%\textcolor{red}{?????????describe chose best split point, 
%we add the reference here or add some text ????????????????} Finally, 
the data of the original leaf are moved to the new leaves, and finally the OutBufs corresponding to the leaves of the subtree currently processed are flushed to disk (line~\ref{al3:flushlbl}).

We note that 80\% of the leaves (and therefore also 
the corresponding OutBufs) have size less than the block size.
So, flushing them to disk results in disk random accesses. 
For this reason,  the use of a lock 
to synchronize disk accesses of threads during leaf materialization
would cause performance degradation.
%without any performance benefit. The system can handling 
%the write command automatic and the store command will keep 
%in memory at first, and the change will be synchronized later and in background.}
%having each thread flushing the leaves of their subtree on disk
%in a sequential way using a lock, does not result in better performance. 

\begin{algorithm}[tb]
	{\footnotesize
		\SetAlgoLined
		\KwIn{\textbf{Index} $index$} 
		\vspace*{.1cm}
		%        \textbf{Thread\_lock}  $lock\_write$\;
		\textbf{Shared integer} $n_{b}=0$\;  
		\vspace*{.1cm}
		\While{(TRUE) }
		{
			$i\leftarrow${\em Atomically} fetch and increment $n_{b}$\;\label{al3:gotnode}
			%$lock$ ($glock\_RecBuf$)\;
			%i = $n_{b}$\;  
			%$n_{b}$++\; 
			%$unlock$ ($glock\_RecBuf$)\;
			\textbf{if} ($i \geq 2^w$) \textbf{then} break\; 
			\For{\textbf{{\em every}} $\langle isax, pos \rangle$ {\em pair} $\in index.RecBuf[i]$}
			{
				$targetLeaf \leftarrow$ Leaf of $index$ tree to insert $\langle isax, pos \rangle$\;
				\While{$targetLeaf$ {\em is full}}
				{
					SplitNode($targetLeaf$)\;\label{al3:split}
					$targetLeaf \leftarrow$ New leaf to insert $\langle isax, pos \rangle$\;
				}
				Insert $\langle isax, pos \rangle$ in $targetLeaf$'s OutBuf buffer\;\label{al3:putdata}
			}
			%		acquire $lock\_write$\;\label{al3:loc}
			Flush $targetLeaf$'s OutBuf buffer to disk\;\label{al3:flushlbl}
			%%		release $lock\_write$\;\label{al3:unloc}
			%		set $targetLeaf$ to be in PARTIAL mode\;\label{al3:full}
			Clear this node OutBuf\;
		}
	}
	\caption{$IndexConstruction$ }
	\label{algo3}
\end{algorithm}

%-----------------------------------------------------------------------------------------

\subsection{Index Construction: ParIS+}

\begin{figure}[tb]
	\centering
	\subfigure[{\em IndexBulkLoading} worker after synchronization barrier\label{fig:inc2a}] {
		\includegraphics[page=1,width=0.77\columnwidth]{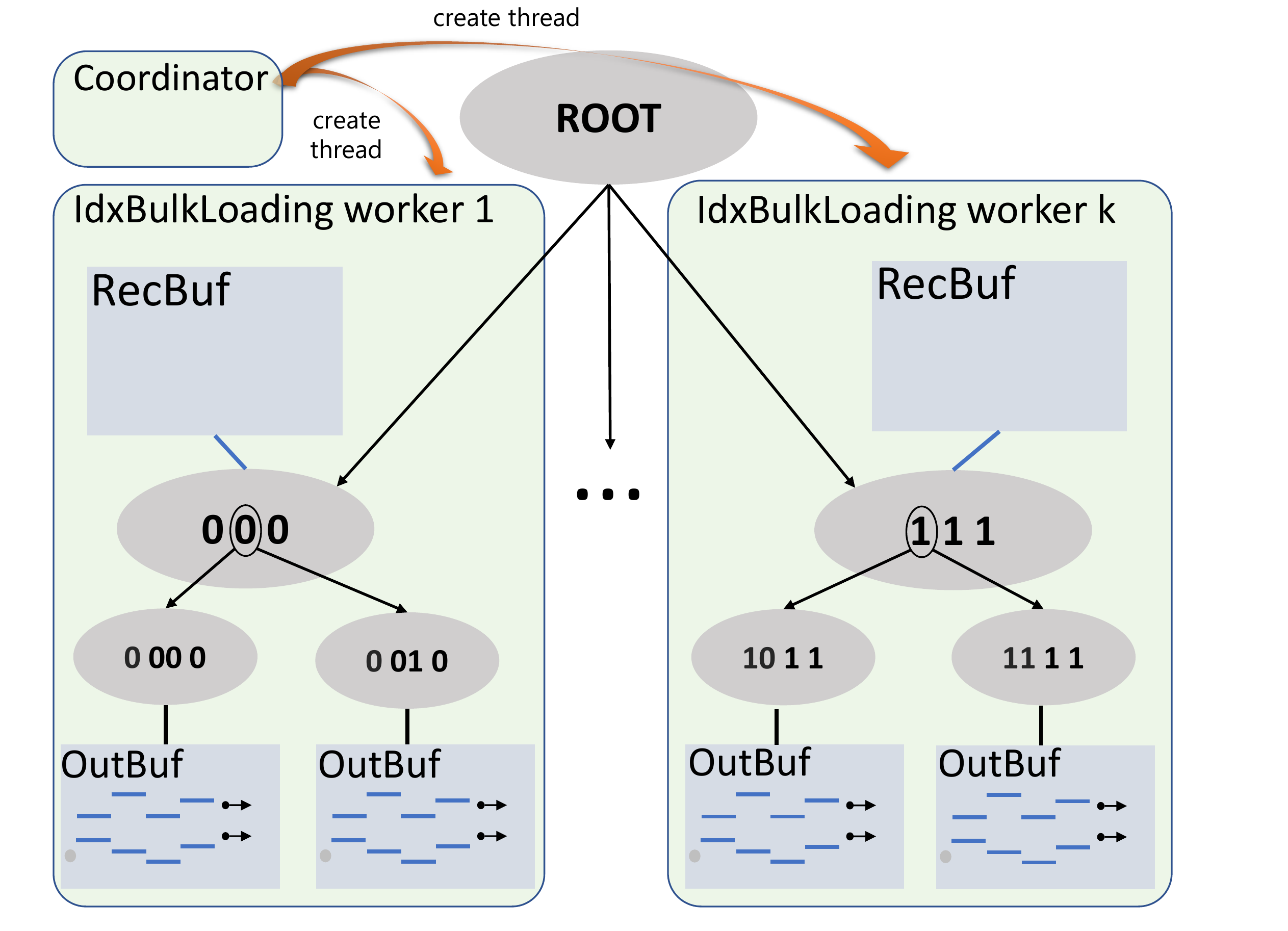}
	}
	\subfigure[index construction \label{fig:inc2b}] {
		\includegraphics[page=1,width=1.02\columnwidth]{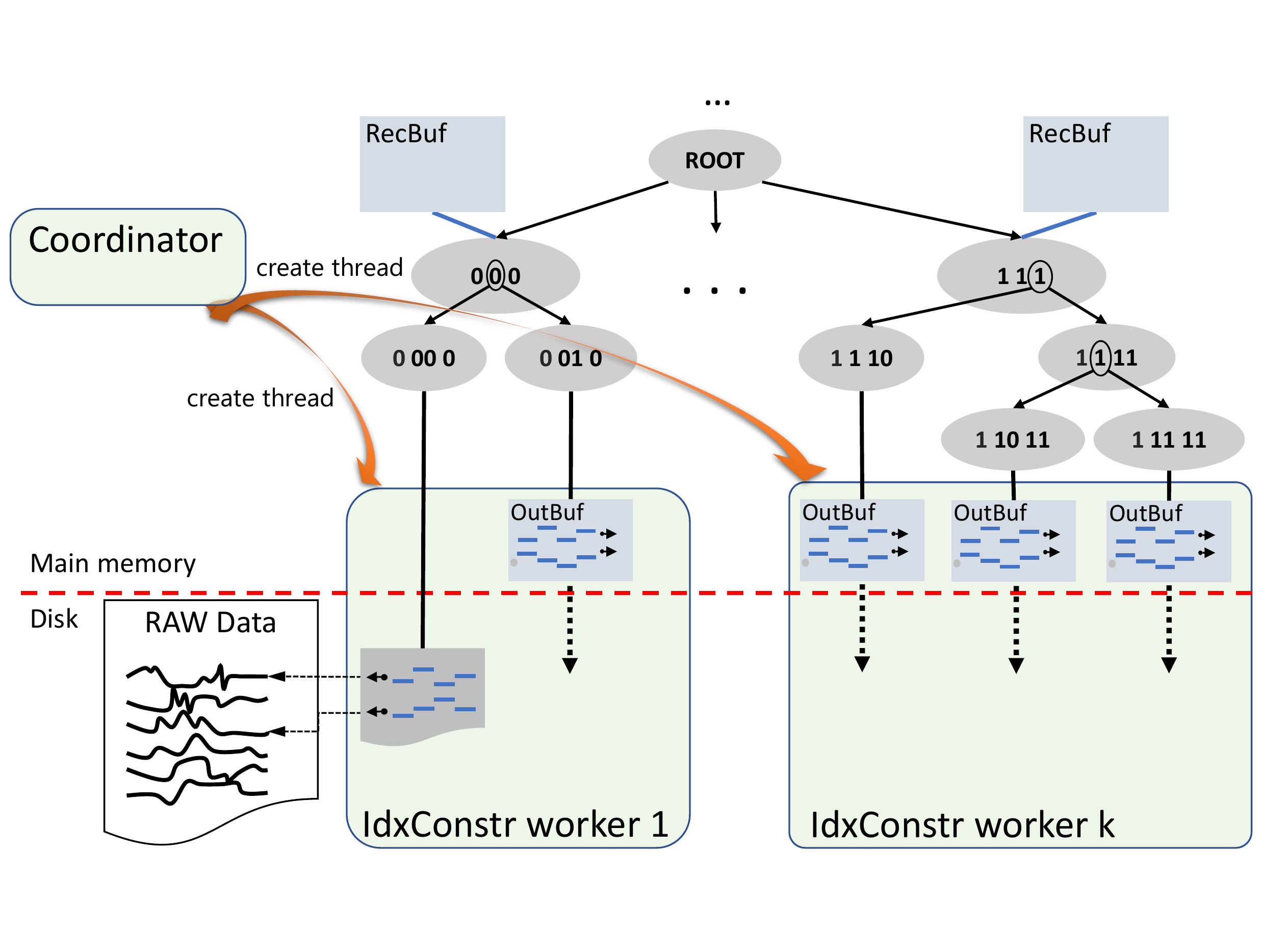}
	}
	\caption{Workflow and algorithms relevant to ParIS+ index building.}
	\label{fig:inc2}
\end{figure}

In this section, we present Paris+, which improves ParIS by completely masking out the CPU cost when creating the index.
This is not true for ParIS, whose index creation (stages 1-3) is not purely I/O bounded (as we will see in Figure~\ref{fig:res1}).  
The reason for this is that, in ParIS, the IndexConstruction workers do not work concurrently with the Coordinator worker.
%, although they perform a lot of CPU computations to build the index. 
Moreover, the IndexBulkLoading workers do not have enough CPU work to do to fully overlap the time needed by the Coordinator worker to read the raw data file.

ParIS+ (Algorithms~\ref{algo1new}-\ref{algo3new}) is an optimized version of ParIS, which achieves a complete overlap of the CPU computation with the I/O cost.
In ParIS+, the IndexBulkLoading workers have undertaken the task of building the index tree, in addition to performing the tasks of stage 2. 
The IndexConstruction workers now simply materialize the leaves by flushing them on disk.

In ParIS+, the Coordinator worker (Algorithm~\ref{algo1new}) creates the IndexBulkLoading workers 
%at the beginning of the execution, 
right after it has finished filling in one part of the
raw data buffer for the first time (line~\ref{al1n:1stread}).
Note that before starting to fill in this part again, the Coordinator reaches a barrier (line~\ref{al1n:barrier1} of Algorithm~\ref{algo1new}), which ensures that the IndexBulkLoading workers have finished processing it (line~\ref{al2n:barrier1} of Algorithm~\ref{algo2new}).
An additional barrier (line~\ref{al1n:barrier2} of Algorithm~\ref{algo1new}) between the Coordinator and each of the IndexBulkLoading workers is necessary to ensure that no IndexBulkLoading worker accesses the OutBuf buffers as long as the IndexConstruction workers operate on them. 

%before it starts storing new data in it.
Algorithm~\ref{algo2new} provides pseudocode for the IndexBulkLoading workers. 
Note that the IndexBulkLoading workers have to reach a barrier (line~\ref{al2n:barrier1}) after they finish processing the part of the raw data buffer they have been assigned. 
This barrier is necessary since more than one IndexBulkLoading worker adds items in each of the RecBufs, and the tree index construction should start only after all of them have finished their current phase of adding items in the RecBufs.
%As a result, the operation of switch the double buffer, and the start of index bulkloading
%can be executed independent and efficient (without the overhead of creating
%and destroying threads like ParIS). } 
Additional barriers (lines~\ref{al2n:barrier2} and~\ref{al2n:barrier3}) ensure the necessary synchronization of the IndexBulkLoading workers with the Coordinator thread (as discussed above).

The {\em IndexConstruction} workers (Algorithm~\ref{algo3new}) simply flush the Outbufs of the leaves of each subtree of the index tree on disk.

Finally, we note that we also considered an alternative, where the IndexBulkLoading workers were performing their work concurrently with the tree index construction phase performed by the IndexConstruction workers.
However, this design did not have any positive impact in performance, and thus, do not discuss it further.

\begin{algorithm}[tb]
	{	\footnotesize
		\SetAlgoLined
		\KwIn{\textbf{File*} $file$, \textbf{Index} $index$, \textbf{Integer} $n_b$, \textbf{Integer} chunksize}
		% 	\KwOut{$index$}
		% 	\tcp*[f]{create data/iSAX buffers for worker threads}\\
		\vspace*{.1cm}
		\textbf{Pointer} $B_{1} \leftarrow index.TS[0], B_{2} \leftarrow index.TS[1]$\;
%		\textbf{Integer} p  = 0\;
		\vspace*{.1cm}
		% 	\textbf{shared iSAX summarizations} $SAX$\;
		%\textbf{sax\_type} $B_{s1s}$[],$B_{s2}$[]\;
		%\tcp*[f]{prepare the data of first round}\\
		%Move data from $file$ to part of raw data buffer pointed to by $B_{1}$\;\label{al1:doublebuffer1}

		%{\footnotesize store in $B_1$, pairs of the form $\langle ts, p \rangle$, where $ts$ is a time series from $file$ and $p$ is its position in it} \\  
		
			$B_{1}\leftarrow$ read data from $file$\; \label{al1n:1stread}
	\For{$i$ $\leftarrow$ $0$ \emph{\KwTo} $n_w-1$} 
			{
				%\tcp*[f]{call workers to generate index}\\
				create a thread to execute an instance of 
				$IndexBulkLoading$($index$,$i$, $n_w$, $chunksize$); \label{al1:callworkers}}
%			\y{	write in shared variable $offset_i$ the value $p+i*chunksize$;}
%		        \y{Barrier to synchronize with the {\em IndexBulkLoading} workers\;}
			
		\While{not reached end of $file$}
		{
			
			$B_{2}$ $\leftrightarrow$ $B_{1}$\;\label{al1n:exd}
			$B_{1}\leftarrow$ read data from $file$ \; \label{al1n:readd2}
		        Barrier to synchronize with the {\em IndexBulkLoading} workers\;\label{al1n:barrier1}
			\If{ main memory is full}
			{
				\label{al1n:fullmemory}
				\For{$i$ $\leftarrow$ $0$ \emph{\KwTo} $n_w$} 
				{
					create a thread to execute an instance of $IndexConstruction$($index$)\; \label{al1n:flushfbl}
				}
				Wait for IndexConstruction workers to finish\;
			}
		%	$p \leftarrow p + n_t * chunksize$\;
		%	\y{write in shared variable $offset_i$ the value $p+i*chunksize$\;} 
			\For{$i$ $\leftarrow$ $0$ \emph{\KwTo} $n_w-1$} 
			{
		        Barrier to synchronize with {\em IndexBulkLoading} worker $i$\;\label{al1n:barrier2}
			}
		}
		%\tcp*[f]{store iSAX summaries of last round}\\
		%disk$\leftarrow$ $B_{s1}$\;\label{al1:writes2}
		Kill IndexBulkLoading workers\;
	}
	\caption{$Coordinator$ $(ParIS+)$}
	\label{algo1new}
\end{algorithm}

\begin{algorithm}[tb]
	%\DontPrintSemicolon
	{	\footnotesize
		\SetAlgoLined
			\KwIn{\textbf{Index} $index$, \textbf{Integer} $id$, \textbf{Integer} $n_w$, \textbf{Integer} $chunksize$}
		%, \textbf{iSAX summarizations} $SAX$[]}
		\vspace*{.1cm}
		\textbf{Shared integer} $f_{w}=0$\;  
		\textbf{Integer} $p = id * chunksize$, $cnt = 1$\;
		\textbf{Boolean} $toggle = 0$\;
		\vspace*{.1cm}
%	\y{	Barrier to sychronize the {\em IndexBulkLoading} workers with the Coordinator worker\;\label{al2n:barrier1}}
%		\y{$p = offset_i$\;}
		\While {(TRUE)} {
			\For{$i$ $\leftarrow$  $0$ \emph{\KwTo} $chunksize-1$}
			{
				%\y{		\tcp*[f]{convert the data series to its iSAX summarization and store  this summarization in RecBuf and SAX}}\\
				$index.SAX[p+i]$ = $ConvertToSAX$ ($index.TS[toggle][i]$)\;\label{al2n:cover}
				%\tcp*[f]{insert the (isax,position) pair into RecBuf}\\
				acquire appropriate  lock from $index.RecBufLock[]$\;
				$InsertIntoRecBuf$ ($\langle index.SAX[p + i], p+i \rangle$)\;\label{al2n:insert}
				release the acquired lock; \\
			}
			Barrier to synchronize the {\em IndexBulkLoading} workers with one another\;\label{al2n:barrier1}
			\While{(TRUE) }
			{
				$i\leftarrow${\em Atomically} fetch and increment $f_{w}$ \;\label{al3n:gotnode}
				\textbf{if} ($i \geq cnt * 2^w$) \textbf{then} break \; 
				\For{\textbf{{\em every}} $\langle isax, pos \rangle$ {\em pair} $\in index.RecBuf[i]$}
				{
					$targetLeaf \leftarrow$ Leaf of $index$ tree to insert $\langle isax, pos \rangle$\;
					\While{$targetLeaf$ {\em is full}}
					{
						SplitNode($targetLeaf$)\;\label{al3n:split}
						$targetLeaf \leftarrow$ New leaf to insert $\langle isax, pos \rangle$\;
					}
					Insert $\langle isax, pos \rangle$ in $targetLeaf$'s OutBuf buffer\;\label{al3n:putdata}
				}
			}
			$p = p + n_b * chunksize$\;
			$toggle = 1 -toggle$\;
			Barrier to synchronize the {\em IndexBulkLoading} workers with one another and with the Coordinator worker\;\label{al2n:barrier2}
			Barrier to synchronize this {\em IndexBulkLoading} worker with the Coordinator worker\;\label{al2n:barrier3}
		}
	}
	\caption{$IndexBulkLoading$ $(ParIS+)$}
	\label{algo2new}
\end{algorithm}

\begin{algorithm}[tb]
	{\footnotesize
		\SetAlgoLined
		\KwIn{\textbf{Index} $index$} 
		\vspace*{.1cm}
		%        \textbf{Thread\_lock}  $lock\_write$\;
		\textbf{Shared integer} $f_{c}=0$\;  
		\vspace*{.1cm}
		\While{(TRUE) }
		{
			$i\leftarrow${\em Atomically} fetch and increment $f_{c}$\;
			\textbf{if} ($i \geq 2^w$) \textbf{then} break\; 
			%		acquire $lock\_write$\;\label{al3:loc}
			For each leaf in the subtree rooted at the $i$-th root child\; 
			\hspace*{1cm}	Flush leaf's OutBuf buffer to disk\;\label{al3n:flushlbl}
			%%		release $lock\_write$\;\label{al3:unloc}
			%		set $targetLeaf$ to be in PARTIAL mode\;\label{al3:full}
			\hspace*{1cm}	Clear the OutBuf buffer\;
			%Flush $targetLeaf$'s OutBuf buffers to disk\;\label{al3n:flushlbl}
			%%		release $lock\_write$\;\label{al3:unloc}
			%		set $targetLeaf$ to be in PARTIAL mode\;\label{al3:full}
			%Clear this node OutBuf\;
		}
	}
	\caption{$IndexConstruction$ $(ParIS+)$}
	\label{algo3new}
\end{algorithm}

\subsection{Query-Answering}
\label{sec:query}
In this section, we describe methods for parallel query-answering. 

The algorithm first performs an \emph{approximate search} 
to obtain the first Best-So-Far (BSF) answer, 
and then proceeds with a sequential scan of the raw data 
that could not be pruned using the BSF, in order to produce the exact, final answer to the query. 
The approximate search is really fast, 
requiring only a negligible percentage (a few msec) 
of the (mostly) on-disk sequential scan cost. 
It is a simple, in-memory path traversal from the index root 
to the leaf with the iSAX representation that is the most similar to that of the query.
Once a leaf is reached, 
%it is materialized and 
the distance between the query and each of the leaf's data series is calculated. 
The minimum distance found is the first BSF answer (see left part of Figure~\ref{fig:qa2}).
This BSF is used to prune the candidate series by computing lower bound distances to their summarizations. 
The series that are not pruned will be visited in the raw file, and the true distance will be computed (the BSF may be updated during this phase).

In the following, we concentrate on our algorithm for parallelizing the scan phase. 
We first describe how we exploit SIMD for
performing the lower bound distance calculations.
Then, we present, in Section~\ref{nb-paris}, a simple technique for further parallelizing this phase (nb-ParIS+, which stands for non-balanced ParIS+), which however does not result in optimal performance, because of the lack of load balancing. 
Finally, we present in Section~\ref{sec:paradsplus} our proposed method for exact search in ParIS+
(the exact search algorithm for ParIS is the same).

%, and we then explain our parallel exact search algorithm.

\subsubsection{Lower-Bound Distance Calculation}
\label{sec:lowerboundsimd}

The algorithm starts by calculating the lower bound distance 
between the query series and the iSAX summarizations 
of all series in the index. 
This is a main memory operation, 
since the iSAX summarizations are small enough 
to fit in the memory of modern servers\footnote{%This is true even for very large datasets: e.g., the 
The highest granularity 
iSAX summarizations for 1 billion data series (occupying 1TB on disk) 
only need about 10GB of space in main memory.}.
This is a procedure that we execute using SIMD, 
since both the queries and the index series are vectors, 
on which we perform the same operation (i.e., a distance calculation).

Using SIMD, we can perform eight calculations in parallel, using a single instruction 
(we assume 256-bit SIMD vectors, containing $8$ 32-bit float elements). 
%\textcolor{blue}{Sec. V.A, 2nd par. So the word-size is assumed to be 32?}
We need to implement a conditional branch in SIMD, 
but contrary to previous solutions~\cite{tang2016exploit}, 
this is a complex branch: not only do we have to use 
different conditional branches for different positions in the SIMD vector, 
but also need to make different assignments for different branches. 

\begin{figure*}[tb]
	\begin{minipage}[p]{0.85\columnwidth}
		\centering
		\includegraphics[width=1.05\columnwidth]{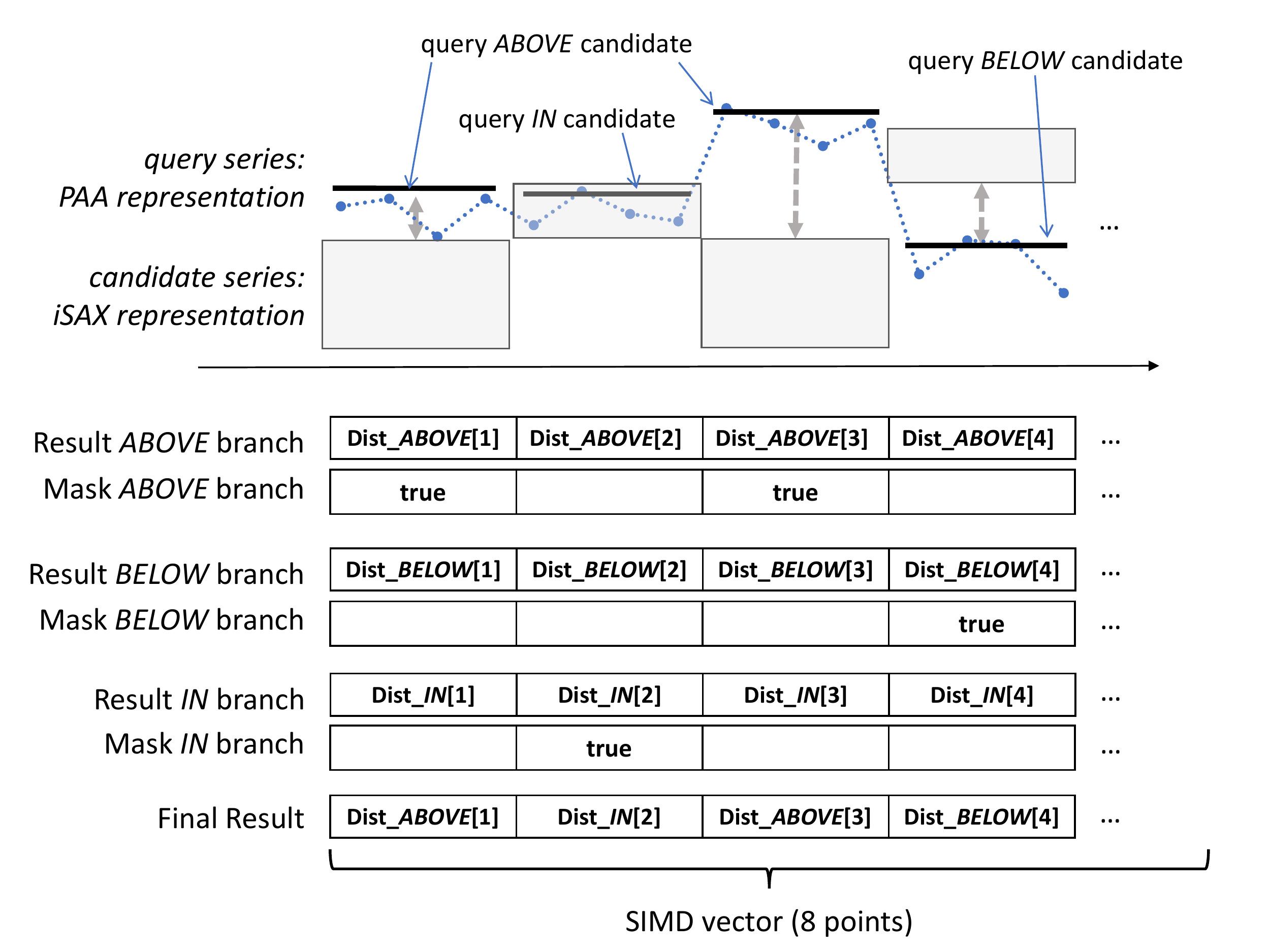}
		\vspace*{-0.7cm}
		\caption{SIMD conditional branch calculation.}
		\label{fig:SIMDcondi}
	\end{minipage}
	\hspace*{0.5cm}
	\begin{minipage}[p]{1.15\columnwidth}
		\hspace{-0.2em}
		\includegraphics[page=1,width=0.98\columnwidth]{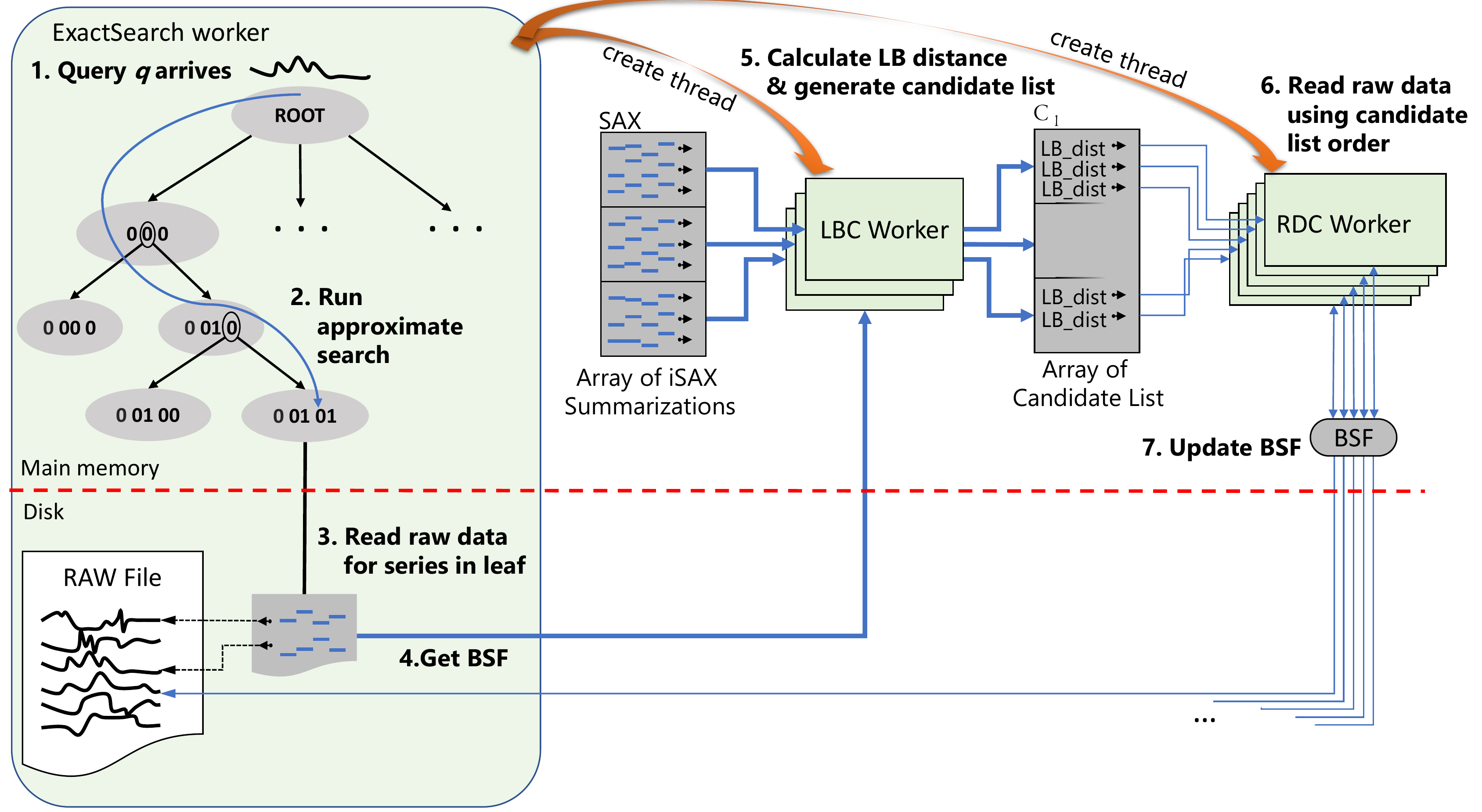}
		\caption{Workflow and algorithms for query answering with ParIS+ (balanced).}
		\label{fig:qa2}
		\end{minipage}
	\end{figure*}

In our case, the calculation of the lower bound distance between the 
PAA of the query series and an iSAX summarization has 3 branches (conditions): checking whether
the PAA lies (i)~\emph{ABOVE}, (ii)~\emph{BELOW}, or (iii)~\emph{IN} the iSAX interval.
%, (ii) PAA lies below the iSAX interval, 
%and (iii) PAA lies within the iSAX interval. 
Thus, we need to choose different values from different dictionaries 
in order to perform the distance computations in SIMD (see Figure~\ref{fig:SIMDcondi}).
We first calculate the distance results of the above 3 branches for every point in the SIMD vector.
We then use a conditional mask to extract the results in the correct branch.

In particular, we generate 3 branch masks, i.e., \emph{ABOVE}, \emph{BELOW}, and \emph{IN}. 
These masks contain a value of \emph{true} (i.e., $1$) only in the SIMD vector positions for which the corresponding branch is true. 
In Figure~\ref{fig:SIMDcondi} for example, the first query PAA segment is above the corresponding candidate series iSAX representation, which means that only the \emph{ABOVE} mask will be true for this position; consequently we will consider the \emph{Dist\_ABOVE} distance value for this position of the SIMD vector.
Using the appropriate SIMD instruction (AVX, AVX2 and SSE3)~\cite{coorporation2009intel}, 
we can efficiently calculate the value of the 3 branch masks. % as the result of \textcolor{red}{if statements.} 
Next we apply a logical "AND" between the 3 branch results and their masks. 
After that, all bits of the branch result in the wrong branch will be zero. 
Now there is only one value at the same position in those 3 branch results. 
Finally, we merge all possible branches in one vector, which is the correct final result.

%\begin{figure}[tb]
%\centering
%\includegraphics[width=\columnwidth]{newsimdbranch}
%\caption{SIMD conditional branch calculation.}
%\label{fig:SIMDcondi}
%\end{figure}

In this way, we have a SIMD version of the distance computation function, 
which is a frequent and (CPU) time-consuming operation. 
Our solution renders all computations vectorial, which can not only accelerate the calculations, but also reduce the time spent for changing register types (the registers used for vector and normal values are different).

%In order to evaluate the effect on performance of our SIMD lower bound distance calculation function, we measured the execution time of exact similarity search when all data are loaded in main memory (thus, factoring out the disk I/O cost).
%We compared our solution to the case where all computations are performed using Single Instruction Single Data (SISD).
%%Table~\ref{fig:res4} shows 
%The results showed that the average time cost per lower-bounding calculation when using SIMD is 2.6x faster than the SISD solution.
%This is a non-negligible speedup, which is attributed to the large number of vectorial computations that need to be executed in the context of data series similarity search. % (refer to Algorithm~\ref{algo8}).
%(We omit these results for brevity.)

\ignore{
\commentnote{1. Their workload balancing method is a kind of task parallelism. In detail, they divide works in each thread (Alg 6) in unit of task (Each loop of Alg 9), then execute each task in parallel. They need to discuss their workload balancing in further detail, or in the context of task parallelism (such as [1]).[1] Subhlok, Jaspal, et al. "Exploiting task and data parallelism on a multicomputer." ACM SIGPLAN Notices. Vol. 28. No. 7. ACM, 1993.\\
. The quality of presentation should be significantly improved.
Concrete descriptions for challenges, goals and motivations of workload balance are required, as this is one of the important contributions.}
}

\subsubsection{Exact Search in nb-ParIS+}
\label{nb-paris}

We now present nb-ParIS+ that served as an intermediate step for developing ParIS+, using a simple design with no communication among the distance computation worker threads (see also \textsection~\ref{discussion-paris-nb-paris}).

\begin{figure}[tb]
	\centering
	\hspace*{-0.5cm}
	\includegraphics[width=0.9\columnwidth]{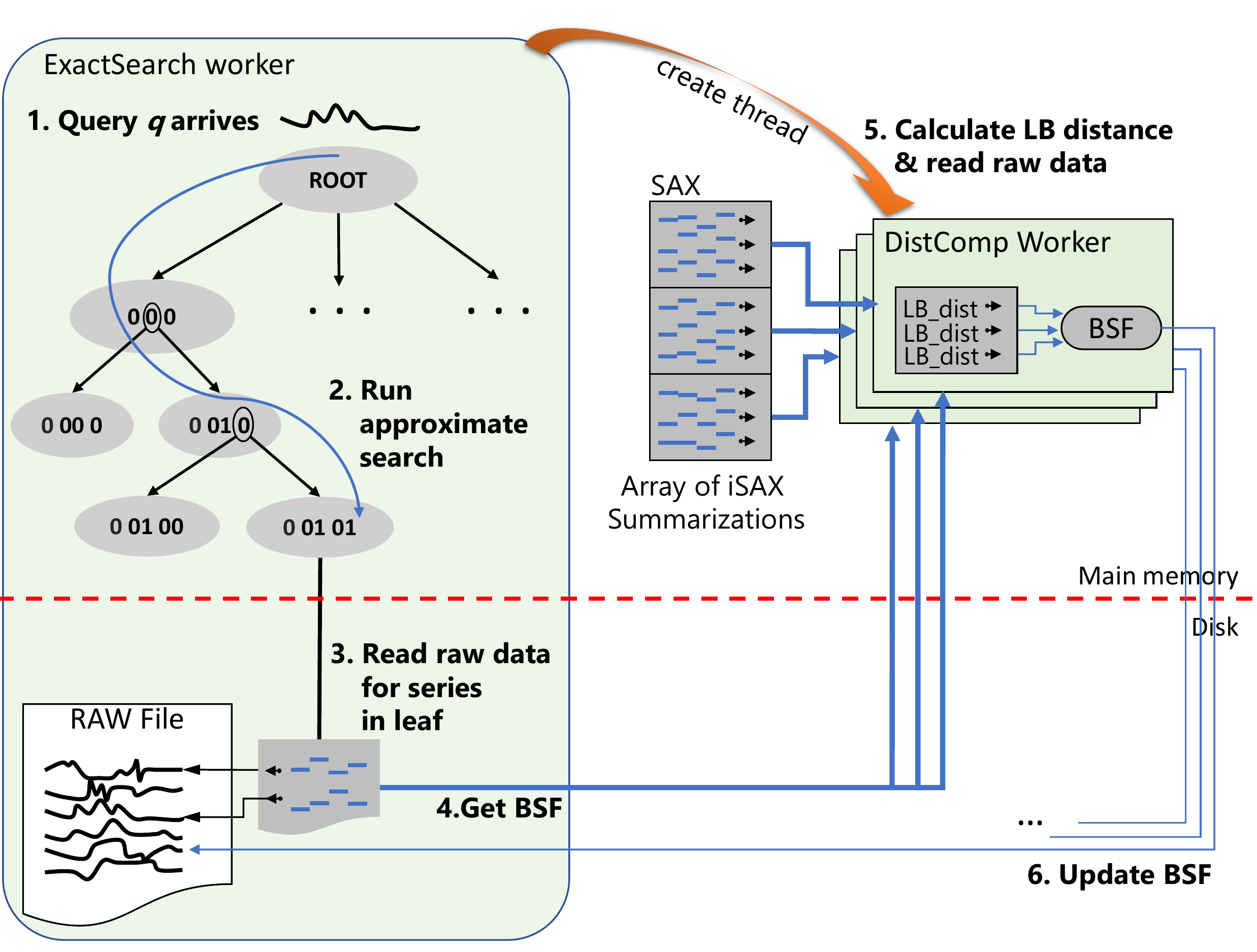}
	\caption{Workflow and algorithms for query answering with nb-ParIS+ (non-balanced).}
	\label{fig:parisnb}
\end{figure}

Exact Search in nb-ParIS+ is illustrated in Figure~\ref{fig:parisnb}, and shown in Algorithm~\ref{algo5}.
It employs a standard parallelization technique, which splits SAX in blocks and has different workers, called {\em Distcomp workers}, work on different blocks independently.
When a thread $t$ executes an ExactSearch (Algorithm~\ref{algo5}), it first performs an approximate search to get the initial BSF value (line~\ref{al5:aps}).
BSF is used for pruning. Each Distcomp worker updates its own copy of BSF to store the minimum distance it has calculated so far. 
This copy is stored in an appropriate element of vector $V_{bsf}$.
Note that since each worker calculates its own estimate of BSF, no synchronization is needed among them.

When $t$ creates the Distcomp workers (line~\ref{al5:workers}),
it informs them about the initial BSF value  
through vector $V_{bsf}$.
%$V_{bsf}$ has one element for each DistComp worker. 
Each such worker computes the lower bound distance between the query PAA and each iSAX summarization in its SAX part (Algorithm~\ref{algo6}, line~\ref{al6:cal}).
It does so using the SIMD approach we described in Section~\ref{sec:lowerboundsimd}. 
If this distance is higher than the current value of the worker's copy of BSF, then the data series is pruned. Otherwise, the Distcomp worker reads the required data from disk, calculates the real distance (line~\ref{al6:rcal}), 
and if necessary, updates  the appropriate element of $V_{bsf}$ (line~\ref{al6:update}).
Finally, $t$ waits for all DistComp workers to finish, 
calculates the minimum value stored by these workers in $V_{BSF}$, and returns this value. 
We use one DistComp Worker thread per core (thus resulting in 24 DistComp workers in total). 
Note that nb-ParIS+ does not necessarily balance the work among the DistComp workers, since the pruning degree may turn out to be different for each worker.
Moreover, different threads produce disk requests concurrently, which results in random accesses to disk. 
ParIS+ improves upon nb-ParIS+ to address these problems.

\begin{algorithm}[tb]
	{	\footnotesize
		\SetAlgoLined
		\KwIn{\textbf{querySeries} $QTS$, \textbf{query iSAX} $isax$, \textbf{Index} $index$, %\textbf{queryLeafSize} $qs$, 
			\textbf{File*} $file$}
		\KwOut{$realDistance$}
		%	\textbf{SAX Summarizations} $SAX$\;
		\textbf{float} $BSF$,$V_{bsf}[]$\;
		%	\tcp*[f]{Read iSAX summary in memory}\\
		%	  \If{$SAX$ = $\emptyset$} {
		%	$SAX$ = $loadSAXFromDisk$ ();\label{al5:readsax}
		%  }
		
		\tcp*[f]{ Perform an approximate search}\\
		$BSF$ = $approxSearch$ ($QTS$, $isax$, $index$)\;\label{al5:aps}
		\tcp*[f]{ distribute BSF into $V_{bsf}$}\\
		$V_{bsf}\leftarrow BSF$\;
		%\vspace{0.5em}
		\tcp*[f]{calculate minDist and realDist in parallel}\\
		
		create a number of threads, each executing an instance of DistCompWorker($TS$, $isax$, $V_{bsf}$, appropriate part of $index.SAX$, $file$)\;\label{al5:workers}
		
		Wait for all threads to finish\; \label{al5:finish}
		\Return ($min$ ($V_{bsf}$))\;\label{al5:return}
	}
	\caption{nb-ParIS+: $ExactSearch$}
	\label{algo5}
\end{algorithm}

\begin{algorithm}[tb]
	{	\footnotesize
		\SetAlgoLined
		\KwIn{\textbf{querySeries} $TS$, \textbf{query iSAX} $isax$, \textbf{float} $V_{bsf}$[], \textbf{iSAX summarizations} $SAX\_part$[], \textbf{Index} $index$, \textbf{File*} $file$} \label{al5_2:receive}
		\KwOut{$BSF$}
		\textbf{float} $BSF$ = read initial BSF value from $V_{bsf}$\;
		\For{$i$ $\leftarrow$  $1$ \emph{\KwTo} size of $SAX\_part$}
		{
			$minDist$ = $LowerBound\_SIMD$ ($TS$, $SAX\_part[i]$)\;\label{al6:cal}
			\If{$minDist<BSF$}
			{
				Move file pointer to appropriate position in $file$\;
				$rawData$ = read raw data series from $file$\;
				$realDist$ = $Dist$ ($rawData$, $TS$)\;\label{al6:rcal}
				\If {$realDist < BSF$}
				{
					$BSF\leftarrow  realDist$\;\label{al6:update}
				}
			}
		}
	}
	\caption{nb-ParIS+: $DistComp$}
	\label{algo6}
\end{algorithm}

%\subsubsection{Balanced Exact Search}
\subsubsection{Exact Search in ParIS+}
\label{sec:paradsplus}

%In the worst case, all disk accesses will be executed by a single thread. 
%Therefore, the benefit of using multi-core CPUs is wasted.
%In addition, the density of access disk isn't as much as the disk can afford so that we need to increase the number of thread to read time series on disk and make sure that the quantity of access disk not rise too much. 
%This part will introduce the deeply parallel function. we resolve the problem that  unbalance distribution of access disk during the parallel query answering and the low density of access the disk.
As in nb-ParIS+, the exact search algorithm in ParIS+\footnote{Note that the exact search algorithm of ParIS is the same as in ParIS+.} employs approximate search as a first step and uses the approximate answer as the initial BSF (see Algorithm~\ref{algo7}). 
Unlike to nb-ParIS+ though, BSF is now stored in a variable
shared by all workers. 
We note that the state-of-the-art {\em sequential} index similarity search algorithm spends more than 95\% of its time on I/O (on our server, described below) and in particular, on reading data from disk.
In order to achieve better parallelism,
%benefit by parallel I/O and skip sequential reading, \here{Botao: I do not understand the previous phrase.}
the ExactSearch algorithm separates the phase of the lower bound calculation from that of the real distance calculation and has two types of worker threads,
namely the Lower Bound Computation ({\em LBC}) and the Real Distance Computation ({\em RDC}) workers, respectively, 
executing each type of calculation (see right part of Figure~\ref{fig:qa2}).

When a thread $t$ executes an ExactSearch (Algorithm~\ref{algo7}), it first performs an approximate search to get the initial BSF answer (line~\ref{al7:aps}),
and then it initiates a number of LBC workers (line~\ref{al8:list}).
Different LBC workers work on different parts of SAX. 
Each such worker computes the lower bound distance between the query PAA and each iSAX summarization in its SAX part and records the data series for which this distance is less than the current BSF in a local candidate list, which it eventually returns to $t$ (see Algorithm~\ref{algo8}). 
This list contains the position and the lower-bound distance, needed to read the raw data and to calculate 
the real distance for the data series.
Once all LBC workers have finished, $t$ merges the candidate lists they have created (Algorithm~\ref{algo7}, line~\ref{al7:merge}) and initiates the RDC Workers (line~\ref{al7:real}).

Each RDC Worker (Algorithm~\ref{algo9}) repeatedly retrieves a (minDistance, position) pair from the merged candidate list ($C_l$) in an atomic way (line~\ref{al9:pair}). 
Atomicity is achieved with the use of a lock which all RDC workers share. 
%If necessary (line~\ref{al7:juge}), the worker 
The worker then reads the required data from disk, 
calculates the real distance (line~\ref{al9:rcal}), 
and if necessary, updates the shared BSF variable (line~\ref{al9:bsf}).
A thread lock ensures that the BSF modification is done atomically.
%
%Observe that 
Storing BSF in shared memory and updating it during the course of the execution
contributes towards reducing the number of calculations that RDC workers perform.

In this study, we use 1 LBC Worker thread per core, and 5 RDC Worker threads per core.
Oversubscribing the RDC Workers (that are involved in expensive I/O operations) ensures that we saturate the disk I/O bandwidth and the CPU remains busy.
%Note that we instantiate more RDC Workers than LBC Workers, (our experiments demonstrate that five times more is enough), in order to make sure that we saturate the disk I/O bandwidth. 
Our experiments showed that time performance remains relatively stable as we vary the number of RDC Worker threads per core (especially between 3-5 threads for the HDD server, and 4-10 threads for the SSD server), %(as shown in Figure~\ref{fig:readernumbertest}), 
while 1 LBC Worker thread was enough to achieve the best performance (results omitted for brevity). 
%(For brevity, we omit these experiments.)
%\commentnote{Is lock contention an issue here?}
%\begin{figure}[tb]
%\centering
%\includegraphics[width=\columnwidth]{queryanswering.pdf}
%\caption{Example of ParIS Exact Search.}
%\label{fig:sims2}
%\vspace{-1em}
%\end{figure}

\begin{algorithm}[tb]
{	\footnotesize
    \SetAlgoLined
	\KwIn{\textbf{querySeries} $QTS$, \textbf{query iSAX} $isax$, \textbf{Index} $index$, \textbf{File*} $file$}
%,\textbf{queryLeafSize} $qs$
	\textbf{candidate list} $C_l$, $subC_l[]$\;
%	\textbf{Thread Lock} $lock_{BSF}$\;
	\textbf{float} $BSF$\;
%	\tcp*[f]{ Read iSAX summary in memory}\\
%	  \If{$SAX$ = $\emptyset$} {
%	$SAX$ = $loadSAXFromDisk$ ()\;\label{al7:readsax}
%  }
 %\tcp*[f]{Perform an approximate search}\\
BSF = approxSearch($QTS$, $isax$, $index$)\;\label{al7:aps} %, $qs$

	%\tcp*[f]{initiate the Lower Bound Compute workers}\\
	create a number of threads, each executing 
              $subC_l \leftarrow LBCWorker$($QTS$, proper part of $index.SAX$, BSF)\;

        Wait for all threads to finish\;
	$C_l\leftarrow$ merge all sublists ($subC_l$) returned by the LBCWorker threads\;\label{al7:merge}
	%\tcp*[f]{initiate the Real Distance Compute Workers} \\
	create a number of threads, each executing an instance of $RDCWorker$ ($QTS$, $C_l$, BSF, $file$)\;\label{al7:real}
        Wait for all threads to finish\;
	\Return (BSF)\;\label{al7:return}
}
\caption{ParIS+: $ExactSearch$}
\label{algo7}
\end{algorithm}

\begin{algorithm}[tb]
{	\footnotesize
    \SetAlgoLined
	\KwIn{\textbf{querySeries} $QTS$, \textbf{iSAX summarizations} $SAX\_part[]$, \textbf{float} BSF}
 \textbf{local candidate list} $subC_l$\;
	\For{$i$ $\leftarrow$  $1$ \emph{\KwTo} size of $SAX\_part$ }
	{
		$minDist \leftarrow LowerBound\_SIMD$ ($QTS$, $SAX\_part[i]$)\;\label{al8:simddistancecalculation}
		\If{$minDist <$ BSF}
		{
			add ($minDist$, Raw Data file position of $SAX\_part[i]$) pair in $subC_l$\;\label{al8:list}
		}
	}
        \Return{($subC_l$)}
}
\caption{ParIS+: $LBC-Worker$}
\label{algo8}
\end{algorithm}

\remove{
\yy{ Each of them computes the lower bound distance between the query PAA 
and each iSAX summarization stored in its part of SAX. For those that 
the distance is less than the BSF, ++++}

The pseudocode for \textcolor{red}{LBC Worker} is shown in Algorithm~\ref{algo8}.
Each \textcolor{red}{LBC Worker} calculates the lower bound distance (line~\ref{al8:simddistancecalculation})
for the positions in SAX that has been associated to it, 
and adds a pair in its local candidate list if needed.
}

\begin{algorithm}[t]
{	\footnotesize
    \SetAlgoLined
	\KwIn{\textbf{querySeries} $QTS$, \textbf{candidate list} $C_l$, \textbf{float} BSF, \textbf{File*} $file$}
	\While{not reached end of $C_l$}
	{	
		{\em Atomically} read the next ($minDist$,$position$) pair from $C_l$\;\label{al9:pair}
		\If{$minDist < $BSF\label{al9:juge}} 
		{
			Move file pointer to the proper position in $file$\;
			$rawData \leftarrow$ read raw data series from file\;
			$realDist \leftarrow Dist$ ($rawData$, $QTS$)\;\label{al9:rcal}
			\If{$realDist < $ BSF}
			{
%				lock ($lock_{BSF}$)\;
					{\em Atomically} update BSF to  the value of $realDist$\;\label{al9:bsf}
%				unlock ($lock_{BSF}$)\;
			}
		}
	}
}
\caption{ParIS+: $RDC-Worker$}
\label{algo9}
\end{algorithm}

\subsubsection{Discussion of nb-ParIS+ and ParIS+}
\label{discussion-paris-nb-paris}

Using nb-ParIS+, we were able to identify some design choices that 
resulted in bad performance during query answering. 
Specifically, nb-ParIS+ needs synchronization between the different threads only for computing the minimum BSF value, but may result in load imbalance in terms of real distance calculations performed in each chunk. 
Since each real distance calculation performs I/O (to read the raw data series), some threads may finish much later than others. 
Moreover, as the requests of different threads are interleaved, 
nb-ParIS+ may perform random I/Os. 
ParIS+ addresses these problems in common cases (when the pruning ratio is large), by separating the phase of lower bound distance calculations from that of real distance calculations through the use of the candidate list. 
The candidate list is sorted to ensure that random accesses to disk are minimized. 
Moreover, a fetch\&add is used to assign entries of the candidate list to threads for processing, in order to achieve load balancing. 
In this way, it is ensured that all threads finish at about the same time.

\section{Experimental Evaluation}
\label{sec:experiments}
%In this section, we describe our experimental evaluation. 

%\subsection{Setup}

\noindent\textbf{[Setup]} 
We ran the experiments on two servers, whose physical memory was limited 
to 75GB\footnote{We used GRUB to limit the amount of RAM, so that all methods 
are forced to use the disk. Note that GRUB prevents the operating system 
from using the rest of the RAM as a file cache.}. %, which is what we wanted for our experiments.}.% except for the memory-resident experiments, where we use no limit and all datasets fit comfortably in RAM (in this case, the total RAM is 384GB). 
The first server (default) comprises two Intel Xeon E5-2650 v4 2.2Ghz processors with 12 cores 
%(24 hyper-threads, 30M shared L3-cache) 
each, and has 10.8TB (6 x 1.8TB) 10K RPM SAS HDD drives in RAID0,
with sequential access throughput of the RAID0 array being 1200MB/sec and random access throughput 12MB/sec. %3164 iops (I/O operations per second).
The second server, with the same setup for CPUs and memory, has 3.2TB (2 x 1.6TB) SATA
%SATA2 
SSD drives in RAID0, with 500MB/sec sequential throughput and 450MB/sec random access throughput. %112626 iops(I/O per second).
%Unless otherwise noted, experiments were run using HDDs. 

All algorithms were implemented in C, and compiled using the GCC6.2.0 with the O3 optimization flag % (our SIMD code was better than the one produced by using the O3 flag)
on Ubuntu Linux 16.04.
Unless otherwise mentioned, in our experiments we use one socket for index creation and two sockets for query answering. 
%The number of sockets that we use is determined by the number of threads that we need to perform the computation. 
%We need much less threads for creating the index than for query answering (so in the index creation, one socket is enough).

\noindent\textbf{[Datasets]} 
In order to evaluate the performance of the proposed approach, 
we use several synthetic datasets for a fine grained analysis, 
and two real datasets from diverse domains.
Unless otherwise noted, the series have a size of 256 points, 
which is a standard length used in the literature,
and allows us to compare our results to previous work.

We used synthetic datasets of sizes 50GB-250GB (default size: 100GB),
%For the synthetic datasets, we used 
and a random walk data series generator that works as follows: 
a random number is first drawn from a Gaussian distribution N(0,1), 
and then at each point in time a new number is drawn from this distribution 
and added to the value of the last number. 
This generator has been extensively used in the past (and
has been shown to model real-world financial data)~\cite{yi2000fast,shieh2008sax,wang2013data,isax2plus,zoumpatianos2016ads}.
We used this process to generate 100 query series. 

%100GB(100 million) synthetic data set as default during index creation. In query answering, we use 25GB(25 million) synthetic dataset during vary core test and 50GB(50 million) - 250GB(250 million) dataset on scalable test. 

For the \emph{Seismic} real dataset, 
we used the IRIS Seismic Data Access repository~\cite{iris}
to gather 100M series representing seismic waves from various locations,  
% using a sliding window with a resolution of 1 sample per second, sliding every 4 seconds. 
%The complete dataset size was 110 GB. 
for a total size of 110GB.
%We used 100 series outside of the dataset as the query workload (produced using our synthetic series generator).
The \emph{SALD} real dataset includes neuroscience 
MRI data series~\cite{url:SALD}, for a total of 200M series of length 128 points each, and total size 100 GB.
%We obtained an additional 100 data series from the raw dataset using the same technique to be used as a query workload.
%We once again used 100 synthetic data series as the query workload.
In both cases, we used as queries 100 series that were not part of the datasets (produced using our synthetic series generator, since these datasets do not come with query workloads).

In all cases, we ran the experiments 5 times and report the mean values. 
We omit reporting error bars, since all runs gave results that were very similar ($<$3\% difference).
Queries were always run in a sequential fashion, one after the other, in order to simulate an exploratory analysis scenario, where users formulate
new queries after having seen the results of the previous one.

\noindent\textbf{[Algorithms]} 
We experiment with our ParIS and ParIS+ algorithms, and compare those to the sequential state-of-the-art data series index, 
ADS+~\cite{zoumpatianos2016ads}.
We also compare to (i) the UCR Suite~\cite{rakthanmanon2012searching}, the state-of-the-art, 
optimized serial scan technique 
%which implements all the known optimizations 
for exact 
%data series 
similarity search,
and (ii) DS-Tree~\cite{wang2013data}, a modern data series index that stores the raw data in the leaves.
All algorithms are available online~\cite{paradssources}.
%
%We also developed a parallel version, UCR Suite-P, that represents the optimal serial scan memory-resident solution.
%In UCR Suite-P, each thread works on a partition of the data; when all threads finish their work, we just need to merge the results and identify the most similar match.
%Note that UCR Suite-P always needs to read all data, so when the dataset is disk-resident, it performs random accesses, as a result of multi-plexing the disk read requests of the multiple threads, which slow the algorithm down. 
%This is not a problem for memory-resident datasets, since in this case, each core reads data from the local memory cache at maximum speed.
%It is for this reason that we compare to UCR Suite for the disk-resident experiments, and to UCR Suite-P for the in-memory experiments.
%
%
%Unless otherwise stated, ADS+ uses 1 thread, 
%UCR Suite 1, 
%\textcolor{red}{UCR Suite-P 24, ParIS-nb 6 for index creation and 24 for query answering, and ParIS 6 for index creation, 24 for in-memory computations, and 120 for reading raw data on HDD, 360 for reading raw data on SSD. {\bf ??? Botao let's discuss this! ???}}
%
For the disk-resident experiments, we never load the datasets in main memory. 
In order to mitigate the effects of caching, we clear the caches before each experiment 
(i.e., before running index creation and before executing each query).
%The query answering experiments are repeated 100 times (using 100 different queries, as detailed above), and we report the mean value.

\subsection{Results}
We present the performance results for %the index creation and query answering in 
ParIS/ParIS+, 
and compare them to %the results of 
two modern data series indices, 
ADS+ and DS-Tree. %the current state-of-the-art index creation algorithm, as well as with the DS-Tree data series index.
%We perform experiments on index creation, as well as query answering. 

\begin{figure}[tb]
	\includegraphics[page=1,width=0.9\columnwidth]{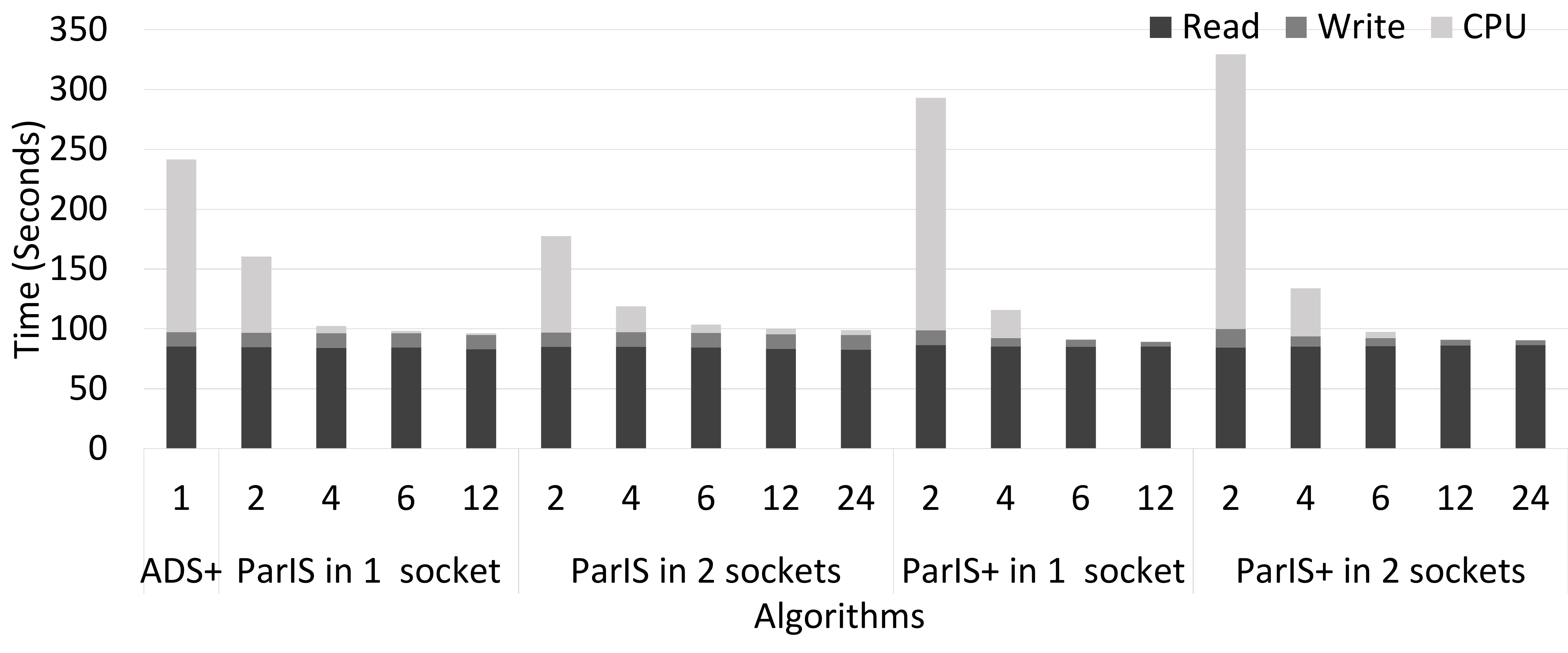}
	\caption{Index creation time (HDD) as the number of cores increases.}
	\label{fig:res1}
\end{figure}
%\begin{figure}[tb]
%	\includegraphics[page=1,width=\columnwidth]{ssdindexcreation}
%	\caption{\textcolor{red}{Index creation time (SSD) as the number of cores increases.}}
%	\label{fig:ssdindex}
%\end{figure}

\begin{figure}[tb]
	\subfigure[ParIS on 1 Socket]{
	\includegraphics[page=1,width=0.47\columnwidth]{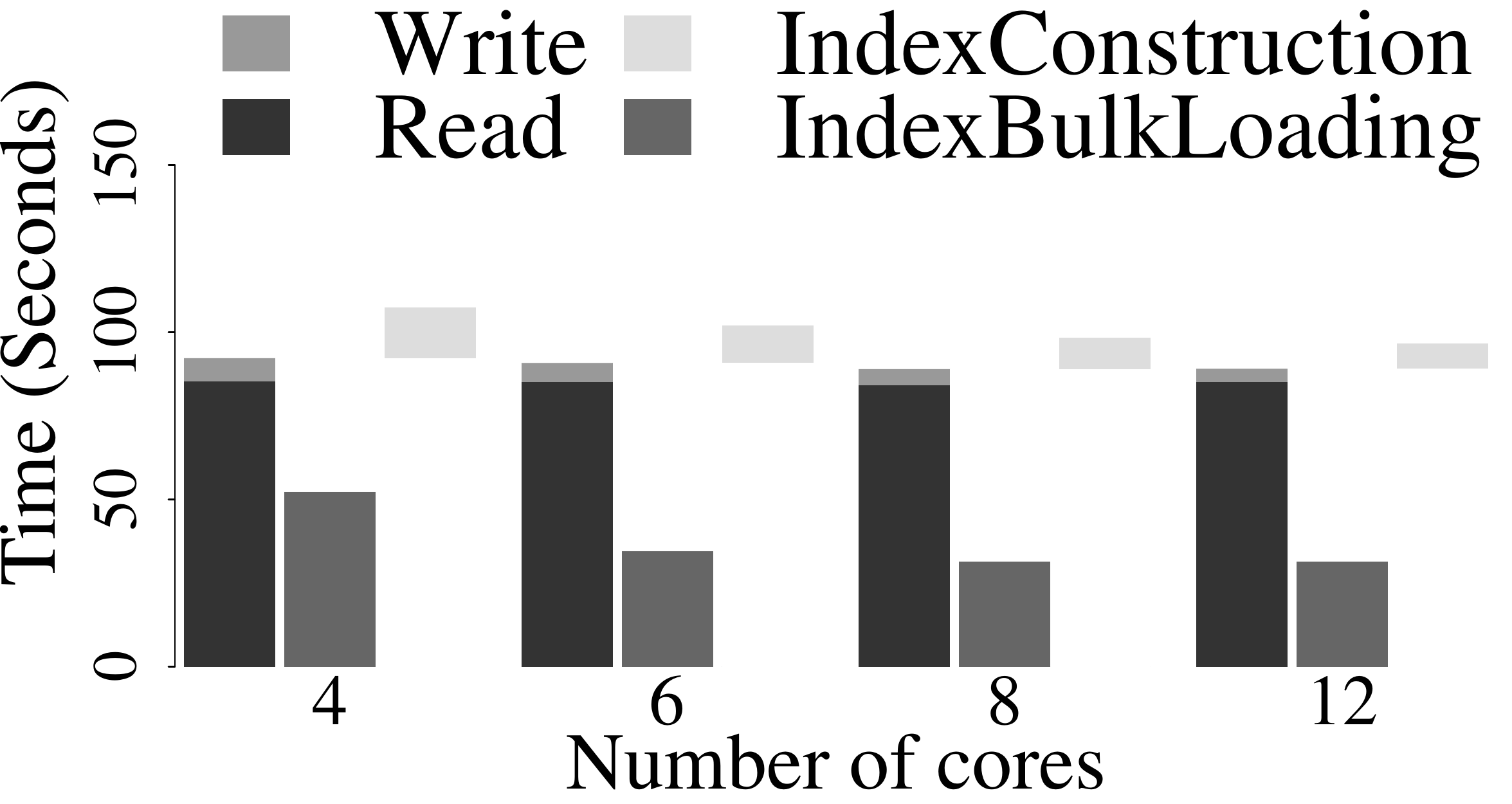}
}
\subfigure[ParIS on 2 Sockets]{
	\includegraphics[page=1,width=0.47\columnwidth]{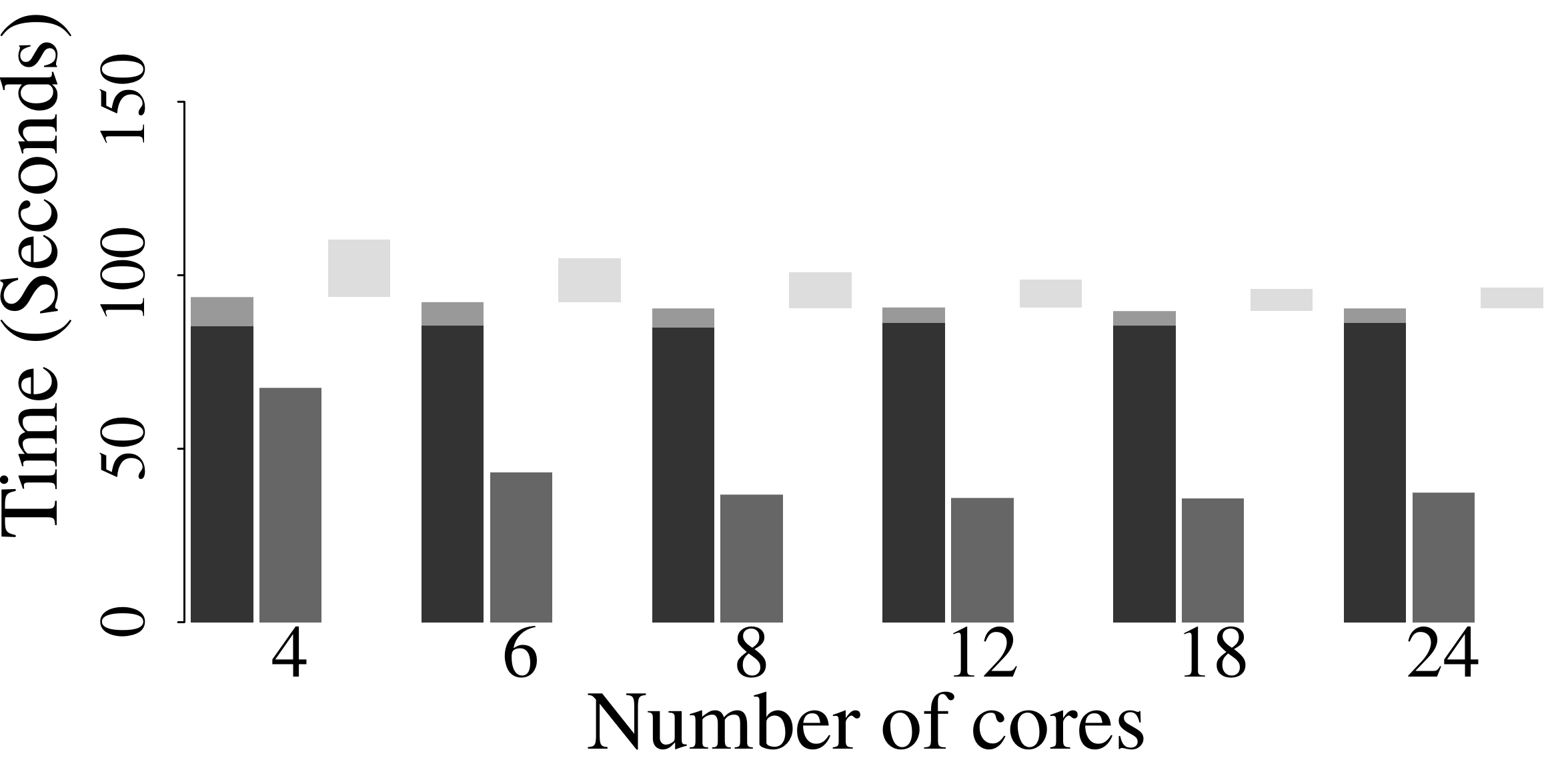}
}
\subfigure[ParIS+ on 1 Socket]{
	\includegraphics[page=1,width=0.47\columnwidth]{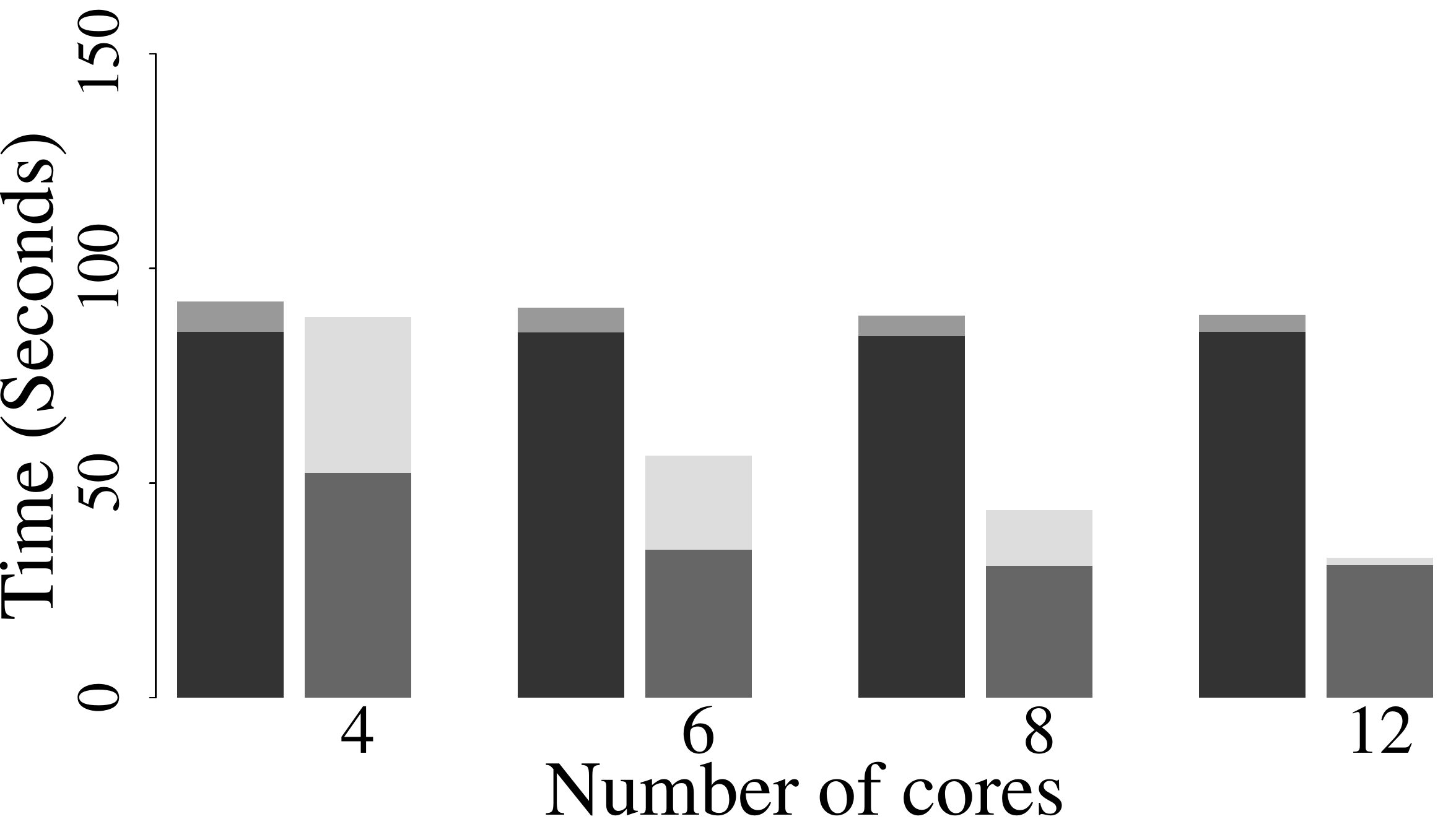}
}
\subfigure[ParIS+ on 2 Sockets]{
	\includegraphics[page=1,width=0.47\columnwidth]{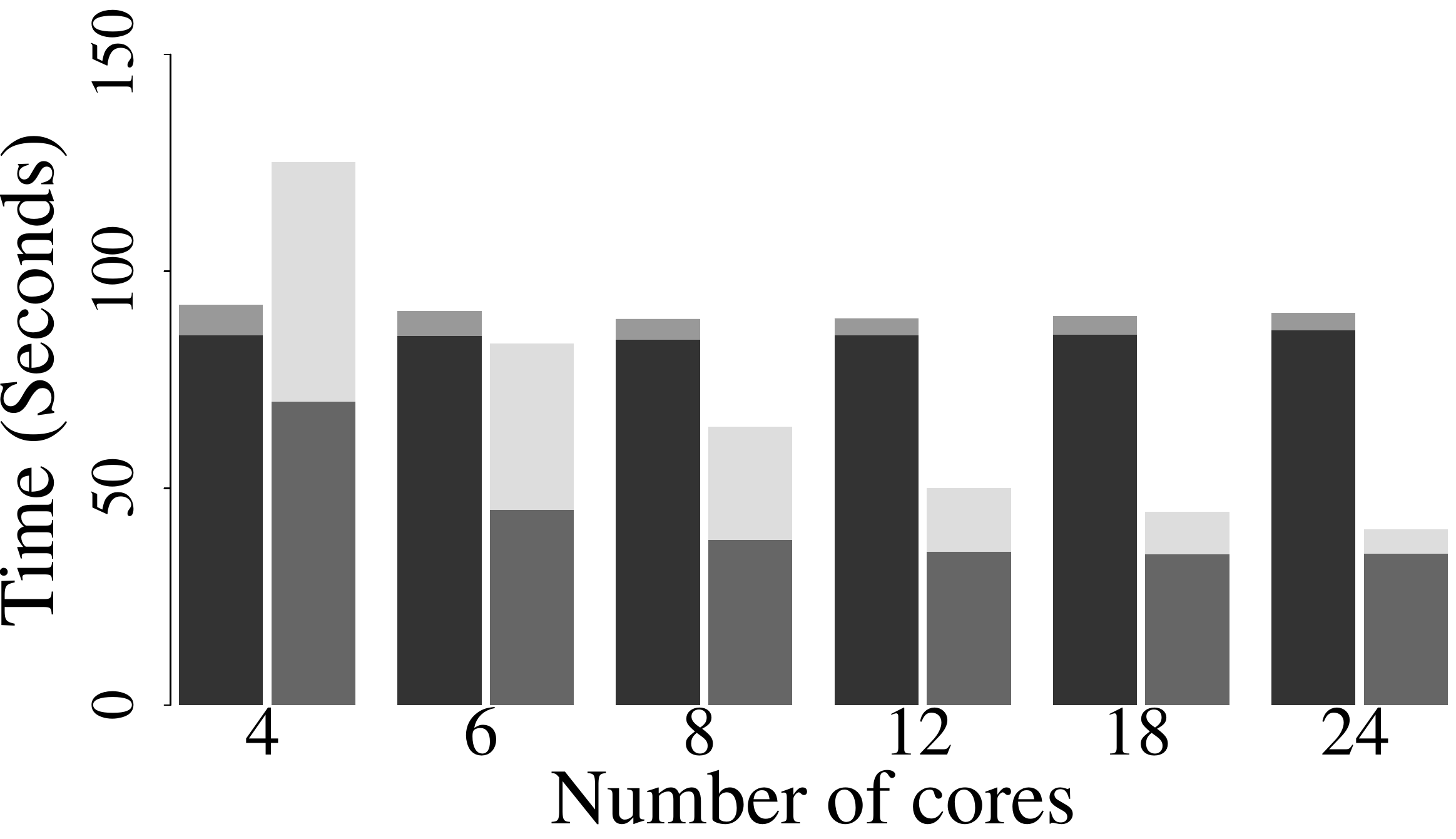}
}
	\caption{Overlap of I/O time and CPU time during index creation (HDD).}
	\label{fig:res12}
\end{figure}

\subsubsection{Index Creation Performance Evaluation}
%\commentnotecorrected{(W2) Some important details of the mechanism is missing.\\
%	For example, it isn't clear how many threads for each stage of the algorithm would be good in practice to side-step some of the I/O bottleneck.\\
%	Or it isn't clear how ParIS determines memory is almost full.\\}
In our first experiment (Figure~\ref{fig:res1}), we evaluate the time it takes to create the tree index for a synthetic dataset of 100M series.
The figure illustrates that the performance of ParIS and ParIS+ improves 
as the number of cores grows from 4 to 6 (note that a single thread runs on each core); 
after 6 cores the improvement is rather small.  
The reason for this behavior is illustrated in Figure~\ref{fig:res12}.
Note that there are 4 types of time costs: (i) read raw data from disk; (ii) write raw data to disk; 
(iii) CPU time by IndexBulkLoading workers; and
(iv) CPU time by IndexConstruction workers. 
%manage data, which involves processing the raw data and inserting the iSAX representation in the correct 
%RecBuf; and (v) create node, which takes place when we grow a subtree during the flushing of a RecBuf into the OutBufs. 
%If we look at the single core case (Figure~\ref{fig:exp1}), we can see that all types of computation are executing in a sequential manner. 
When we use more than one core, 
the time to read the data from disk and the management of data series are performed concurrently.
Moreover, the time cost for the management of data series decreases with the number of cores, since the data that each core needs to process gets reduced.
The cost of the index construction
also reduces.
However, 
%even though we increase the number of cores, 
the time cost to read data is always the same, since we access the same disk. 

Figure~\ref{fig:res1} shows that ParIS
%completely masks out the CPU latency 
%(using only 6 threads)
%and 
results in performance which is up to $2.4$x faster than ADS+. 
Still, ParIS does not completely hide the CPU latency.
This is achieved by ParIS+, when $6$ or more threads are used, as can be seen in Figures~\ref{fig:res1} and~\ref{fig:res12}. 
%Figure~\ref{fig:res12} shows the exact I/O cost (for read and write),
%as well as the CPU cost by each of the two types
%of worker threads, and their overlap.
Note that in ParIS+, more work is performed than in ParIS, because the IndexBulkLoading workers traverse the tree more than once. 
This cost is more evident in the $2$ sockets case, where
the threads do not benefit from the use of the L3 cache. 
%(as in the one socket case).
However, ParIS+ achieves better overlap of CPU time with I/O cost (Figure~\ref{fig:res12}).
Therefore, the time to execute the additional work completely overlaps with the I/O cost when the number of threads is at least $6$, and ParIS+ achieves better performance than ParIS. 

%Figure~\ref{fig:res12} also shows that ParIS and ParIS+
%have the potential to deliver more benefit as we move to faster storage media.
%This is particularly so for ParIS+ since ParIS will always pay the CPU
%cost to execute the non-overlapping part of the computation. 

Overall, these results demonstrate that not only does the proposed solution completely hide the CPU latency (using $\ge 6$ cores), but it will continue to do the same %(as shown by the trend of the CPU cost with increasing cores in Figure~\ref{fig:res5}) 
when the storage medium of the dataset becomes much faster, e.g., with NVRAMs.
In the following, we use 6 cores by default.
The results with SSD follow the same trends (in this case ParIS+ completely hides the CPU latency when using $\ge 4$ cores), and we omit them for brevity.

\begin{figure*}[tb]
	\remove{
	\begin{minipage}[b][5cm]{0.4\columnwidth}
		\includegraphics[page=1,height=3.5cm]{shadeillustration}
		\caption{Time cost illustration of index creation process}
		\label{fig:exp1}
	\end{minipage}
	}
\hspace*{-0.2cm}
	\begin{minipage}[b]{0.37\columnwidth}
%		\centering
		\includegraphics[page=1,height=3.0cm]{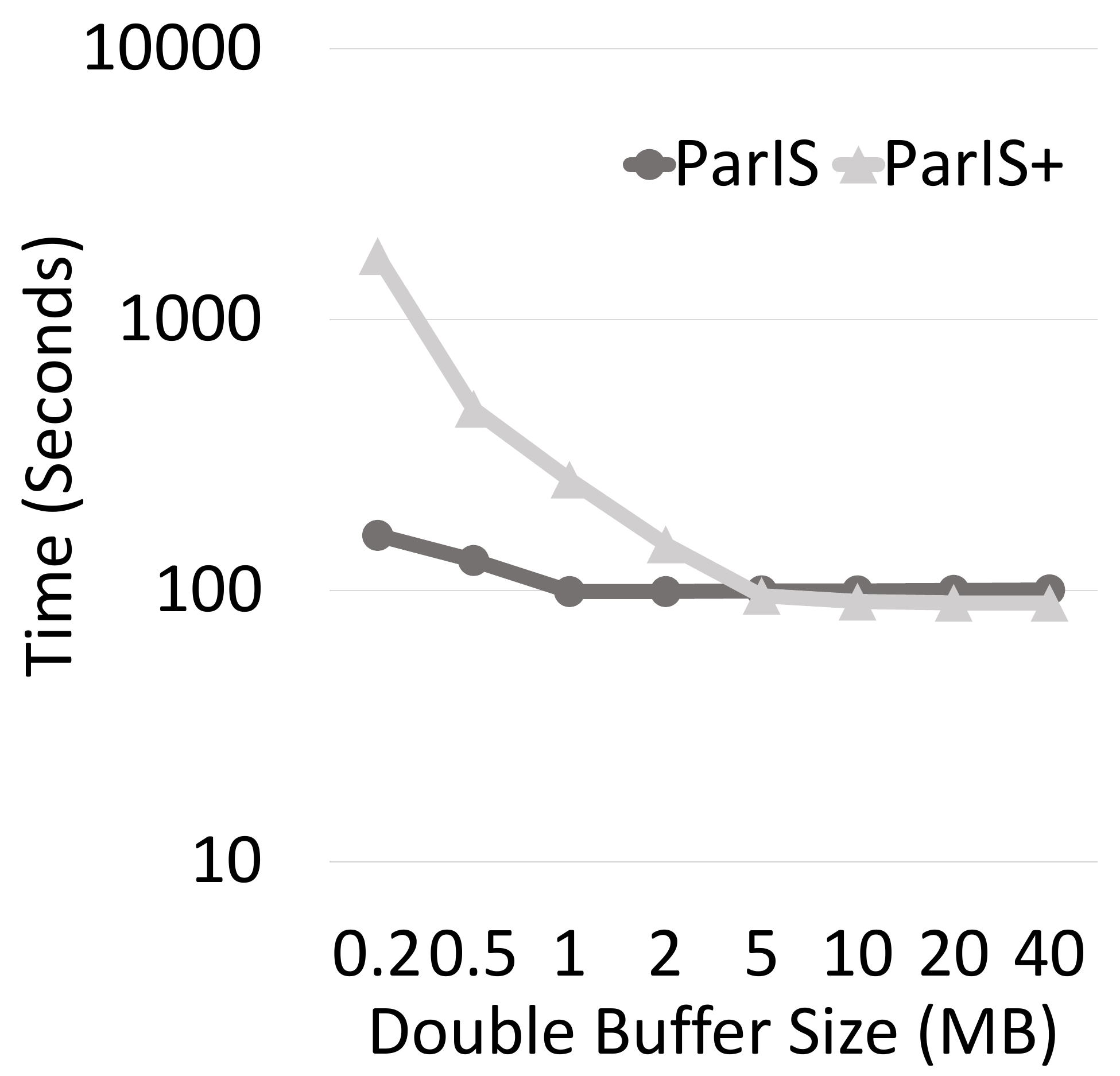}
		\caption{Index creation time vs double buffer size.}
		\label{fig:newin3}
	\end{minipage}
\hspace*{0.05cm}
		\begin{minipage}[b]{0.25\textwidth}
	\includegraphics[width=\columnwidth]{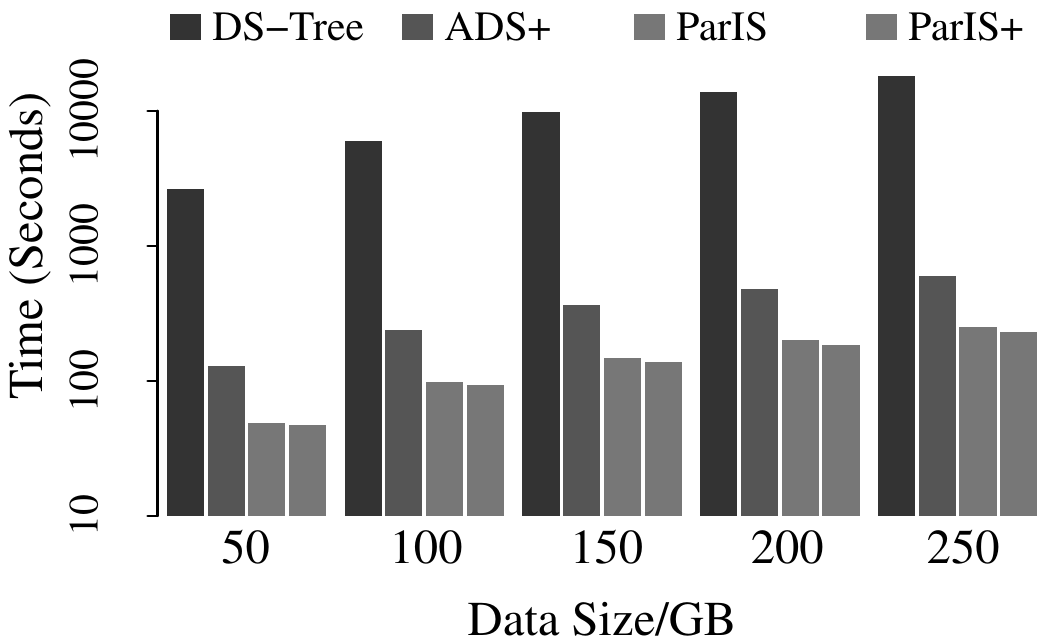}%scalableindex}
	\caption{Index creation time (HDD) vs dataset size.} % for varying dataset size} % (12 cores)}
	\label{fig:res6}
\end{minipage}
\hspace*{0.05cm}
\begin{minipage}[b]{0.25\textwidth}
	\includegraphics[width=\columnwidth]{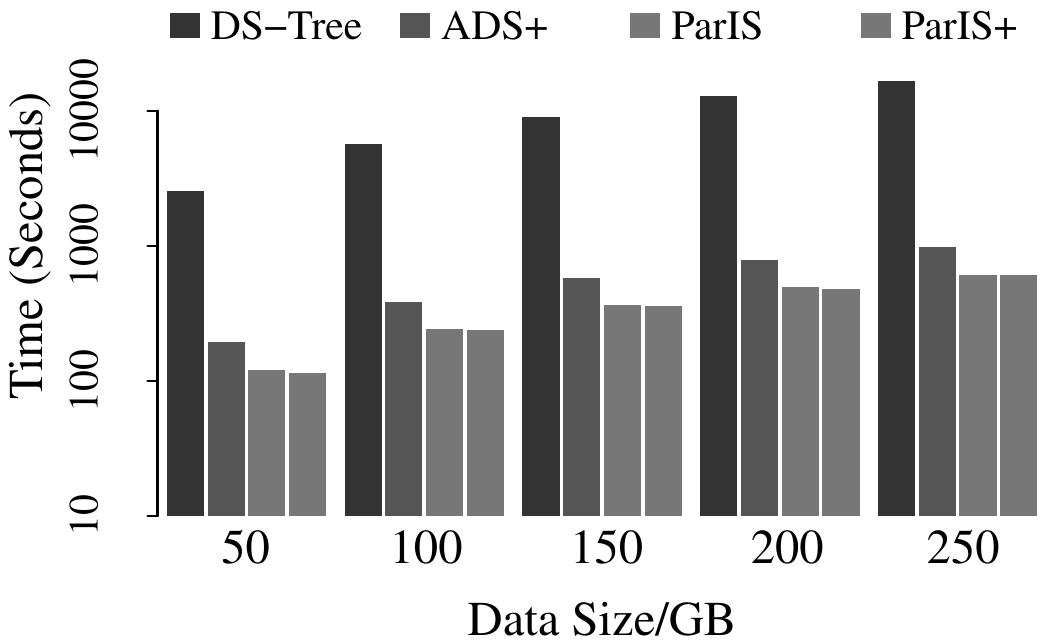}%scalableindex}
	\caption{Index creation time (SSD) vs dataset size.} % for varying dataset size} % (12 cores)}
	\label{fig:res6_2}
\end{minipage}
\hspace*{0.05cm}
	\begin{minipage}[b]{0.6\columnwidth}
		\hspace*{-0.15cm}
		\includegraphics[page=1,height=3.1cm]{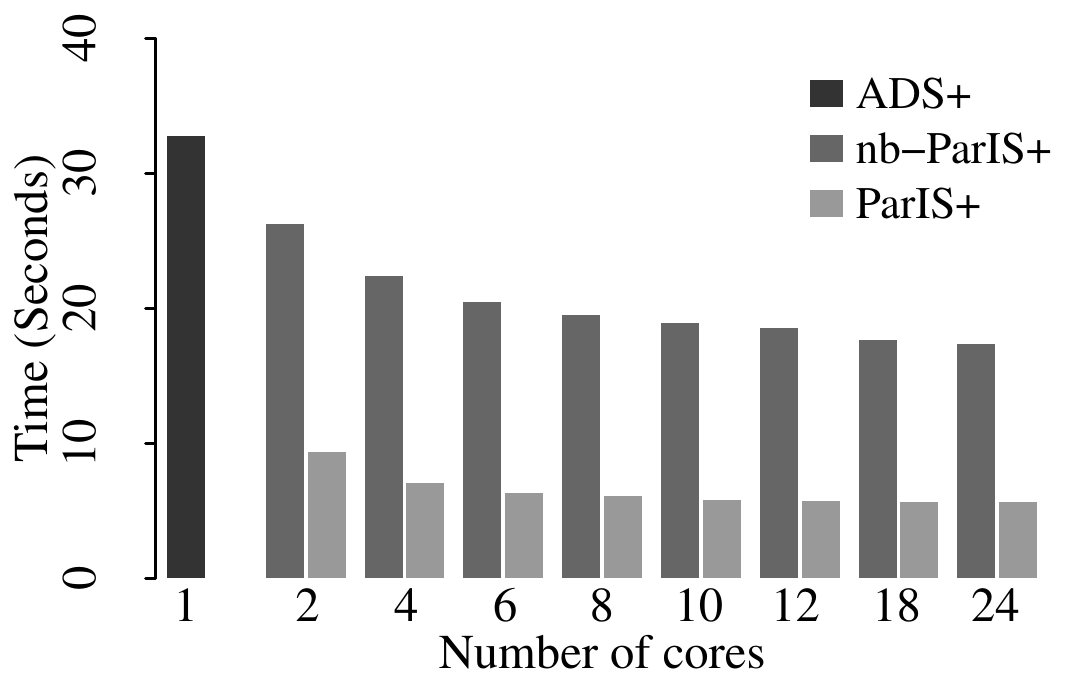}
		\caption{Exact query answering time vs number of cores (HDD).} %(5 worker threads per core)}
		\label{fig:res3_2}
	\end{minipage}

\end{figure*}
%\begin{figure}[h]
%	\centering
%	\includegraphics[page=1,width=\columnwidth]{newparisvarycore}
%	\caption{\textcolor{red}{Time cost of ParIS+ index creation in vary core}}
%	\label{fig:newin}
%\end{figure}

\begin{figure*}[tb]
	\subfigure[Lower Bound Calculation (LBC) worker\label{fig:detailofLBDW}]{
	\includegraphics[page=1,width=0.64\columnwidth]{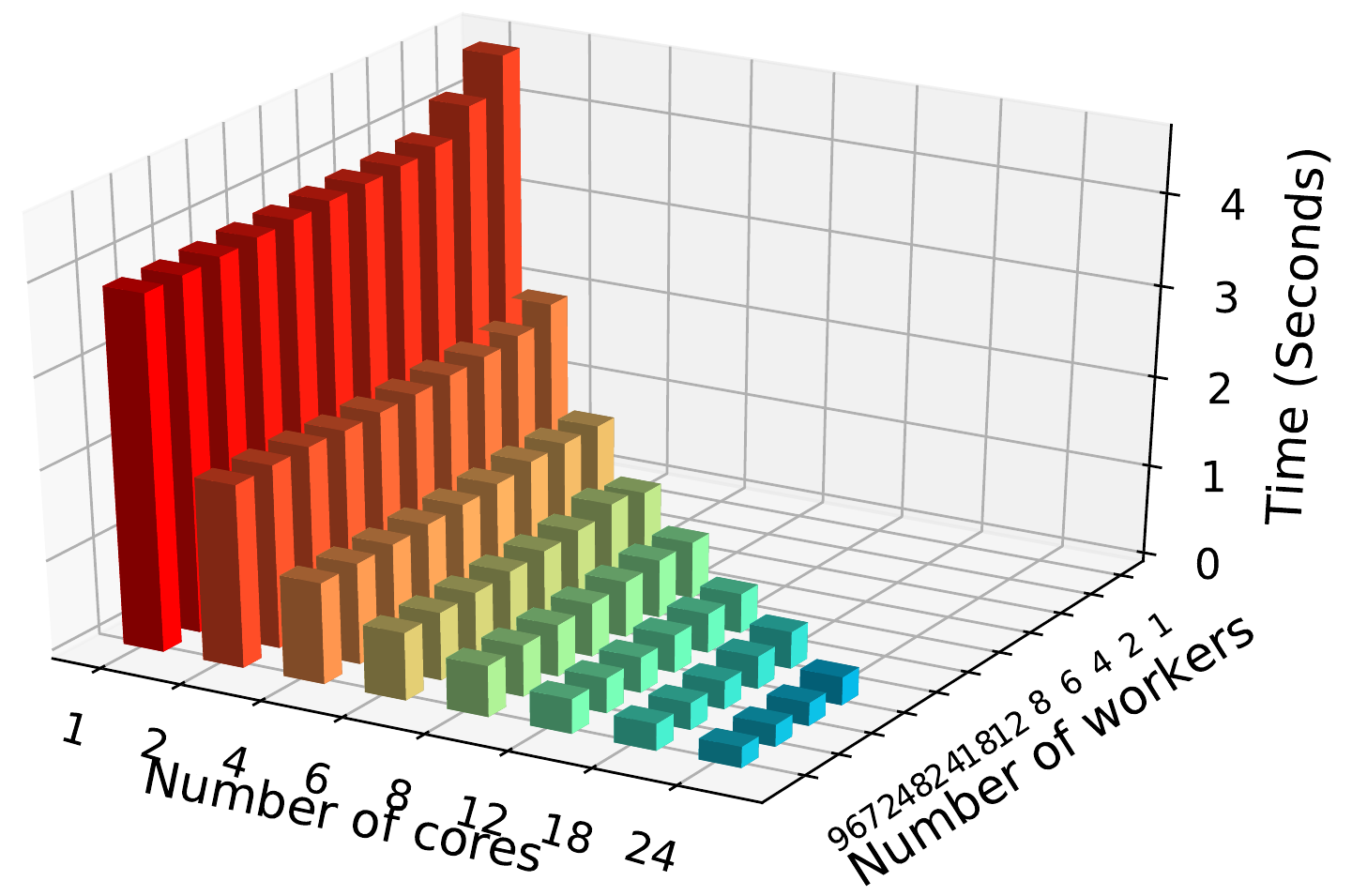}
	}
	\subfigure[Real Dist. Calculation (RDC) worker - HDD\label{fig:detailofRDCWHDD}]
	{
	\includegraphics[page=1,width=0.64\columnwidth]{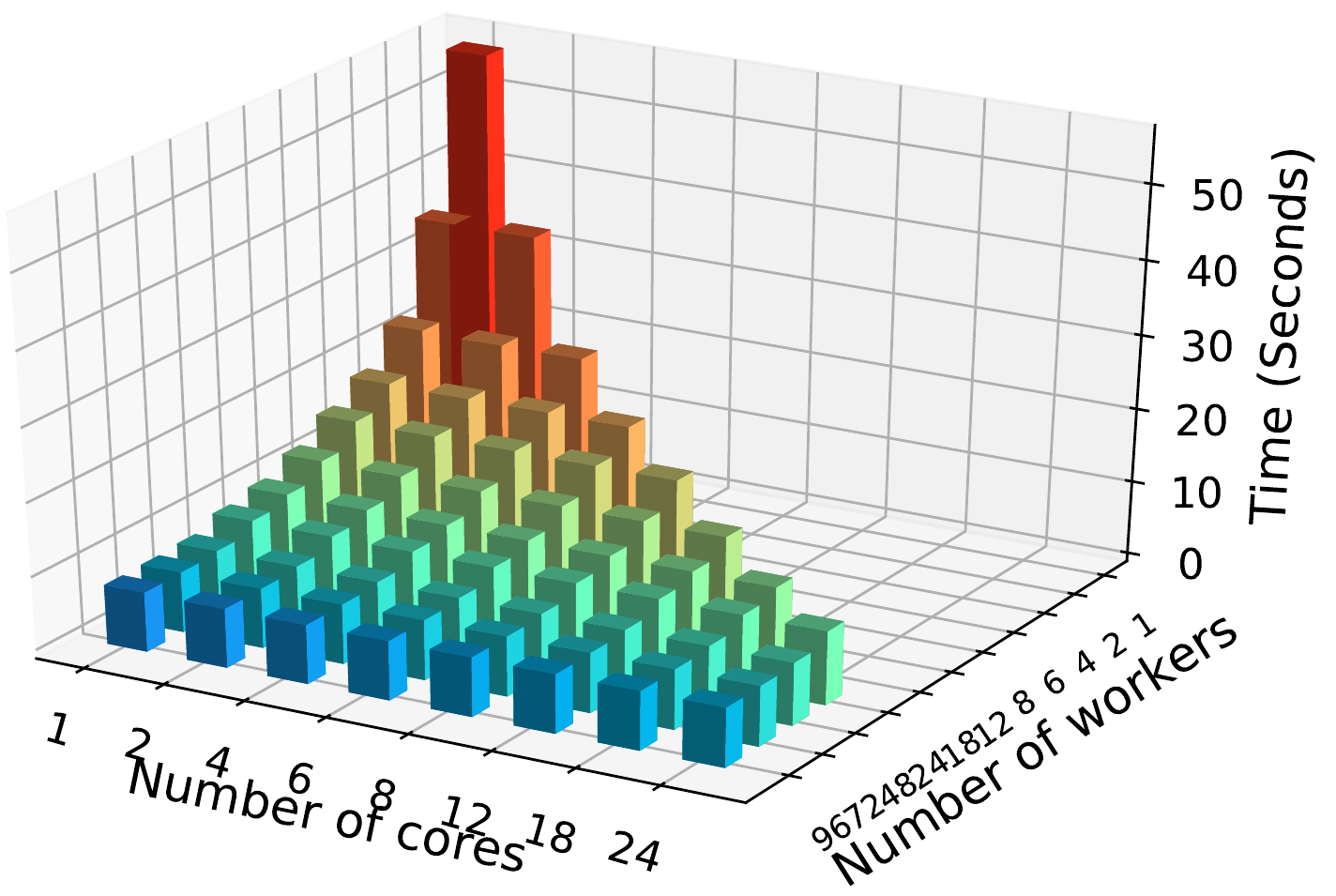}
}
\subfigure[Real Dist. Calculation (RDC) worker - SSD\label{fig:detailofRDCWSSD}]
{
	\includegraphics[page=1,width=0.64\columnwidth]{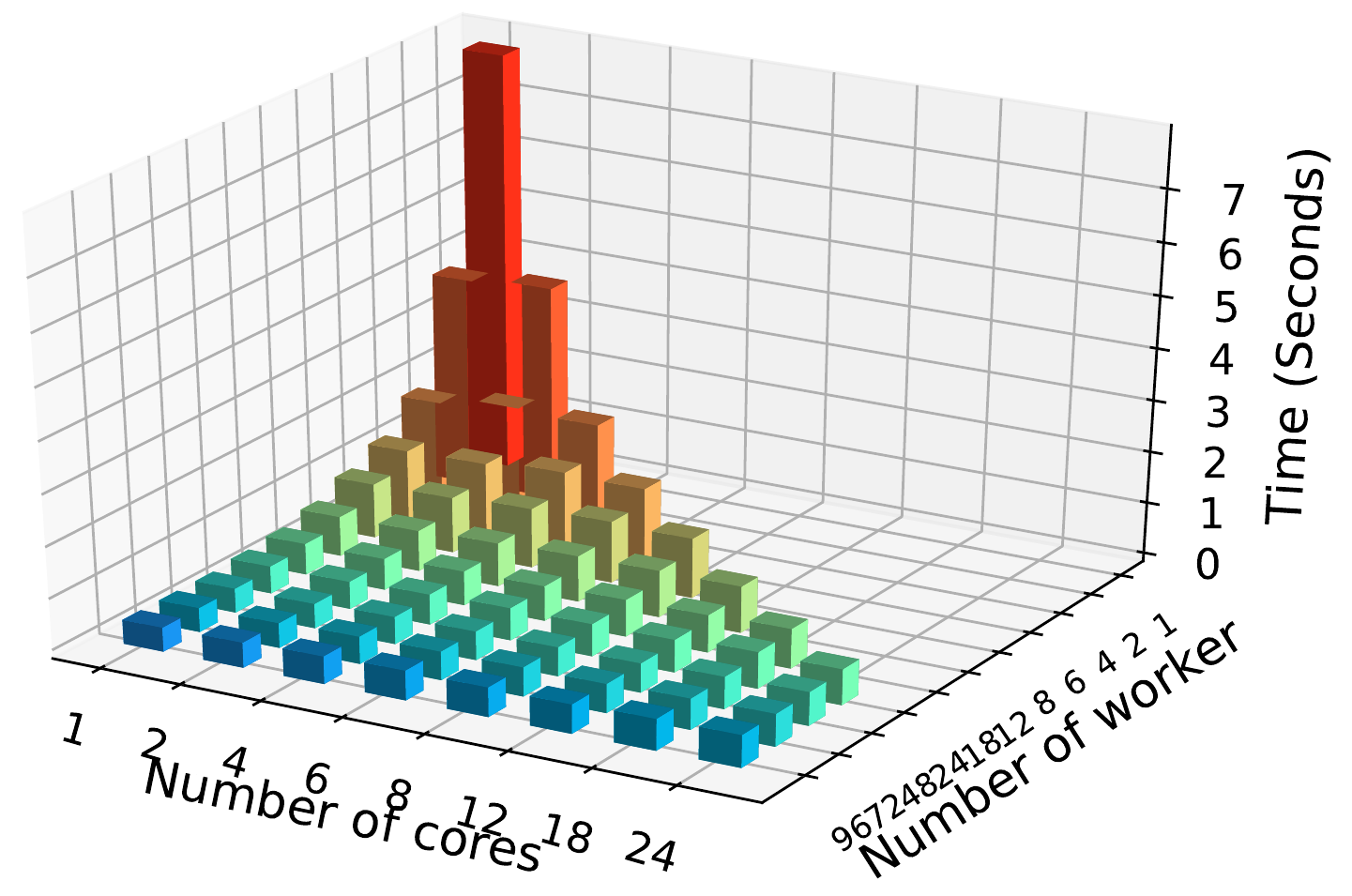}
	}
\caption{Time cost of ParIS+'s query answering workers, varying the number of cores and the number of workers.}
\label{fig:detailofLbcRdc}
\end{figure*}

Figure~\ref{fig:newin3} shows the impact of the double buffer size on performance (for the same experiment as in Figure~\ref{fig:res1}).
The results show that a good choice for the size of the raw data buffer is 1MB for ParIS, whereas it is 5MB for ParIS+. 
The reason for this difference is that as the buffer size increases, the IndexBulkLoading workers in ParIS+ traverse the index tree fewer times, and achieve better overlap with the work performed by the coordinator.

\begin{figure*}[tb]
\hspace*{-0.15cm}
%\begin{minipage}[h]{\columnwidth}
%	\includegraphics[page=1,width=1.03\columnwidth]{readernumbertest}
%	\caption{Time cost for varying number of RDC workers (HDD and SSD). }
%	\label{fig:readernumbertest}
%\end{minipage}
\begin{minipage}[tb]{0.25\textwidth}
	\includegraphics[width=1.03\columnwidth]{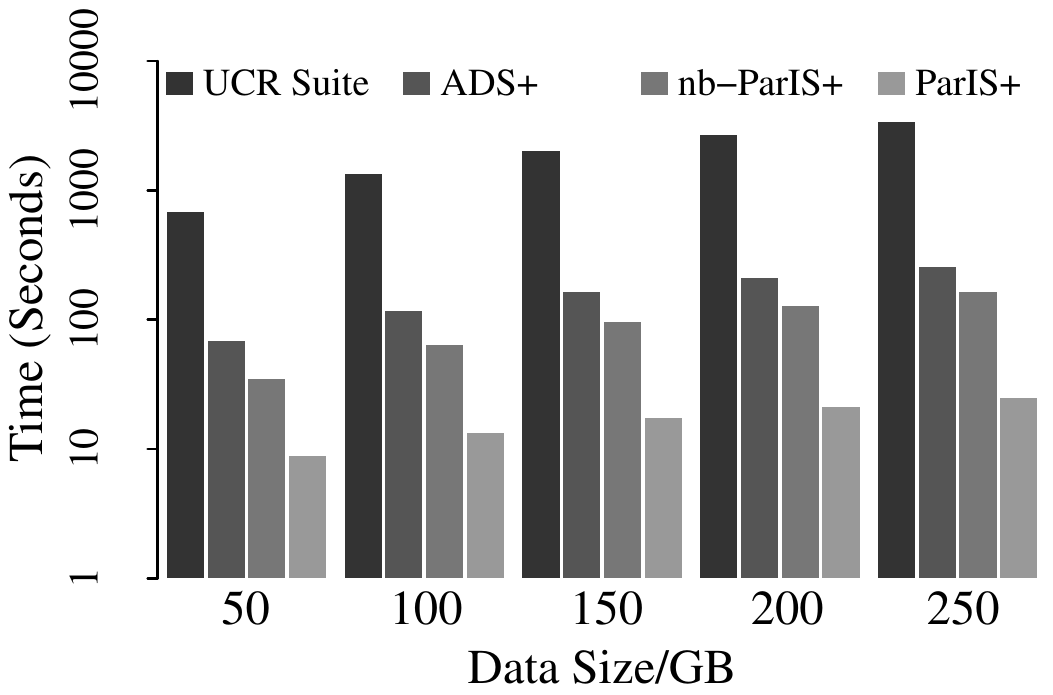}%parissacalabletesterato}
	\caption{Exact query answering time (HDD), varying dataset size.}
	\label{fig:newscalablequeryhdd}
\end{minipage}
\begin{minipage}[tb]{0.25\textwidth}
	\includegraphics[width=1.03\columnwidth]{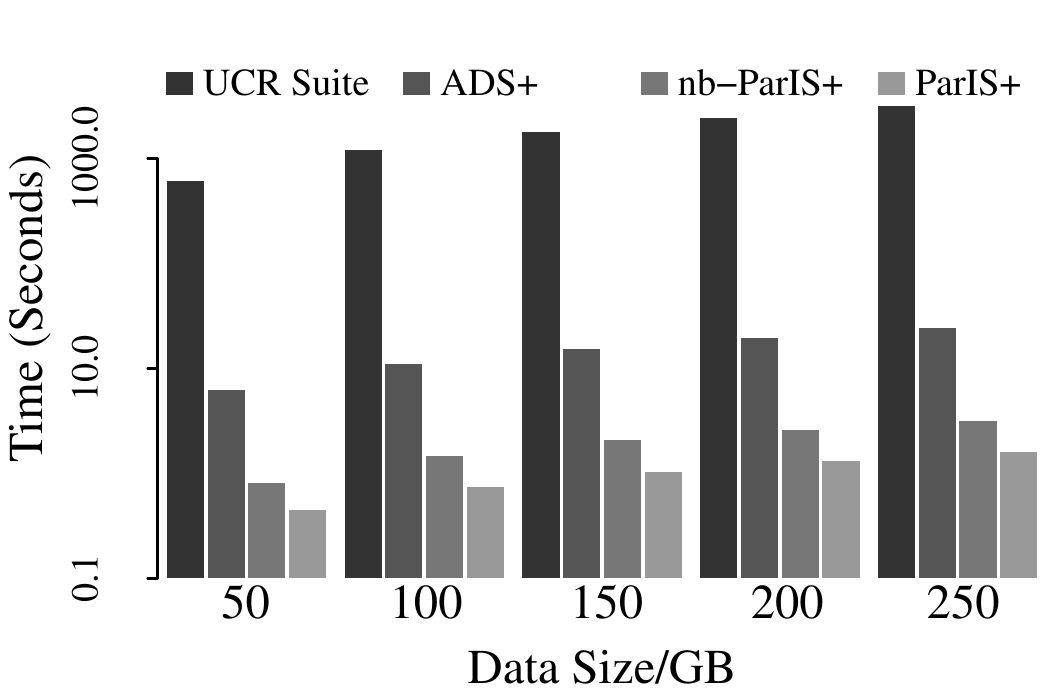}%ssdtest}
	\caption{Exact query answering time (SSD), varying dataset size.}
	\label{fig:res9}
\end{minipage}
\hspace*{0.05cm}
\begin{minipage}[tb]{\columnwidth}
	\subfigure[HDD\label{fig:knnhdd}]
	{
		\includegraphics[page=1,width=0.47\columnwidth]{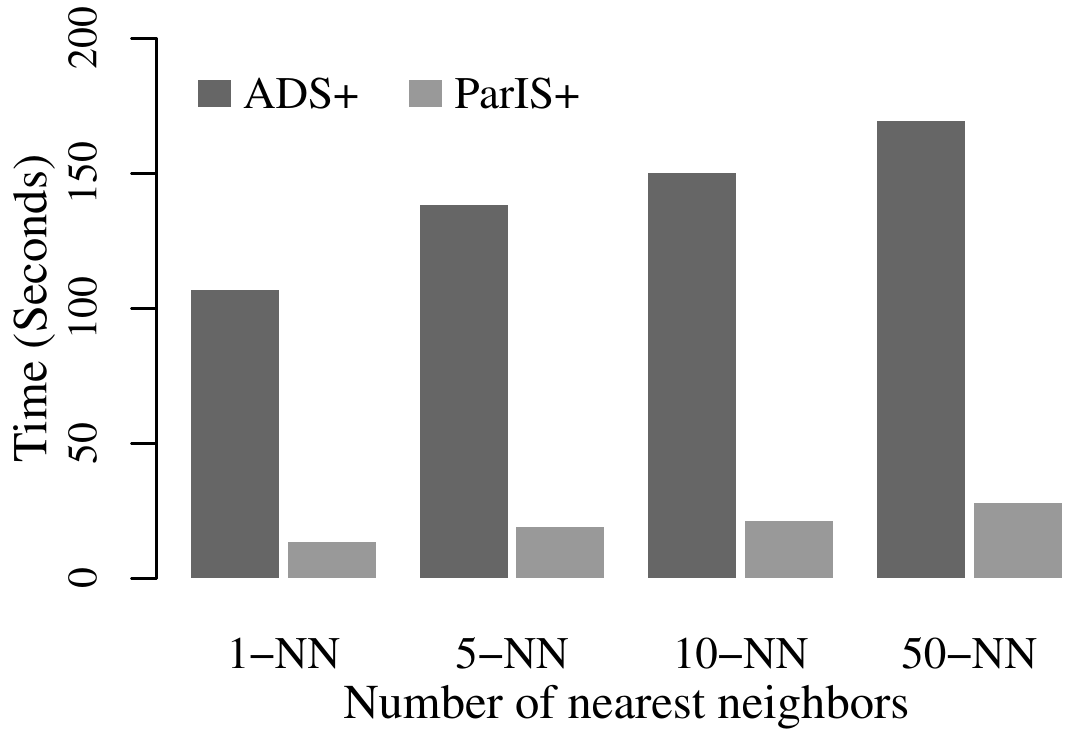}
	}
	\subfigure[SSD\label{fig:knnssd}]
	{
		\includegraphics[page=1,width=0.47\columnwidth]{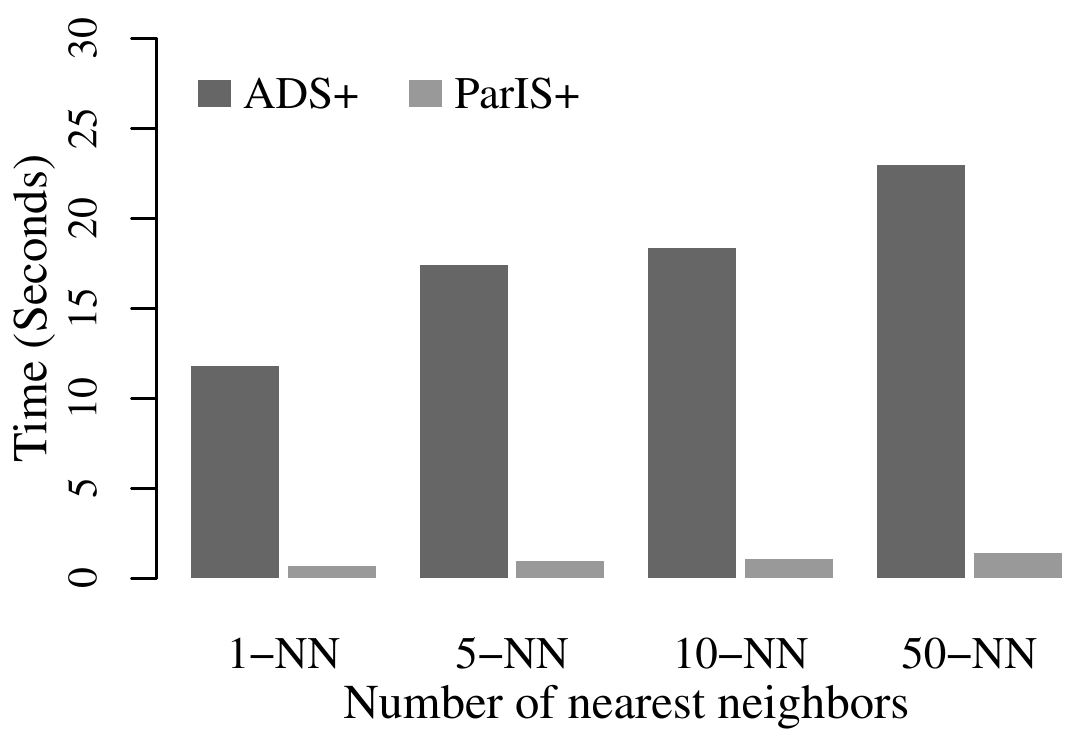}
	}
	\caption{Time for a k-NN Classifier that uses ADS+/ParIS to classify one object (100GB dataset).} 
\end{minipage}
\begin{minipage}[tb]{0.58\textwidth}
	\subfigure[Index creation time\label{fig:realindex}] {
	\includegraphics[width=0.25\columnwidth]{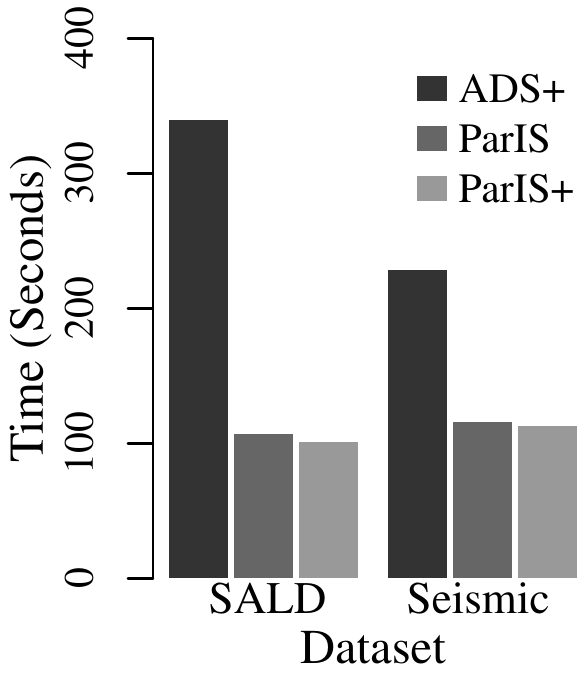}%parisrealindex}
		}
	\subfigure[Exact query answering time (HDD)\label{fig:allcompared}] {
	\includegraphics[width=0.32\columnwidth]{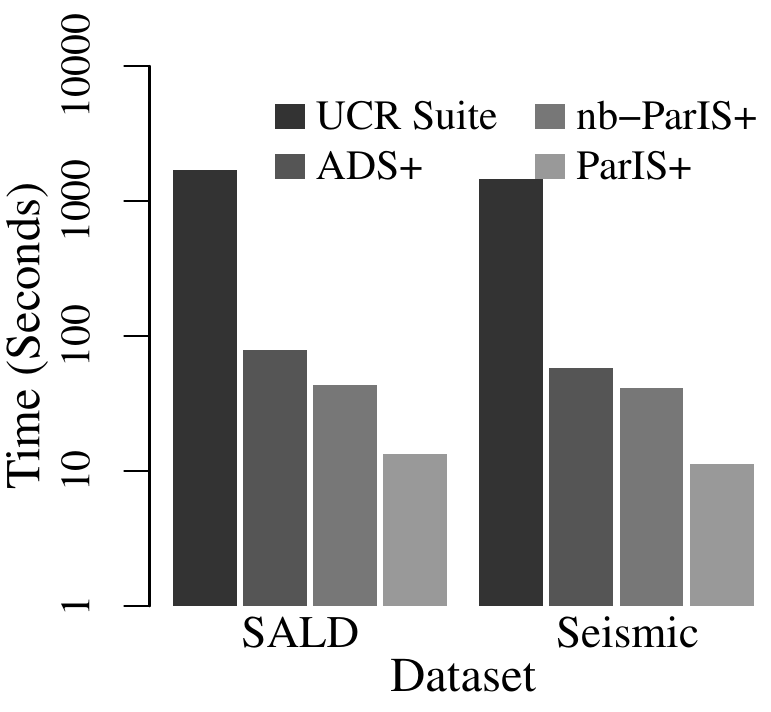}%realquery}
		}
	\subfigure[Exact query answering time (SSD) \label{fig:ssdallcompare}] {
		\includegraphics[width=0.32\columnwidth]{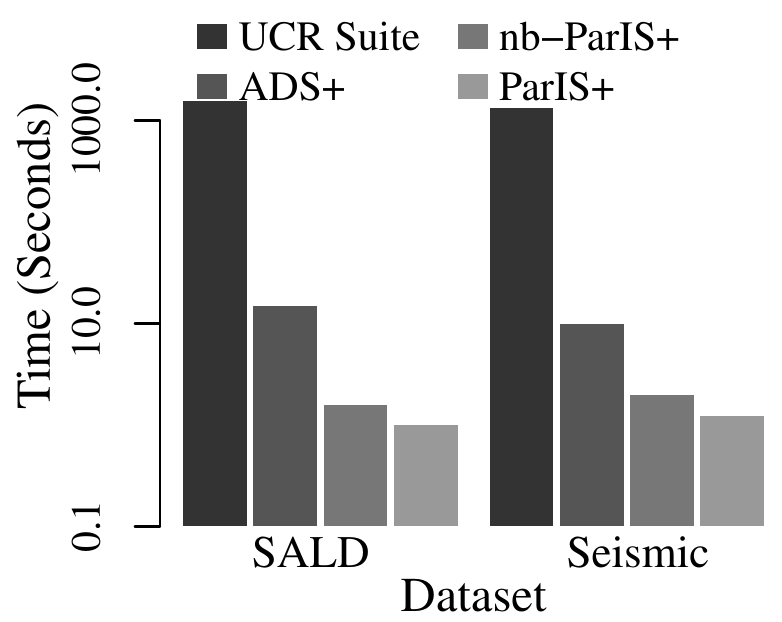}%smallssdrealtest}
		}
%	\subfigure[Query answering time
%\label{fig:faster1}]{
	%\includegraphics[page=1,width=\columnwidth]{newSALDquery}
%}
	\caption{Time cost for index creation and similarity search for real data.} % (24 cores in 2 CPUs)}
\end{minipage}
\hspace*{-0.15cm}
\begin{minipage}[tb]{0.88\columnwidth}
			\subfigure[BSF Updates\label{fig:updatebsfnumber}] {
			\includegraphics[width=0.45\columnwidth]{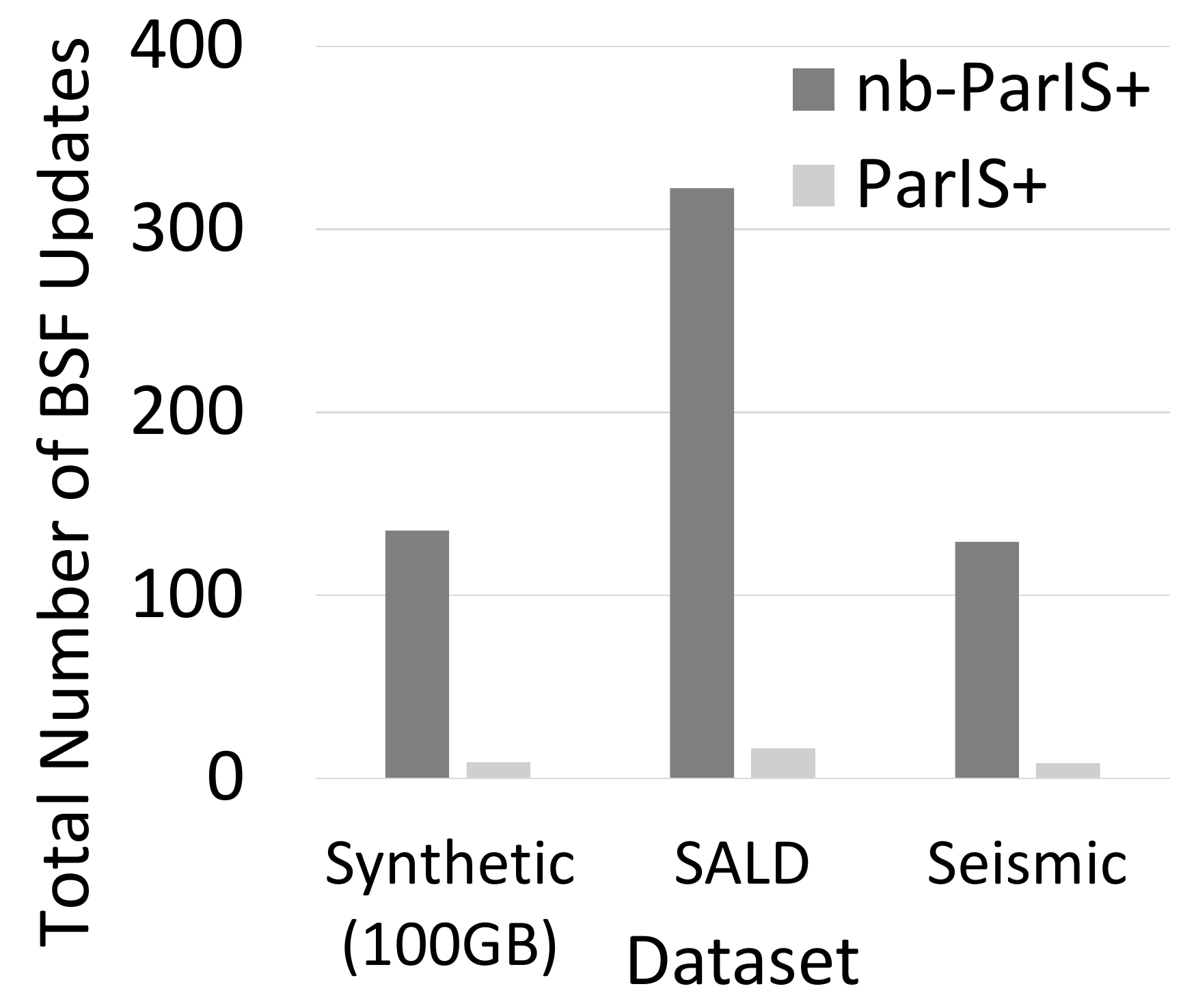}
				}
				%		\hspace*{0.5cm}
		\subfigure[Raw Series to Read\label{fig:notprunednumber}] {
			\includegraphics[width=0.45\columnwidth]{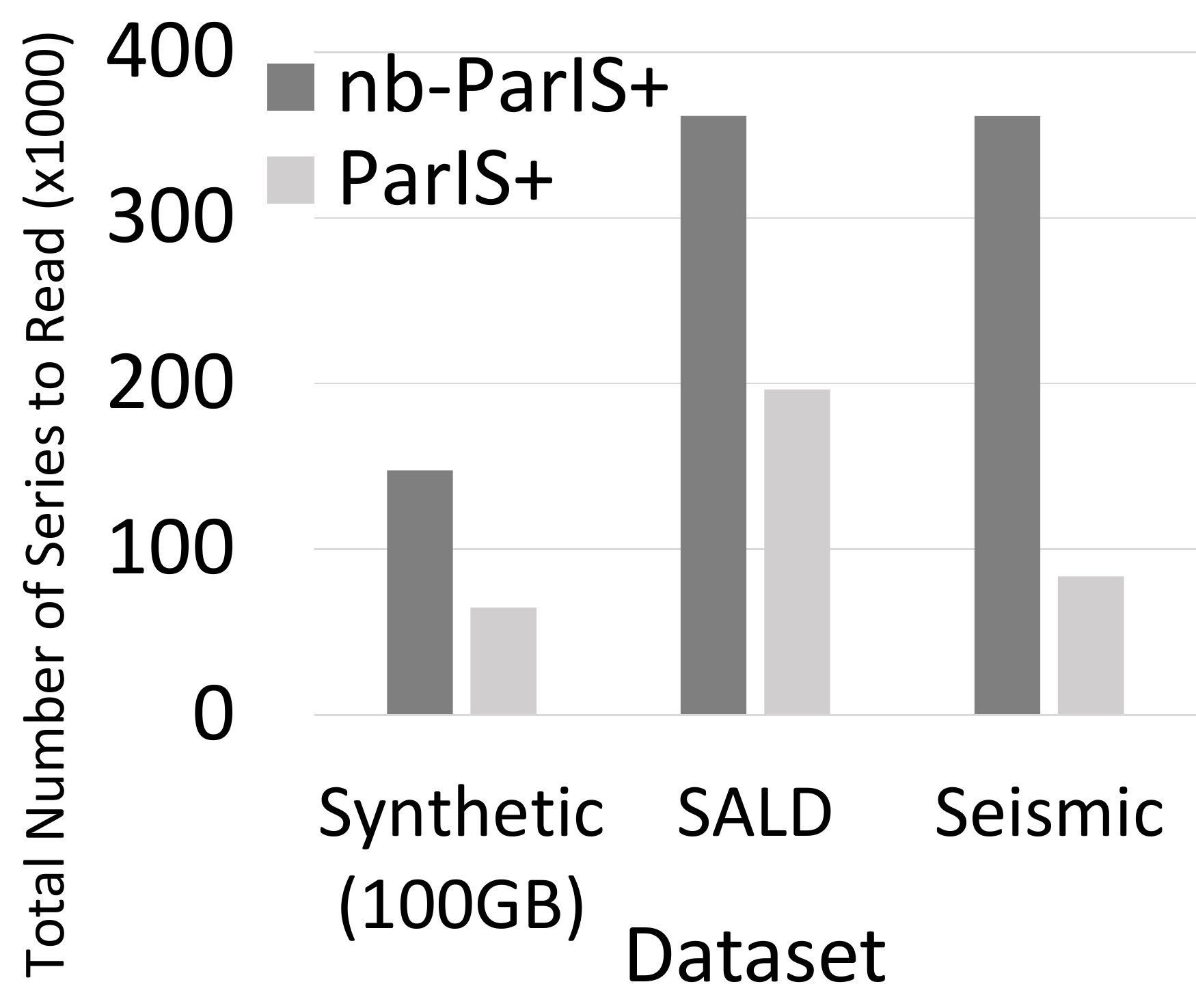}
				}
			\caption{Effort of ParIS and nb-ParIS+ (number of non-pruned raw data series).}
			\label{fig:notpruned}
	\end{minipage}

\end{figure*}

We now turn our attention to 
%the evaluation of the performance of our approach on 
datasets of increasing size,
% (which demostrate the dataset scalability characteristics of ParIS).
%In this case, we 
and additionally compare ParIS to another competitive data series index, DS-Tree.
%, as well as to our extension, UCR Suite-P (described earlier). 
%\here{Y: The description of UCR Suite-P is commented out. Shall we bring it back or the reference to UCR Suite-P 
%should be removed from here?}
Figures~\ref{fig:res6} and~\ref{fig:res6_2} depict the results for HDD and SSD, respectively.
%, for dataset sizes ranging from 50GB to 250GB.
The results show that the performance of ParIS and ParIS+ is always better than that of ADS+ and DS-Tree. Moreover, 
ParIS+ is always faster than ParIS. 
This improvement is up to 7\% on HDD. 
However, it is smaller on SSD because the SSD I/O bandwidth in our server is smaller than that of the HDD, resulting in higher read cost. 
However, the time to build the tree index does not change, and therefore, it now accounts for a smaller percentage of the I/O time.
%with the improvement increasing for larger datasets. 
Note that the DS-Tree is always one order of magnitude 
slower than the other approaches, 
so we do not consider the DS-Tree in the rest of our experiments.

%Finally, we studied the influence of the double buffer size on index creation performance. %Figure~\ref{fig:doublebuffer} show the result, 
%The results (which we omit for brevity) show that the time cost reduces and then stabilizes once the buffer size gets larger than 2MB. 
%In all our experiments, we set 2MB as the double buffer size. 

\subsubsection{Query Answering Performance Evaluation}

We now present results on %that demonstrate 
ParIS+'s efficiency 
in query answering. %; in summary, ParIS+ is more than 10x faster than ADS+, and up to 1000x faster than UCR Suite. 

\remove{
%
%\subsubsection{Scalability Evaluation}
%
We 
%now turn our attention to scalability experiments, in order to 
evaluate the performance of our approach on datasets of increasing size. 
In this case, we additionally compare ParIS+ and \textcolor{red}{ParIS+nb} to the state-of-the-art serial scan solution, UCR Suite, which is often used as a baseline.
%, as well as to our extension, UCR Suite-P (described earlier). 
%\here{Y: The description of UCR Suite-P is commented out. Shall we bring it back or the reference to UCR Suite-P 
%should be removed from here?}
\textcolor{red}{Figure~\ref{fig:res6} and \ref{fig:res6_2} compares the index creation performance between DS-Tree, ParIS,ParIS+ and ADS+ on HDD and SSD , for dataset sizes ranging from 50GB to 250GB.
The ParIS algorithm is up to 2.5x faster than ADS+ on index creation, with the improvement increasing for larger datasets. By the way, DS-Tree always one order slower than others so that we do not consider DS-Tree any further. We can see that ParIS+ is faster than ParIS. }
}

%\begin{figure}[tb]%
%	\centering
%	\includegraphics[page=1,width=\columnwidth]{parisvarycoretest}
%	\caption{\textcolor{red}{Time cost of ParIS-nb exact query answering}}
%	\label{fig:res2}
%\end{figure}

Figure~\ref{fig:res3_2} shows the exact query answering time for ParIS+, nb-ParIS+, and ADS+, as we vary the number of cores. % (where just one thread runs on each core). 
We observe that the performance improves as we increase the number of cores 
(though the improvement is rather small when we go beyond 6 cores). 
For example, for 24 cores, nb-ParIS+ is no more than 2 times faster than ADS+, whereas ParIS+ is almost 6 times faster than ADS+. 

Figure~\ref{fig:detailofLbcRdc} shows how the time
for executing the two stages of query answering in ParIS+ is influenced as we increase the number of cores and the number of threads running on each core.
%Specifically, Figure~\ref{fig:detailofLBDW} focuses on the time needed for the computation performed by the LBC workers, whereas Figures~\ref{fig:detailofRDCWSSD} and~\ref{fig:detailofRDCWHDD} show the influence on the time required by the RDC workers for HDD and SSD, respectively. 
The results show that the LBC workers execution time decreases as the number of cores increases, with the degree of oversubscribing not playing an important role in performance (Figure~\ref{fig:detailofLBDW}). 
On the contrary, for the RDC execution time 
the degree of oversubscribing is crucial, both for HDD (Figures~\ref{fig:detailofRDCWHDD}) and SSD (Figure~\ref{fig:detailofRDCWSSD}).
%(whereas the number
%of cores does not play any significant role in performance). 
The reason is that the LBC workers perform in memory computations, for which it is important to use more cores to execute them faster. 
On the other hand, the RDC workers perform I/O to read the required data from disk, and thus, oversubscribing is useful to keep the CPU busy at all times. 
These diagrams justify the use of $1$ LBC worker per core and $5$ RDC workers per core, which are the default values we have used here. % in our experiments. 

%\textcolor{red}{In Figure~\ref{fig:res3_2}, we can also see the average time cost of query answering. 

Figure~\ref{fig:newscalablequeryhdd} (log-scale y-axis) shows the performance of query answering 
for UCR Suite, ADS+, nb-ParIS+, and ParIS+ as the dataset size increases. 
%, for disk-resident data. 
We observe that nb-ParIS+ is about 2 times faster than ADS+ and about 20 times faster 
than UCR Suite in general. 
ParIS+ is much better than this: it is one order of magnitude faster than ADS+, 
and more than two orders of magnitude faster than UCR Suite. 
%\textcolor{red}{ParIS+ is better than nb-ParIS+}. % for the 250GB dataset.
%The ParIS-nb algorithm is 1.5x faster that ADS+, while ParIS achieves a speedup of 10x. 
We also note that the performance improvement of ParIS+ gets larger with increasing dataset sizes, 
so ParIS+ is able to scale better than UCR Suite. 
%
%ADS+ is 10x faster than UCR Suite. 
%ParIS-nb and ParIS are 20x and 139x faster than the serial scan. 
%The results also show that ParIS is affected much less by the increasing dataset size than UCR Suite.
This is because ParIS+ can effectively prune the search space, while UCR Suite always has to read all the data from disk. 
%Note that ParIS+ also exhibits better scalability than ADS+. 
%This is because the degree of acceleration of some of the computations performed by ParIS+ depends on the dataset size. 

%
% XXXXXXXXXXXXXXXXXXXXXXXXXX       SSD query answering       XXXXXXXXXXXXXXXXXXXXXXXXX
Figure~\ref{fig:res9} (log-scale y-axis) shows the performance of exact query answering for the SSD server. 
All three algorithms, ADS+, nb-ParIS+, and ParIS+ benefit from the SSD's low random access latency. 
%\textcolor{red}{Especially for nb-ParIS+, it's the improvement is lager than before be compare with ADS+.}
% of the SSD disks. 
%ParIS is always more than one order of magnitude faster than ADS+.
The performance improvement of ParIS+ is increasing with the size of the dataset (since the number of random disk accesses increases, too), achieving in our experiments performance up to 15x faster than ADS+, and 2000x faster than UCR Suite. 
(Note that nb-ParIS+ results in lower numbers: 
it is about 7.5x faster than ADS+ and up to 1000x faster than UCR Suite.)

\noindent\textbf{[Vectorial (SIMD) Lower Bound Distance Calculation]}
In order to evaluate the effect on performance of our new lower bound distance calculation function that uses SIMD, we conducted an experiment that factors out the disk I/O cost: 
we measured the execution time of exact similarity search when all data are loaded in main memory.
We compared our solution to the case where all computations 
are performed using Single Instruction Single Data (SISD).
%Table~\ref{fig:res4} shows
The results (refer to Table~\ref{table1} show that the average time cost per lower-bounding 
calculation when using SIMD is 3.5x faster than the SISD solution.
This is a non-negligible speedup, attributed to the large number 
of vectorial computations %that need to be 
executed in %the context of 
data series similarity search (refer to Algorithm~\ref{algo8}).

\begin{table}[tb]
	\centering
		\caption{Time cost of lower bound distance calculations. }\label{table1}
	\begin{tabular}{cc}
		\toprule
		Implementation technology& Time (nanoseconds)\\
		\midrule
		SISD& 107.5\\
		SIMD& 31.142\\
		\bottomrule
	\end{tabular}
\end{table}

\subsubsection{Real Datasets}

%In this set of experiments, we test the algorithms using real datasets.
Figure~\ref{fig:realindex} shows the result of index creation time cost %for ADS+, ParIS and ParIS+ 
on the SALD and Seismic real datasets.
Similar to our previous results, ParIS+ is faster than ADS+ during index creation: ParIS+ is up to 2.4x faster for SALD, and 2x faster for Seismic. 
Moreover, ParIS+ is slightly faster than ParIS, as expected.
Figure~\ref{fig:allcompared} (log scale y-axis) reports the exact similarity search time cost on HDD for UCR Suite, ADS+, nb-ParIS+, and ParIS+. 
The results differ for the two real datasets.
For SALD, ParIS+ is 140x faster than UCR Suite and 4x faster than ADS+, while for Seismic, ParIS+ is 130x faster than UCR Suite and 5x faster than ADS+.
The SSD experiments show similar, yet more pronounced trends (Figure~\ref{fig:ssdallcompare}, log scale y-axis): ParIS+ is almost 1 order of magnitude faster than ADS+, and 3 orders of magnitude faster than UCR Suite.

We also observe that nb-ParIS+ is much slower than ParIS+ on HDD.
Apart from the fact that nb-ParIS+ does not manage in an optimal way the disk read operations and the corresponding load in the individual worker threads, the major reason for its low performance is the lack of communication among the nb-Paris+ worker threads during the execution of the similarity search algorithm.
(Recall that during query answering in nb-ParIS+, the workers have a local copy of the BSF, while in ParIS+, all workers share a common copy of BSF.)
Consequently, when one worker finds a better BSF value that can help prune more data series, this value (contrary to ParIS+) is not shared with the rest of the workers, who perform unnecessary expensive disk read operations. 
%However, the performance of ParIS+ and nb-ParIS+ is similar on SSD in most cases.
%
%It is interesting to note that ParIS-nb performs faster than ADS+ for Seismic, but not for Astronomy. 
%Apart from the fact the ParIS-nb does not manage in an optimal way the disk read operations and the corresponding load in the individual worker threads, the major reason for its low performance on Astronomy is the lack of communication among the worker threads during the execution of its similarity search algorithm.
%Consequently, when one worker finds a better BSF value that can help prune more data series, this value (contrary to ParIS) is not shared with the rest of the workers, who perform unnecessary expensive disk read operations. 
%This situation is less prominent with the Seismic dataset, where lower bound pruning is more effective.
%
%\subsubsection{ParIS+ vs nb-ParIS+}
\label{sec:battle}
%We now study in more detail these performance differences of ParIS+ and nb-ParIS+. 
%Our previous experiments showed that they perform equally for SSD-resident data, but ParIS is faster than ParIS for HDD-resident data.
Figures~\ref{fig:updatebsfnumber} and~\ref{fig:notprunednumber}, illustrate the above observation.
%In these experiments, we measure the number of BSF updates, as well as the number of raw series that were not pruned, and had to be read and processed. 
The results show that the extra work for sharing a common BSF pays off for ParIS+, since it leads to both a smaller number of BSF updates (i.e., we arrive to a better BSF earlier), and a reduced number of raw data to read (i.e., we prune more). 
This gives ParIS+ an edge that is more pronounced on the HDD server, rather than on the SSD one: having to read fewer raw data translates to a smaller number of the expensive HDD seek/rotation operations.

We note that it is possible for nb-ParIS+ to perform better than ParIS+: this happens when the queries are very hard~\cite{zoumpatianos2018generating} and the resulting pruning ratio is small.
In such cases, the creation and manipulation of the (long) candidate list results in high overheads, while the benefit of having all RDC worker threads of ParIS+ communicating in order to update the BSF, which leads to saving some real distance computations, is not significant (results omitted for brevity).
Usually though, the query workload is not very hard overall, which justifies the use of ParIS+ as the method of choice.

\section{Related Work}
\label{sec:related}
\noindent{\bf [Data series summarization and indexing]}
Various dimensionality reduction techniques exist 
%to transform data series into summarizations that enable approximating and lower bounding the distance between any two data series.  Examples include generic Discrete Fourier Transforms (DFT)~\cite{Agrawal1993,DBLP:conf/sigmod/FaloutsosRM94,Rafiei1997,Rafiei1998}, Piecewise Linear Approximation (PLA)~\cite{Keogh1997}, Singular Value Decomposition (SVD)~\cite{Korn97,RaviKanth1998}, Discrete Haar Wavelet Transforms (DHWT)~\cite{ChanF99, DBLP:conf/kdd/KashyapK11}, Piecewise Constant Approximation (PCA), and Adaptive Piecewise Constant Approximation (APCA)~\cite{Chakrabarti2002}, as well as data series specific techniques such as Piecewise Aggregate Approximation (PAA)~\cite{Keogh2000}, Symbolic Aggregate approXimation (SAX)~\cite{Lin2003} and the indexable Symbolic Aggregate approXimation ($i$SAX)~\cite{shieh2008sax,Camerra2010}.
%These smaller summarizations can 
for data series, which can then 
be scanned and filtered~\cite{DBLP:conf/kdd/KashyapK11,Li1996} or indexed and pruned~\cite{Guttman1984, Assent2008, wang2013data, shieh2008sax,Shieh2009, zoumpatianos2016ads,DBLP:journals/pvldb/KondylakisDZP18,ulissevldb,evolutionofanindex}
%,DBLP:conf/icdm/YagoubiAMP17,ulisse} 
%to avoid accessing parts of the data that do not contain the nearest neighbor. 
during query answering.
We follow the same approach of indexing the series based on their summaries,
%a smaller summarization to enable pruning, 
though our work is the first to exploit the parallelization opportunities offered by multi-core architectures, 
%namely, the SIMD, multi-core and multi-socket architectures, 
in order to accelerate data series index construction and similarity search. 
FastQuery is an approach used to accelerate search operations in scientific data~\cite{chou2011fastquery}, 
%This approach is geared towards scientific applications that produce and consume huge volumes of data, and avoids the time-consuming process of uploading these data into a data management system.
%%FastQuery exploits the many-core architecture, and more specifically the threading technology, for parallelizing processing using the array data model.
%%Instead, the focus of our work is on indexes for data series similarity search.
%FastQuery is 
based on the construction of bitmap indices.
% on different features of the data. 
In essence, the iSAX summarization used in our approach is an equivalent solution, though, specifically designed for sequences. % (which have high dimensionalities).

%\subsection{Distance calculation in SIMD}
%\textbf{Distance calculation in SIMD: }
\noindent{\bf [Data structures for SIMD]}
While the interest in using SIMD for improving %the 
performance %of data management solutions 
is not new~\cite{zhou2002implementing}, there are still many algorithms that do not take advantage of this hardware characteristic. 
%\subsection{SIMD-friendly B+-Tree index}
%\textbf{SIMD-friendly B+-Tree Index: }
The problem of developing a SIMD-friendly B+-Tree index was 
%only very 
recently studied~\cite{ZeuchFH14}, 
%The authors of this study focused 
with a focus on a basic B+-Tree method, the k-ary search algorithm. 
%The main idea was to make the inner node of the index as large as the SIMD register, and use bit-mask conditional branch instructions to perform the search in the SIMD. 
%%\textcolor{red}{And they present two algorithms for linearizing a sorted list of keys to discovery the good search strategies base on SIMD. 
%%In my paper, I use the idea of bit mask to do the conditional branch calculation in lower bounding distance calculation.}
%%
For data series in particular, previous work has used SIMD for Euclidean distance computations~\cite{tang2016exploit}. 
In our work, we go beyond this straightforward use of SIMD, and we propose an algorithm that uses SIMD for the computation of lower bounds, which involve branching operations.

%\subsection{Multi-core}
%\textbf{Multi-Core Systems: }
\noindent{\bf [Modern Hardware]}
Multi-core CPUs offer thread parallelism through multiple cores and simultaneous multi-threading (SMT).
Thread-Level Parallelism (TLP) methods, like multiple independent cores and hyper-threads are commonly used to increase algorithm efficiency~\cite{gepner2006multi}.
%Modern multi-core CPUs have individual L1- and L2-caches, and shared L3-cache~\cite{prinslow2011overview}, which is beneficial for the cooperation among cores~\cite{zhang2010does}.
%\subsection{High performance temporal indexing}
%\textbf{High Performance Temporal Indexing: }
A recent study proposed a high performance temporal index similar to time-split B-tree (TSB-tree), called TSBw-tree, which focuses on transaction time databases~\cite{LometN15}. 
%\textcolor{red}{It use LLAMA subsystem~\cite{levandoski2013llama} which only recycle the previously flushed part of main memory(PS: the original sentence is: LLAMA can reclaim main memory by dropping only previously flushed portions of pages from memory,thus not requiring any I/O, even when swapping out “dirty” pages.) and use log-structuring to cut the data swapping between memory and storage, avoid random write and reduce the cost of write to handle the conflict of reader and writer.
%It remove the conflict of reader and writer because it layers on LLAMA subsystem. 
%So that }this system can avoid the splitting during moving historical node and trouble splitting index pages by their mapping table architecture. {\bf ??? rewrite the last two sentences; it is not clear what we mean; readers do not know LLAMA; explain the idea that makes it work ???}
%The proposed solution is based on a latch-free, page-oriented cache and storage manager, which is optimized for modern hardware.
However, this is designed for temporal data, which are 2-dimensional, while in our case, data series can have thousands of dimensions (i.e., the length of the sequence).
%
%\subsection{Similarity Searches on the GPU}
%\textbf{Similarity Searches on the GPU: }
Graphics Processing Units (GPUs) are another modern hardware option, which allows for massively parallel computations. 
A recent study described the use of GPUs for accelerating similarity search in a Trajectory Indexing system~\cite{GowanlockC16}. 
%This study proposed an efficient way to handle the memory and the communication between CPU and GPU. 
%It also proposed an indexing technique for efficient distance threshold search in the GPU. 
In this work, we do not use GPUs. %; though, it is a very interesting research direction, and deserves to be studied in its own right.

\noindent\textbf{[Scans vs indexing]}
Even though recent works have shown that sequential scans can be performed  efficiently~\cite{rakthanmanon2012searching,DBLP:conf/icdm/MueenHE14}, such techniques are applicable when the dataset consists of a single, very long data series, and queries are looking for potential matches in small subsequences of this long series. 
Such approaches, in general, do not provide any benefit when the dataset is composed of 
a large number of small data series, like in our case. 
Therefore, indexing is required in order to efficiently support data exploration tasks, 
%which involve ad-hoc queries, i.e., 
where the query workload is not known in advance.

%To solve the balancing problem, we introduce a new, balanced ExactSearch algorithm.  

\section{Conclusions}
\label{sec:conclusions}

%Data series are a very common data type, with increasingly larger collections being generated by applications in many and diverse domains.
%In this work, 
%we observed that no data series indexing approach exploits the added functionality of modern hardware.
%As a result, 
We presented ParIS and ParIS+, 
%an extension of the current state-of-the-art data series index, 
the first data series indices that exploit %the parallelism opportunities of 
multi-core architectures,
%We proposed parallel algorithms for index creation and query answering.
%First, we optimized the process of index creation, using parallelism and double buffering in such way that we %(almost) completely mask the CPU cost. 
%This technology can shade the calculating during the read time. 
%Then, we greatly accelerated the execution time of the query answering procedure, by developing algorithms that perform multiple, ordered disk access requests in parallel. 
%We experimentally validated the efficiency of the proposed approach using several synthetic and real datasets from diverse domains. 
%The results show the significant benefits of our solution, when compared to the state of the art techniques, and demonstrate our ability to manage and analyze the existing, very large data series collections, in a scalable fashion.
%The experimental evaluation with several synthetic and real datasets demonstrates the efficiency of the proposed approach, which is 
leading to performance 
2-3 orders of magnitude faster than previous approaches.
In our future work we will study in more depth parallel I/O techniques~\cite{ghodsnia2014parallel}, combine our approach with solutions developed for distributed systems~\cite{dpisaxjournal}, extend it to support the DTW distance, and study other hardware parallelization opportunities, e.g., GPUs and FPGAs.

\vspace*{0.20cm}
\noindent{\bf Acknowledgments}
Work partially supported by the Chinese Scholarship Council, FMJH Program PGMO, EDF, Thales and HIPEAC 4.
Part of work performed while P. Fatourou was visiting LIPADE, and while B. Peng was visiting CARV, FORTH ICS.

\bibliographystyle{spmpsci} 
\bibliography{ads,vldb_sample}

\begin{thebibliography}{10}
\providecommand{\url}[1]{{#1}}
\providecommand{\urlprefix}{URL }
\expandafter\ifx\csname urlstyle\endcsname\relax
  \providecommand{\doi}[1]{DOI~\discretionary{}{}{}#1}\else
  \providecommand{\doi}{DOI~\discretionary{}{}{}\begingroup
  \urlstyle{rm}\Url}\fi

\bibitem{iris}
{Incorporated Research Institutions for Seismology} -- {Seismic Data Access}.
\newblock http://ds.iris.edu/data/access/ (2016)

\bibitem{paradssources}
Source code and datasets used in this paper.
\newblock http://www.mi.parisdescartes.fr/\~{}themisp/paris/ (2018)

\bibitem{url:SALD}
Southwest university adult lifespan dataset (sald).
\newblock \url{http://fcon_1000.projects.nitrc.org/indi/retro/sald.html} (2018)

\bibitem{Agrawal1993}
Agrawal, R., Faloutsos, C., Swami, A.N.: Efficient similarity search in
  sequence databases.
\newblock In: FODO (1993)

\bibitem{ailamaki2015databases}
Ailamaki, A.: Databases and hardware: The beginning and sequel of a beautiful
  friendship.
\newblock VLDB  (2015)

\bibitem{Assent2008}
Assent, I., Krieger, R., Afschari, F., Seidl, T.: The ts-tree: efficient time
  series search and retrieval.
\newblock In: {EDBT} (2008)

\bibitem{DBLP:journals/dagstuhl-reports/BagnallCPZ19}
Bagnall, A.J., Cole, R.L., Palpanas, T., Zoumpatianos, K.: Data series
  management (dagstuhl seminar 19282).
\newblock Dagstuhl Reports \textbf{9}(7) (2019)

\bibitem{norma}
Boniol, P., Linardi, M., Roncallo, F., Palpanas, T.: {Automated Anomaly
  Detection in Large Sequences}.
\newblock In: {ICDE} (2020)

\bibitem{series2graph}
Boniol, P., Palpanas, T.: {Series2Graph: Graph-based Subsequence Anomaly
  Detection for Time Series}.
\newblock {PVLDB}  (2020)

\bibitem{seriesgoneparallel}
{Botao Peng (supervised by Panagiota Fatourou and Themis Palpanas)}: {Data
  Series Indexing Gone Parallel}.
\newblock In: {ICDE PhD Workshop} (2020)

\bibitem{isax2plus}
Camerra, A., Shieh, J., Palpanas, T., Rakthanmanon, T., Keogh, E.: {Beyond One
  Billion Time Series: Indexing and Mining Very Large Time Series Collections
  with iSAX2+}.
\newblock KAIS \textbf{39}(1), 123--151 (2014)

\bibitem{Shandola2009}
Chandola, V., Banerjee, A., Kumar, V.: Anomaly detection: A survey.
\newblock CSUR  (2009)

\bibitem{chou2011fastquery}
Chou, J., Wu, K., et~al.: Fastquery: A parallel indexing system for scientific
  data.
\newblock In: CLUSTER, pp. 455--464. IEEE (2011)

\bibitem{coorporation2009intel}
Coorporation, I.: Intel 64 and ia-32 architectures optimization reference
  manual (2016)

\bibitem{lernaeanhydra}
Echihabi, K., Zoumpatianos, K., Palpanas, T., Benbrahim, H.: The lernaean hydra
  of data series similarity search: An experimental evaluation of the state of
  the art.
\newblock {PVLDB}  (2019)

\bibitem{lernaeanhydra2}
Echihabi, K., Zoumpatianos, K., Palpanas, T., Benbrahim, H.: {Return of the
  Lernaean Hydra: Experimental Evaluation of Data Series Approximate Similarity
  Search}.
\newblock {PVLDB}  (2019)

\bibitem{gepner2006multi}
Gepner, P., Kowalik, M.F.: Multi-core processors: New way to achieve high
  system performance.
\newblock In: PAR ELEC (2006)

\bibitem{ghodsnia2014parallel}
Ghodsnia, P., Bowman, I.T., Nica, A.: Parallel i/o aware query optimization.
\newblock In: SIGMOD. ACM (2014)

\bibitem{DBLP:conf/edbt/GogolouTPB19}
Gogolou, A., Tsandilas, T., Palpanas, T., Bezerianos, A.: Progressive
  similarity search on time series data.
\newblock In: Proceedings of the Workshops of the {EDBT/ICDT} Joint Conference
  (2019)

\bibitem{GowanlockC16}
Gowanlock, M.G., Casanova, H.: Distance threshold similarity searches:
  Efficient trajectory indexing on the {GPU}.
\newblock {IEEE} Trans. Parallel Distrib. Syst. \textbf{27}(9) (2016)

\bibitem{Guttman1984}
Guttman, A.: R-trees: {A} dynamic index structure for spatial searching.
\newblock In: SIGMOD, pp. 47--57 (1984)

\bibitem{DBLP:conf/kdd/KashyapK11}
Kashyap, S., Karras, P.: Scalable knn search on vertically stored time series.
\newblock In: SIGKDD, pp. 1334--1342 (2011)

\bibitem{keogh2001dimensionality}
Keogh, E., Chakrabarti, K., Pazzani, M., Mehrotra, S.: Dimensionality reduction
  for fast similarity search in large time series databases.
\newblock KIS  (2001)

\bibitem{DBLP:journals/pvldb/KondylakisDZP18}
Kondylakis, H., Dayan, N., Zoumpatianos, K., Palpanas, T.: Coconut: {A}
  scalable bottom-up approach for building data series indexes.
\newblock {PVLDB} \textbf{11}(6), 677--690 (2018)

\bibitem{coconutpalm}
Kondylakis, H., Dayan, N., Zoumpatianos, K., Palpanas, T.: Coconut palm: Static
  and streaming data series exploration now in your palm.
\newblock In: {SIGMOD} (2019)

\bibitem{Li1996}
Li, C., Yu, P.S., Castelli, V.: Hierarchyscan: {A} hierarchical similarity
  search algorithm for databases of long sequences.
\newblock In: ICDE, pp. 546--553 (1996)

\bibitem{ulisseicde}
Linardi, M., Palpanas, T.: Ulisse: Ultra compact index for variable-length
  similarity search in data series.
\newblock In: {ICDE} (2018)

\bibitem{ulissevldb}
Linardi, M., Palpanas, T.: Scalable, variable-length similarity search in data
  series: The ulisse approach.
\newblock {PVLDB}  (2019)

\bibitem{LometN15}
Lomet, D.B., Nawab, F.: High performance temporal indexing on modern hardware.
\newblock In: {ICDE} (2015)

\bibitem{lomont2011introduction}
Lomont, C.: Introduction to intel advanced vector extensions.
\newblock Intel White Paper pp. 1--21 (2011)

\bibitem{DBLP:conf/icdm/MueenHE14}
Mueen, A., Hamooni, H., Estrada, T.: Time series join on subsequence
  correlation.
\newblock In: ICDM, pp. 450--459 (2014)

\bibitem{DBLP:journals/datamine/MueenKZCWS11}
Mueen, A., Keogh, E.J., Zhu, Q., Cash, S., Westover, M.B., Shamlo, N.B.: A
  disk-aware algorithm for time series motif discovery.
\newblock DAMI \textbf{22}(1-2), 73--105 (2011)

\bibitem{MueenNL10}
Mueen, A., Nath, S., Liu, J.: Fast approximate correlation for massive
  time-series data.
\newblock In: SIGMOD (2010)

\bibitem{DBLP:journals/sigmod/Palpanas15}
Palpanas, T.: Data series management: The road to big sequence analytics.
\newblock {SIGMOD} Record  (2015)

\bibitem{DBLP:conf/ieeehpcs/Palpanas17}
Palpanas, T.: The parallel and distributed future of data series mining.
\newblock In: HPCS (2017)

\bibitem{evolutionofanindex}
Palpanas, T.: {Evolution of a Data Series Index}.
\newblock {Communications in Computer and Information Science (CCIS)}  (2020)

\bibitem{Palpanas2019}
Palpanas, T., Beckmann, V.: {Report on the First and Second Interdisciplinary
  Time Series Analysis Workshop (ITISA)}.
\newblock {SIGMOD Rec.} \textbf{48}(3) (2019)

\bibitem{DBLP:conf/bigdataconf/PengFP18}
Peng, B., Fatourou, P., Palpanas, T.: Paris: The next destination for fast data
  series indexing and query answering.
\newblock In: {IEEE} BigData (2018)

\bibitem{messi}
Peng, B., Fatourou, P., Palpanas, T.: {MESSI: In-Memory Data Series Indexing}.
\newblock In: {ICDE} (2020)

\bibitem{rakthanmanon2012searching}
Rakthanmanon, T., Campana, B.J.L., Mueen, A., Batista, G.E.A.P.A., Westover,
  M.B., Zhu, Q., Zakaria, J., Keogh, E.J.: Searching and mining trillions of
  time series subsequences under dynamic time warping.
\newblock In: SIGKDD (2012)

\bibitem{rakthanmanon2011}
Rakthanmanon, T., Keogh, E.J., Lonardi, S., Evans, S.: Time series epenthesis:
  Clustering time series streams requires ignoring some data.
\newblock In: ICDM, pp. 547--556 (2011)

\bibitem{shieh2008sax}
Shieh, J., Keogh, E.: i sax: indexing and mining terabyte sized time series.
\newblock In: SIGKDD (2008)

\bibitem{Shieh2009}
Shieh, J., Keogh, E.: {iSAX: disk-aware mining and indexing of massive time
  series datasets}.
\newblock DMKD (1) (2009)

\bibitem{tan2017indexing}
Tan, C.W., Webb, G.I., Petitjean, F.: Indexing and classifying gigabytes of
  time series under time warping.
\newblock In: {ICDM} (2017)

\bibitem{tang2016exploit}
Tang, B., Yiu, M.L., Li, Y., et~al.: Exploit every cycle: Vectorized time
  series algorithms on modern commodity cpus.
\newblock In: IMDM (2016)

\bibitem{wang2013data}
Wang, Y., Wang, P., Pei, J., Wang, W., Huang, S.: A data-adaptive and dynamic
  segmentation index for whole matching on time series.
\newblock VLDB \textbf{6}(10) (2013)

\bibitem{xiao2013parallelizing}
Xiao, L., Zheng, Y., Tang, W., Yao, G., Ruan, L.: Parallelizing dynamic time
  warping algorithm using prefix computations on gpu.
\newblock In: (HPCC\_EUC), pp. 294--299. IEEE (2013)

\bibitem{DBLP:conf/icdm/YagoubiAMP17}
Yagoubi, D.E., Akbarinia, R., Masseglia, F., Palpanas, T.: Dpisax: Massively
  distributed partitioned isax.
\newblock In: ICDM (2017)

\bibitem{dpisaxjournal}
Yagoubi, D.E., Akbarinia, R., Masseglia, F., Palpanas, T.: Massively
  distributed time series indexing and querying.
\newblock {TKDE} \textbf{32}(1) (2019)

\bibitem{yi2000fast}
Yi, B.K., Faloutsos, C.: Fast time sequence indexing for arbitrary lp norms.
\newblock In: VLDB (2000)

\bibitem{ZeuchFH14}
Zeuch, S., Freytag, J., Huber, F.: Adapting tree structures for processing with
  {SIMD} instructions.
\newblock In: EDBT (2014)

\bibitem{zhou2002implementing}
Zhou, J., Ross, K.A.: Implementing database operations using simd instructions.
\newblock In: SIGMOD. ACM (2002)

\bibitem{zoumpatianos2016ads}
Zoumpatianos, K., Idreos, S., Palpanas, T.: Ads: the adaptive data series
  index.
\newblock {VLDB} J. \textbf{25}(6) (2016)

\bibitem{zoumpatianos2018generating}
Zoumpatianos, K., Lou, Y., Ileana, I., Palpanas, T., Gehrke, J.: Generating
  data series query workloads.
\newblock {VLDBJ 27(6)}  (2018)

\bibitem{fulfillingtheneed}
Zoumpatianos, K., Palpanas, T.: Data series management: Fulfilling the need for
  big sequence analytics.
\newblock In: ICDE (2018)

\end{thebibliography}

\end{document}